\documentclass[10pt,journal,letterpaper]{IEEEtran}
\IEEEoverridecommandlockouts
\usepackage{amsmath,amsfonts}
\usepackage{array}
\usepackage{textcomp}
\usepackage{pifont}
\usepackage{stfloats}
\usepackage{url}
\usepackage{verbatim}
\usepackage{graphicx}
\usepackage{cite}
\usepackage[noend]{algpseudocode}
\usepackage[linesnumbered,ruled]{algorithm2e}
\usepackage{upgreek}
\usepackage{subfigure}
\usepackage{color}
\def\BibTeX{{\rm B\kern-.05em{\sc i\kern-.025em b}\kern-.08em
    T\kern-.1667em\lower.7ex\hbox{E}\kern-.125emX}}
    
\usepackage{amsthm}
\usepackage{amssymb}
\usepackage{makecell}
\usepackage{cleveref}
\usepackage{enumerate}
\newtheorem{theorem}{Theorem}
\theoremstyle{definition}

\theoremstyle{proposition}
\newtheorem{proposition}{Proposition}
\theoremstyle{Assumption}

\theoremstyle{Lemma}

\graphicspath{{figures/}}
\usepackage{enumitem}

\newcommand{\myref}[1]{%
    \ifthenelse{\equal{#1}{proof_proposition_1}}{A}{%
    \ifthenelse{\equal{#1}{proof_lemma_0}}{B}{%
    \ifthenelse{\equal{#1}{proof_lemma_2}}{C}{%
    \ifthenelse{\equal{#1}{proof_remark_2}}{D}{%
    \ifthenelse{\equal{#1}{proof_theorem_3}}{E}{%
    \ifthenelse{\equal{#1}{proof_theorem_4}}{F}{%
    {\textbf{??}}}}}}}}%
}

\ifodd 1
\definecolor{darkgreen}{RGB}{0,200,0}
\else

\fi
\begin{document}
\title{BALANCE: Hybrid Autoregressive-Speculative LLM Inference at the Network Edge
\thanks{Guanqiao Qu, Qian Chen, and Xianhao Chen are with the Department of Electrical and Computer Engineering, The University of Hong Kong, Hong Kong SAR, China (e-mail: gqqu@eee.hku.hk; qchen@eee.hku.hk; xcheneee@hku.hk). Shuo Chen is with the Electrical and Electronic Engineering Department, Imperial College, London SW7 2BT, U.K. (e-mail: shuo.chen22@imperial.ac.uk). Kin K. Leung is with the Electrical and Electronic Engineering, and Computing Departments, Imperial College, London SW7 2BT, U.K. (e-mail: kin.leung@imperial.ac.uk). The work was supported in part by the Research Grants Council of Hong Kong under Grant 27213824, Grant 17207826, and Grant CRS HKU702/24. Kin K. Leung is supported by the EPSRC grant EP/Y037243/1. \textit{(Corresponding author: Xianhao Chen.)}
}
}
\author{Guanqiao Qu,~\IEEEmembership{Graduate Student Member,~IEEE}, Shuo Chen,~\IEEEmembership{Graduate Student Member,~IEEE}, \\
Qian Chen,~\IEEEmembership{Member,~IEEE}, Kin K. Leung,~\IEEEmembership{Fellow,~IEEE}, Xianhao Chen,~\IEEEmembership{Member,~IEEE}}

\maketitle
\pagestyle{empty}
\thispagestyle{empty}

\begin{abstract}
Edge inference is a promising paradigm to provide large language model (LLM) inference services in next-generation mobile networks. LLM inference mainly relies on two approaches: Autoregressive decoding (AD) generates output tokens sequentially, resulting in long latency; Speculative decoding (SD) accelerates inference by using a small language model (SLM) to generate multiple draft tokens for LLM verification, but incurs extra memory costs. Due to this latency-memory tradeoff, neither approach alone can efficiently serve users with heterogeneous demands under limited edge computing resources. To address this challenge, we propose a hy\underline{b}rid \underline{a}utoregressive-specu\underline{la}tive infere\underline{nce} (BALANCE) framework for edge LLM inference. In BALANCE, an edge server hosts both an SLM and an LLM, admits users, assigns each admitted user to the AD or SD mode, and performs the two modes simultaneously. To maximize the number of served users, we formulate a task throughput maximization problem to jointly determine user admission and computing resource allocation between AD and SD under user latency requirements and server memory constraints. Since the problem is NP-hard, we develop a polynomial-time algorithm that transforms the original problem into two sub-problems and obtains a sub-optimal solution with a constant approximation guarantee. Experiments demonstrate that BALANCE consistently outperforms conventional AD and SD and significantly improves task throughput.
\end{abstract}


\section{Introduction}

In next-generation mobile networks, large language model (LLM) inference at the network edge is emerging as an important paradigm for delivering LLM services to users~\cite{10.1145/3809166,chen2026siftmoe}. In a typical edge LLM inference system, users upload their prompts to an edge server, which performs inference using the hosted LLM and returns the generated responses~\cite{qu2024mobile,huang2024d}. By placing LLMs on edge servers closer to end users, edge LLM inference can reduce transmission latency, compared with cloud-based serving~\cite{11044591,11262690}. As such, edge LLM inference is becoming a key enabler for real-time LLM applications~\cite{11123401,qu2024trimcaching}.

\textit{Autoregressive decoding} (AD) and \textit{speculative decoding} (SD) are two representative approaches in LLM inference. In AD, an LLM generates the response token by token, producing one new token in each forward pass based on the input and previously generated tokens~\cite{10.5555/3691938.3691945,vaswani2017attention}. However, the sequential decoding process of AD incurs high response latency~\cite{jaiswal-etal-2024-ffn}. This issue becomes more severe for tasks with long output responses and stringent latency requirements~\cite{jaiswal-etal-2024-ffn}. To reduce decoding latency, SD has been proposed~\cite{xia-etal-2024-unlocking}. In SD, a small language model (SLM) first generates multiple draft tokens via AD, and the target LLM verifies them in parallel within a single forward pass. By accepting multiple verified draft tokens at once, SD reduces the number of serial LLM forward passes. This reduction can significantly reduce response latency, since each forward pass of the SLM is typically faster than the target LLM~\cite{yin2024theoretical}. Meanwhile, SD maintains output quality, since the verification preserves the target LLM's output distribution~\cite{yan-etal-2025-decoding}. When applying SD to edge LLM inference, the SLM and LLM can be co-deployed on the edge server to serve users~\cite{svirschevski2024specexec}. However, despite its latency advantage, SD introduces higher memory overhead due to the coexistence of the SLM and LLM as well as the memory for parallel draft verification~\cite{pmlr-v267-tiwari25b}.

\begin{figure}[!t]
    \centering
	\subfigure[End-to-end latency vs. output length.]{\includegraphics[height=3.1cm, keepaspectratio]{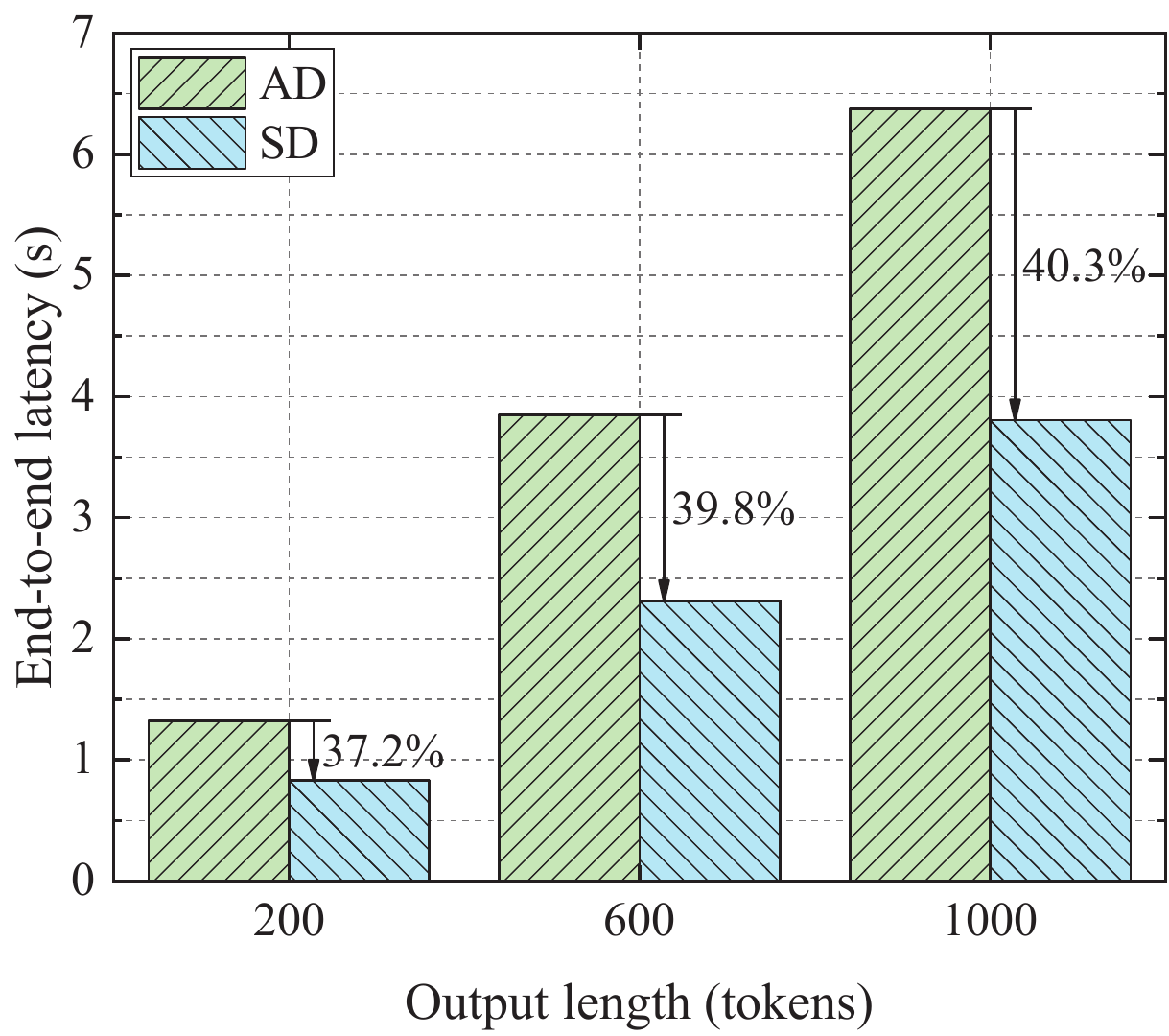}\label{fig:bw}}
    \quad
	\subfigure[Required GPU memory vs. output length.]{\includegraphics[height=3.1cm, keepaspectratio]{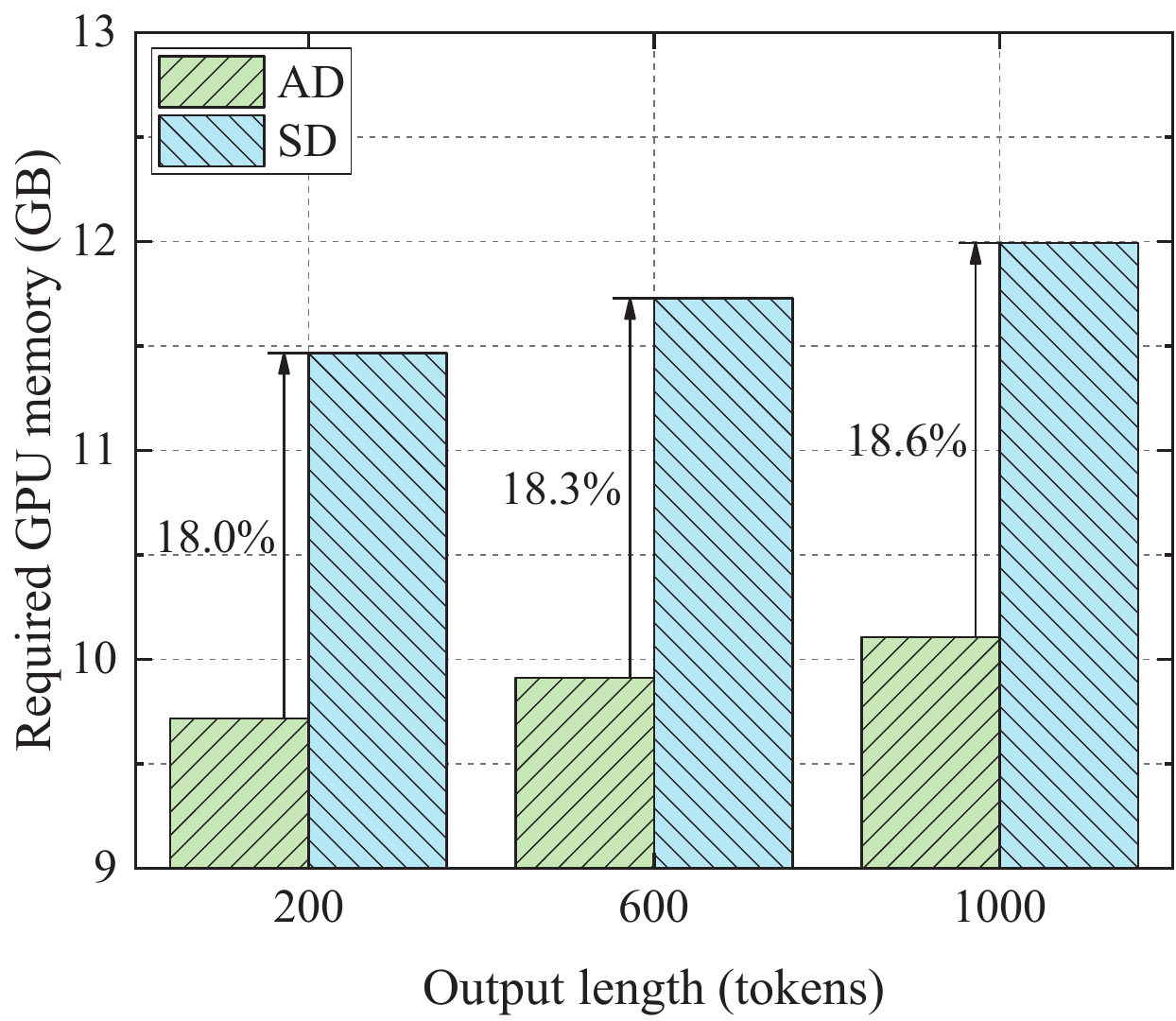}\label{fig:user}}
    \vspace{-0.25cm}
 \caption{Comparison between AD and SD when serving a single user using an NVIDIA GeForce RTX 4090 GPU, where the output length varies, and the input length is fixed at 500 tokens. In this experiment, the end-to-end latency denotes the time from feeding the prompt to the model to completing output generation. Llama-2-7B is used as the LLM in both AD and SD, and TinyLlama-1.1B is used as the SLM in SD.}
\label{fig:intro}
\vspace{-10pt}
\end{figure}

\textbf{Challenge}. Neither AD nor SD alone can efficiently serve users with heterogeneous demands on resource-constrained edge servers. Serving users with AD alone is memory-efficient but may fail to satisfy stringent latency requirements, whereas relying solely on SD reduces latency but may exceed available memory resources~\cite{zhang-etal-2024-draft,xia2024swift}. As illustrated in Fig.~\ref{fig:intro}, when serving a single user, SD reduces the average end-to-end (E2E) latency by 39.1\% compared with AD, but increases the GPU memory consumption by 18.3\%. This latency-memory tradeoff raises a key challenge: \textit{how to efficiently serve users with heterogeneous demands by jointly exploiting AD and SD under constrained computing and memory resources on edge servers.}

\begin{figure}[!t]
    \centering
    \includegraphics[width=0.48\textwidth]{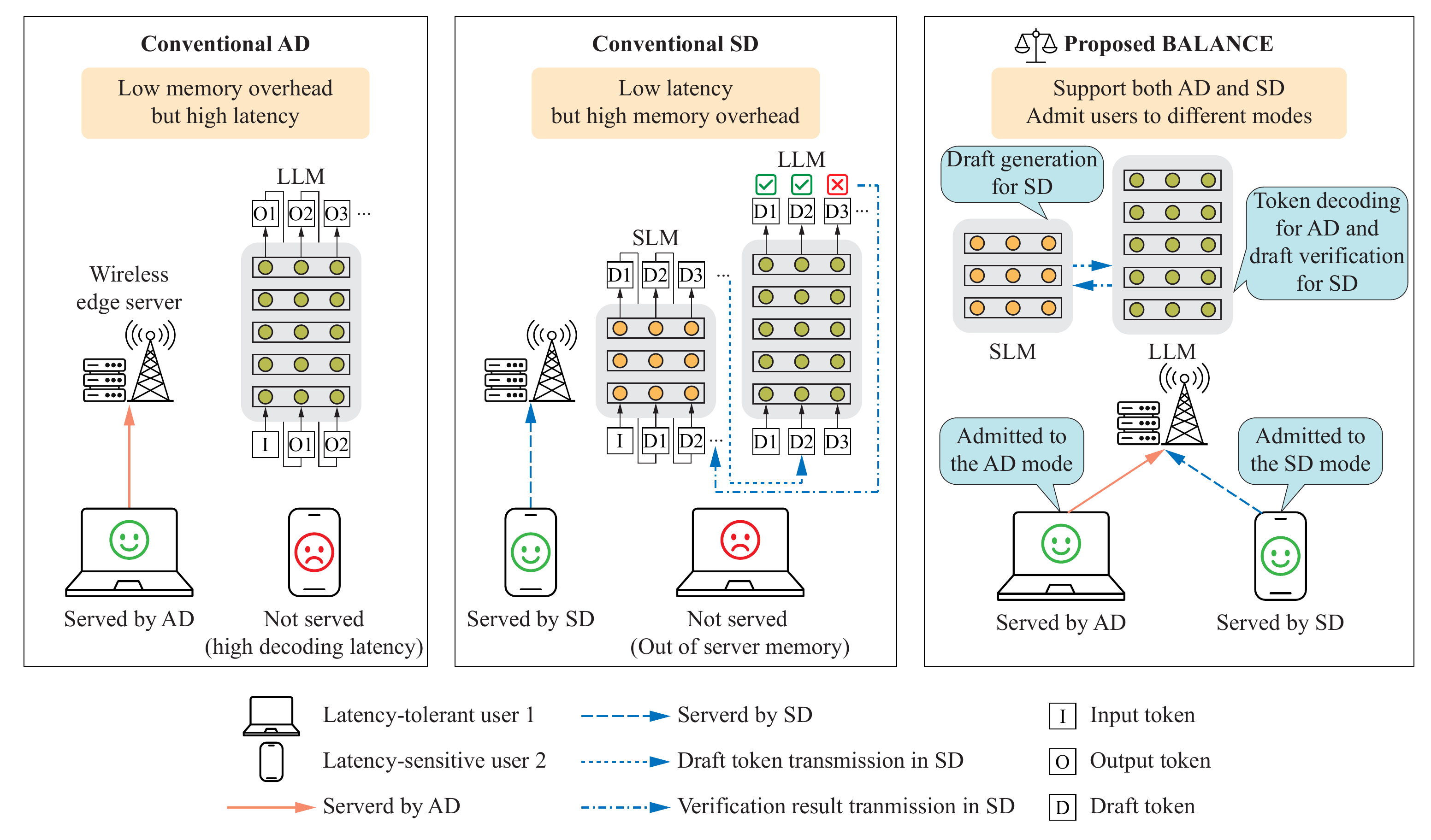}
    \vspace{-0.25cm}
    \caption{Overview of the proposed BALANCE framework and comparison with conventional AD and SD.}
    \label{fig:balance_overview}
    \vspace{-10pt}
\end{figure}

\textbf{Solution}. To address the above challenge, we propose a hy\underline{b}rid \underline{a}utoregressive-specu\underline{la}tive infere\underline{nce} (BALANCE) framework. Fig.~\ref{fig:balance_overview} illustrates the key idea of BALANCE and compares it with conventional AD and SD. In BALANCE, an edge server hosts both an LLM and an SLM, enabling the simultaneous support of AD and SD for different users. To serve users with heterogeneous demands under limited computing resources, the edge server jointly admits users, assigns each admitted user to the AD or SD mode, and allocates computing resources between the two modes. Based on this framework, we formulate a joint user admission and computing resource allocation problem to maximize the number of served users in edge LLM inference while satisfying users' E2E latency requirements and the server's memory constraint. This problem is combinatorial because different assignments of users with heterogeneous demands between AD and SD incur different latency and memory costs, and we show that it is NP-hard. To address this problem, we further design a polynomial-time algorithm with a constant approximation guarantee. To the best of our knowledge, this is the first work to study hybrid autoregressive-speculative inference for edge LLM inference. The main contributions are as follows.

\begin{enumerate}
    \item We propose a hybrid autoregressive-speculative inference framework where an edge server simultaneously supports AD and SD for multiple users. We formulate a joint user admission and computing resource allocation problem to maximize the number of served users under E2E latency requirements and resource constraints. We show that this problem is NP-hard.
    \item To facilitate efficient solution finding, we decompose the original problem into AD and SD sub-problems and design a polynomial-time algorithm with a constant approximation guarantee.
    \item Experimental results show that the proposed BALANCE framework significantly improves the system performance compared with conventional AD and SD.
\end{enumerate}

The rest of this paper is structured as follows. Section~\ref{sec:related_work} reviews the related work, and Section~\ref{sec:preliminary} introduces the preliminaries of conventional AD and SD. Section~\ref{sec:framework} proposes the BALANCE framework and formulates the task throughput maximization problem. Section~\ref{sec:algorithm} designs the algorithm and provides the theoretical analysis. Section~\ref{sec:exp} provides the experimental results, and Section~\ref{sec:conclusion} concludes the paper.

\section{Related Work}\label{sec:related_work}
Existing edge LLM inference systems commonly adopt AD to provide services to end users. To improve system throughput, prior studies have investigated batching techniques that group users requesting the same LLM into one inference batch~\cite{11493570,qu2025partialloading}. Moreover, user admission and resource allocation for multi-user edge inference systems under heterogeneous user demands and constrained server resources have also been studied~\cite{zhang2024beyond,10683673}. Some works further explore \textit{collaborative inference}, where an LLM is partitioned across end devices, edge servers, and cloud servers to exploit distributed computing resources~\cite{11045182,11395617}. These studies provide important foundations for user admission and resource allocation in edge LLM inference under AD.

To reduce response latency, SD has been introduced into edge LLM inference. Similar to AD, existing studies investigate batching, user admission, and resource allocation for SD-based edge inference~\cite{zhu2025efficient,xu2026dip}, as well as collaborative SD at the network edge, where the SLM and the LLM are hierarchically deployed across end devices, edge servers, and cloud servers~\cite{10.1145/3769102.3770608}. To further improve the efficiency of SD, some techniques, such as tree-structured draft generation~\cite{wang2025opt} and parallel draft generation and verification~\cite{liu2025pearl}, have been proposed. Together, these studies advance the efficient deployment of SD in edge networks.

Most existing works focus on either AD or SD alone. Although several studies have explored hybrid LLM inference, they mainly consider distributed deployment of SLMs and LLMs~\cite{11274878}, request routing between SLMs and LLMs~\cite{ding2024hybrid}, and selective LLM involvement during SLM token generation~\cite{oh2026communication}. These hybrid inference methods coordinate SLMs and LLMs rather than jointly supporting AD and SD. Moreover, some works combine different decoding techniques, such as AD for early tokens followed by parallel decoding for subsequent tokens~\cite{huang2025jakiro} and retrieval-based draft generation within SD~\cite{fang2025and}. However, such combinations represent a new inference paradigm rather than joint coordination of AD and SD under resource constraints. Therefore, user admission and resource allocation for hybrid autoregressive-speculative inference remain largely unexplored in edge LLM inference.

\section{Preliminaries}\label{sec:preliminary}
In this section, we introduce the procedures of conventional AD and SD for LLM inference.

\subsection{Conventional AD Procedure}\label{sec:preliminary_AD}
\begin{figure}[!t]
    \centering
    \includegraphics[width=0.19\textwidth]{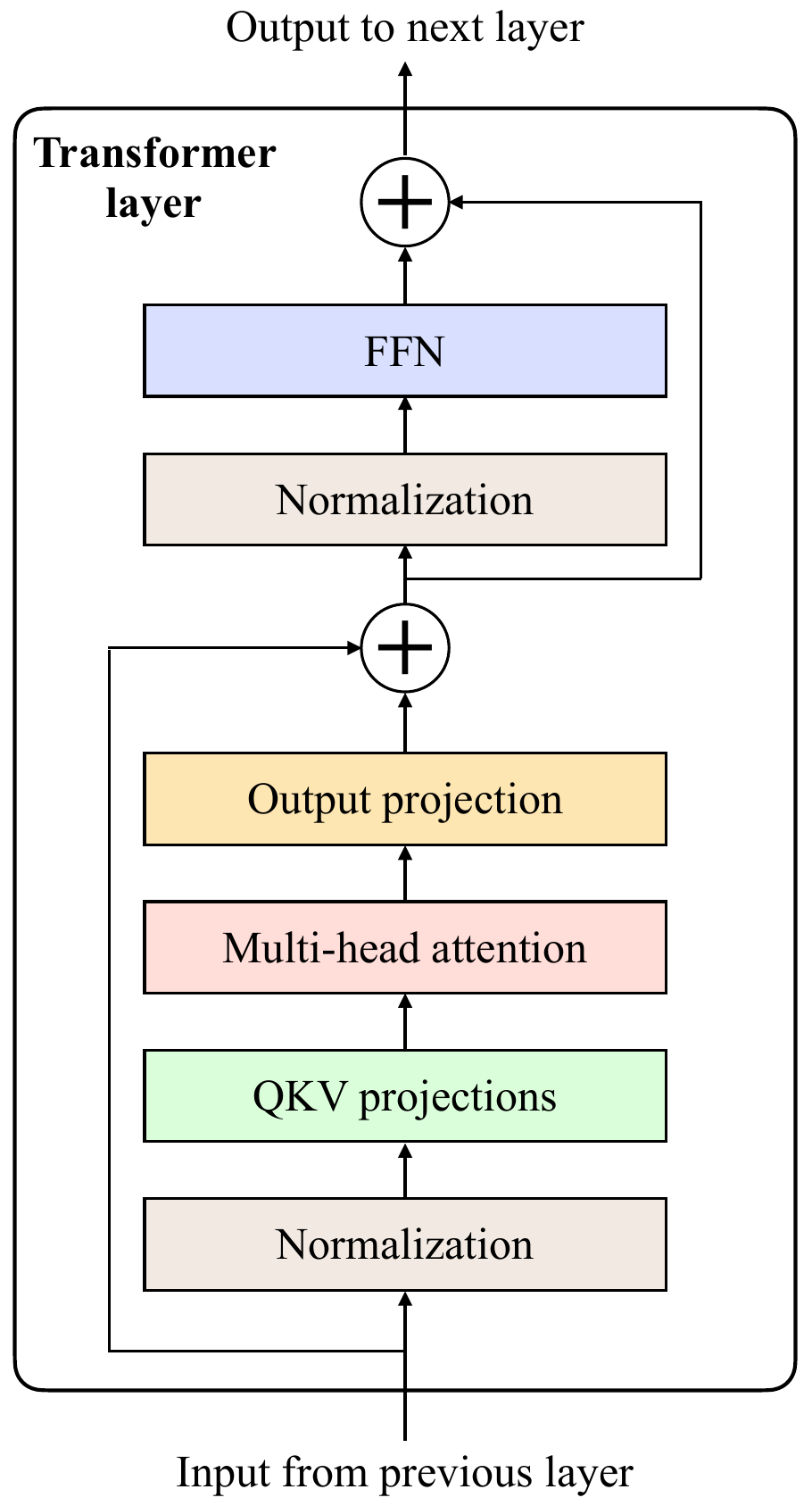}
    \vspace{-0.25cm}
    \caption{Structure of the considered Transformer layer.}
    \label{fig:Transformer_layer}
    \vspace{-10pt}
\end{figure}

Conventional AD consists of prefilling and decoding. We consider a common Transformer architecture for LLMs, with the layer structure shown in Fig.~\ref{fig:Transformer_layer}. The computation in each layer mainly consists of query, key, and value (QKV) projections, multi-head attention (MHA), output projection, and feed-forward computation through a feed-forward network (FFN). In the prefilling stage, the LLM processes the input prompt tokens in a single forward pass through all layers and computes the hidden states of the prompt tokens, from which the first output token is generated. In the decoding stage, the LLM autoregressively generates one new token at each decoding step based on the input prompt tokens and the previously generated tokens until the response generation is completed. To avoid recomputing the key and value states of previous tokens, the LLM maintains a key-value (KV) cache at each layer. At each decoding step, the current token uses its query to attend to the cached key and value states of the previous tokens as well as its own key and value states. Its key and value states are then appended to the KV cache for subsequent decoding steps.

\subsection{Conventional SD Procedure}
Conventional SD employs an SLM and an LLM to generate output tokens through multiple iterations of drafting and verification. Both models are first prefilled with the input prompt tokens, following the prefilling procedure described in Section~\ref{sec:preliminary_AD}. In the first iteration, the SLM autoregressively generates a sequence of multiple draft tokens. The LLM then verifies the draft sequence in \textit{one} forward pass using the KV cache of the input prompt tokens. After verification, the LLM outputs the accepted draft tokens together with one additional token generated by the LLM, and removes the KV cache corresponding to the rejected draft tokens. If some draft tokens are rejected, the LLM generates a resampled token from its output distribution at the first rejected position, and the subsequent draft tokens are not accepted. If all draft tokens are accepted, the LLM samples one bonus token. From the second iteration onward, the SLM first removes the KV cache corresponding to the rejected draft tokens. It then uses the additional token generated by the LLM in the previous iteration as the current input token and generates the next draft sequence based on the retained KV cache. For LLM verification, the additional token from the previous iteration is prepended to the newly generated draft sequence. After verification, the LLM outputs the accepted tokens together with a new additional token and removes the KV cache corresponding to the rejected draft tokens. This process repeats until response generation is completed.

\section{Framework}\label{sec:framework}
This section introduces the BALANCE framework. We first present the multi-user wireless edge network scenario and the BALANCE procedure. Then, we develop the communication, computing, memory, and E2E latency models. Finally, we formulate the task throughput maximization problem.
\subsection{Network Scenario}

\begin{figure}[!t]
    \centering
    \includegraphics[width=0.4\textwidth]{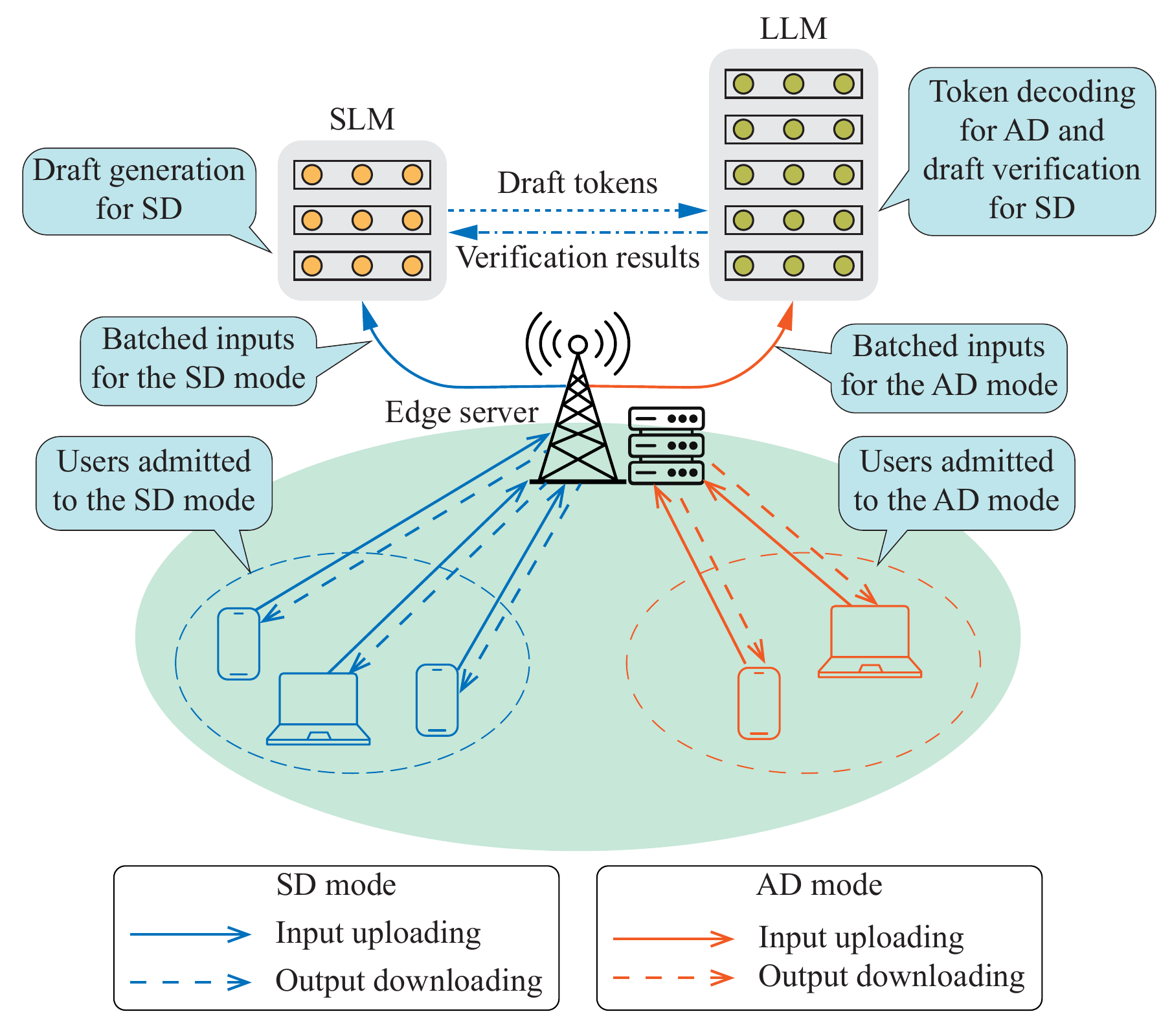}
    \vspace{-0.25cm}
    \caption{Illustration of BALANCE, where an edge server simultaneously supports the AD and SD modes for different users. The blue and orange lines denote the data flows of the SD and AD modes, respectively.}
    \label{fig:framework}
    \vspace{-10pt}
\end{figure}

We consider a multi-user wireless edge network with a single edge server covering a set of users $\mathcal{K}$, as shown in Fig.~\ref{fig:framework}. Each user $k\in\mathcal{K}$ generates an inference task with an E2E latency requirement $T_{k}$ and requests the edge server to perform inference by uploading the task data to the edge server. The edge server hosts both an LLM and an SLM and simultaneously supports the following two inference modes.
\begin{itemize}
    \item \textbf{AD mode}: The edge server performs AD, where the LLM generates output tokens autoregressively.
    \item \textbf{SD mode}: The edge server performs SD, where the SLM generates draft tokens and the LLM verifies them. 
\end{itemize}
To serve heterogeneous user demands, the edge server admits users and assigns each admitted user to the AD or SD mode. To indicate user admission, we define binary variables $x_{k}$ and $y_{k}$, where $x_{k} = 1$ if and only if user $k$ is admitted to the AD mode, and $y_{k} = 1$ if and only if user $k$ is admitted to the SD mode. $x_{k}=y_{k}=0$ indicates that user $k$ is not served. To improve task throughput, users admitted to the same mode are batched for inference. After inference, the generated outputs are returned to the users.

\subsection{BALANCE Procedure}\label{sec:balance_procedure}
\begin{figure}[!t]
    \centering
    \includegraphics[width=0.45\textwidth]{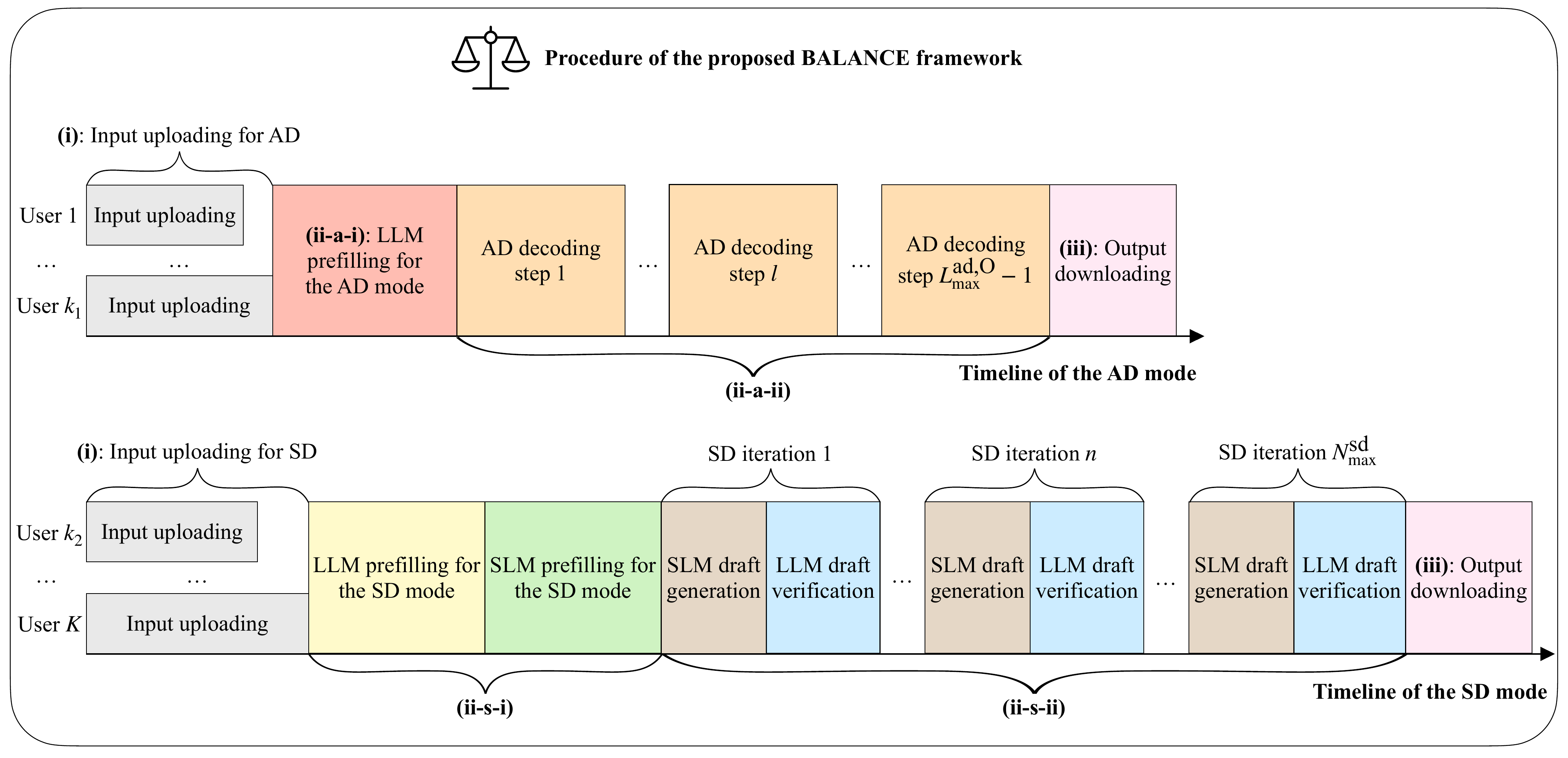}
    \vspace{-0.25cm}
    \caption{The procedure of the proposed BALANCE framework, where the AD and SD modes are executed simultaneously.}
    \label{fig:procedure}
    \vspace{-10pt}
\end{figure}

The procedure of BALANCE is shown in Fig.~\ref{fig:procedure}. Before task execution, each user uploads its input length $L_{k}^{\text{I}}$ and expected output length $L_{k}^{\text{O}}$, measured in tokens, to the edge server through a control signal\footnote{Reporting input and expected output lengths is commonly adopted in user scheduling for LLM inference~\cite{10.1145/3620665.3640383}. $L_{k}^{\text{O}}$ can be specified by the task requirements or obtained by efficient prediction methods, such as lightweight predictors~\cite{aiops2024qiu} and distribution-based prediction~\cite{10.1145/3676641.3716011}. The impact of inaccurate output length prediction on system performance is evaluated in Section~\ref{exp_prediction_impact}.}. Based on the reported information, the edge server determines user admission and sends the admission results to the users. The user admission problem will be formulated in Section~\ref{sec:problem}. \textbf{(i)} After receiving the admission results, the admitted users upload their input prompts. Once the prompts of users admitted to the same inference mode are received, the server processes them in a batch using that mode. \textbf{(ii-a-i)}~In the AD mode, the prompts for AD are first prefilled by the LLM. \textbf{(ii-a-ii)} Then, the LLM generates the output tokens through multiple decoding steps. \textbf{(ii-s-i)} In the SD mode, the prompts for SD are first prefilled by the LLM and then by the SLM. \textbf{(ii-s-ii)} Then, the SLM and LLM generate output tokens over multiple iterations of drafting and verification. \textbf{(iii)} Once one mode is completed, the corresponding outputs are returned to the users admitted to that mode. 

\subsection{Communication Model}
We consider orthogonal frequency-division multiple access for both uplink and downlink transmissions. The data uploading time of user $k$ is 
$\Delta\tau_{k}^{\text{up}}=\frac{\gamma L_{k}^{\text{I}}}{B_{k}R_{k}}$,
where $B_{k}$ and $R_{k}$ denote the uplink bandwidth and uplink spectral efficiency of user $k$, respectively. $\gamma$ is the number of bits per token, and $\gamma L_{k}^{\text{I}}$ is the bit size of user $k$'s input prompt tokens. Thus, the data uploading times of the AD and SD modes are given by
\begin{equation}\label{eq_tau_ad_up}
    \tau^{\text{ad,up}}= \max\limits_{k\in\mathcal{K}}\left\{\Delta\tau_{k}^{\text{up}}\ \middle | \ x_{k}=1\right\},
\end{equation}
and
\begin{equation}\label{eq_tau_sd_up}
    \tau^{\text{sd,up}}= \max\limits_{k\in\mathcal{K}}\left\{\Delta\tau_{k}^{\text{up}}\ \middle | \ y_{k}=1\right\},
\end{equation}
respectively.
The output downloading time of user $k$ is
\begin{equation}\label{eq_tau_down}
    \tau_{k}^{\text{down}}=\left(x_{k}+y_{k}\right)\Delta\tau_{k}^{\text{down}}.
\end{equation}
Here,  
$\Delta\tau_{k}^{\text{down}}=\frac{\gamma L_{k}^{\text{O}}}{\bar{B}_{k}\bar{R}_{k}}$, where $\bar{B}_{k}$ and $\bar{R}_{k}$ denote the downlink bandwidth and spectral efficiency of user $k$, respectively. 

\subsection{Computing Model}
We model the computing procedures and workloads using a common Transformer architecture. Following previous works~\cite{yu2022orca}, we focus on matrix multiplication operations in workload analysis, including query, key, and value (QKV) projections, multi-head attention (MHA) computation, output projection, and feed-forward computation using a feed-forward network (FFN), while ignoring minor contributors.

\subsubsection{Computing procedures and workload analysis of the AD mode}\label{sec:AD_model}
After the input prompts of all users admitted to the AD mode are received, the edge server starts the AD mode. These prompts are batched and prefilled by the LLM in one forward pass, which generates the first output token for each request. Following~\cite{yu2022orca}, we define the LLM workload for prefilling a single input prompt of $L^{\text{I}}$ tokens as $\Gamma^{\text{L,p}}\left(L^{\text{I}}\right)=M^{\text{L}}
    \left[8L^{\text{I}}D^{\text{L,m}}D^{\text{L,h}}H^{\text{L}}
        +4\left(L^{\text{I}}\right)^{2}D^{\text{L,h}}H^{\text{L}}+4L^{\text{I}}D^{\text{L,m}}D^{\text{L,f}}
    \right]$,
where $D^{\text{L,m}}$, $D^{\text{L,h}}$, $H^{\text{L}}$, $D^{\text{L,f}}$, and $M^{\text{L}}$ denote the LLM hidden dimension, the hidden dimension of an attention head, the number of attention heads, the FFN hidden dimension, and the number of layers of the LLM, respectively. The three terms inside the bracket represent the workloads of the QKV and output projections, attention computation, and feed-forward computation in each layer, respectively. Thus, the LLM prefilling workload in the AD mode is given by
\begin{equation}\label{eq_w_ad_p}
\begin{aligned}
    W^\text{ad,p}
    =\Gamma^{\text{L,p}}\left(L_{\max}^{\text{ad,I}}\right)\sum\limits_{k\in\mathcal{K}}x_{k},
    \end{aligned}
\end{equation}
where $\sum\limits_{k\in\mathcal{K}}x_{k}$ is the number of users admitted to the AD mode, and $L_{\max}^{\text{ad,I}}=\max\limits_{k\in\mathcal{K}}\left\{L_{k}^{\text{I}}\ \middle | \ x_{k}=1\right\}$ is the maximum input length of users admitted to the AD mode\footnote{Following previous works~\cite{khattab2020colbert}, we consider padding for efficient GPU matrix multiplication, as supported and recommended by Hugging Face and PyTorch~\cite{padding_hugging,padding_pytorch}.}.

After prefilling, the LLM generates one token for each unfinished user request at each decoding step. Since prefilling has generated the first token, user $k$ requires $L_{k}^{\text{O}}-1$ decoding steps for output generation. Therefore, the total number of decoding steps in the AD mode is $L_{\max}^{\text{ad,O}}-1$, where 
\begin{equation}
    L_{\max}^{\text{ad,O}}=\max\limits_{k\in\mathcal{K}}\left\{L_{k}^{\text{O}} \ \middle | \ x_{k}=1\right\}
\end{equation}
is the maximum output length of users in the AD mode. 
We define the LLM workload for generating one token at the $l$-th decoding step for a single input prompt of $L^{\text{I}}$ tokens as $\Gamma_{l}^{\text{L,a}}\left(L^{\text{I}}\right)=M^{\text{L}}\left[
            \begin{aligned}
                8D^{\text{L,m}}D^{\text{L,h}}H^{\text{L}}+4\left(L^{\text{I}}+l\right)D^{\text{L,h}}H^{\text{L}}
                +4D^{\text{L,m}}D^{\text{L,f}}
            \end{aligned}\right]$,
where the attention workload depends on the sequence length $L^{\text{I}}+l$, since the query of the current token attends to the KV cache of the previous $L^{\text{I}}+l-1$ tokens as well as the key and value of the current token. Thus, the LLM workload at the $l$-th decoding step in the AD mode is given by
\begin{equation}\label{eq_w_l_ad_a}
    \begin{aligned}
        W_{l}^\text{ad,a}
        =\sum\limits_{k\in\mathcal{K}}x_{k}\mathbb{I}_{\left\{l\le L_{k}^{\text{O}}-1\right\}}\Gamma_{l}^{\text{L,a}}\left(L_{\max}^{\text{ad,I}}\right),
    \end{aligned}
\end{equation}
where users with $l>L_{k}^{\text{O}}-1$ are excluded at decoding step $l$.

\subsubsection{Computing procedures and workload analysis of the SD mode}\label{sec:SD_model}
After all input prompts of users admitted to the SD mode are received, the edge server starts the SD mode. These prompts are batched, first prefilled by the LLM, and then prefilled by the SLM. Following~\eqref{eq_w_ad_p}, the LLM prefilling workload in the SD mode is given by

\begin{equation}
    \begin{aligned}
    W^\text{sd,L,p}=\Gamma^{\text{L,P}}\left(L_{\max}^{\text{sd,I}}\right)\sum\limits_{k\in\mathcal{K}}y_{k},
    \end{aligned}
\end{equation}
where $\sum\limits_{k\in\mathcal{K}}y_{k}$ is the number of users admitted to the SD mode, and $L_{\max}^{\text{sd,I}}=\max\limits_{k\in\mathcal{K}}\left\{L_{k}^{\text{I}}\ \middle | \ y_{k}=1\right\}$ is the maximum input length of users admitted to the SD mode. Additionally, following $\Gamma^{\text{L,p}}\left(L^{\text{I}}\right)$, we define the SLM workload for prefilling a single input prompt of $L^{\text{I}}$ tokens as $\Gamma^{\text{S,p}}\left(L^{\text{I}}\right)=M^{\text{S}}
    \left[8L^{\text{I}}D^{\text{S,m}}D^{\text{S,h}}H^{\text{S}}
        +4\left(L^{\text{I}}\right)^{2}D^{\text{S,h}}H^{\text{S}}+4L^{\text{I}}D^{\text{S,m}}D^{\text{S,f}}
    \right]$,
where $D^{\text{S,m}}$, $D^{\text{S,h}}$, $H^{\text{S}}$, $D^{\text{S,f}}$, and $M^{\text{S}}$ denote the SLM hidden dimension, the hidden dimension of an attention head, the number of attention heads, the FFN hidden dimension, and the number of layers of the SLM, respectively. Thus, the SLM prefilling workload in the SD mode is expressed as
\begin{equation}
    \begin{aligned}
    W^\text{sd,S,p}
    =\Gamma^{\text{S,p}}\left(L_{\max}^{\text{sd,I}}\right)\sum\limits_{k\in\mathcal{K}}y_{k}.
    \end{aligned}
\end{equation}

After prefilling, SD proceeds over multiple iterations. In each iteration, the SLM autoregressively generates draft tokens for unfinished requests, and the LLM verifies the batched draft sequences in one forward pass. After verification, the accepted draft tokens and one additional token are retained for each request, and the KV cache of rejected draft tokens is removed. The additional token is used as the input token for the next draft generation. Let $A_k$ denote the expected acceptance rate that a draft token is accepted by the LLM for user $k$'s task. The number of accepted draft tokens for user $k$'s task is given by $L_{k}^{\text{A}} = \frac{A_{k}\left[1-\left(A_{k}\right)^{L^{\text{D}}}\right]}{1-A_{k}}$. User $k$ requires $N_{k} = \lceil \frac{L_{k}^{\text{O}}}{L_{k}^{\text{A}}+1}\rceil$ SD iterations, where term 1 represents the additional token. Therefore, the total number of SD iterations is denoted by
\begin{equation}
    N_{\max}^{\text{sd}}=\max\limits_{k\in\mathcal{K}}\left\{N_{k} \ \middle | \ y_{k}=1\right\}.
\end{equation}
Moreover, let $L_{n}^{\text{sd,P}}$ be the prefix length of the retained tokens before draft generation in iteration $n$, which is given by
\begin{equation}
    L_{n}^{\text{sd,P}}=
    \begin{cases}
        L_{\max}^{\text{sd,I}}, \ n=1,\\
        \begin{aligned}
            L_{\max}^{\text{sd,I}}
            +\left(n-2\right) \max\limits_{\substack{k \in \mathcal{K}\\ y_{k}=1}}\left\{L_{k}^{\text{A}}+1\right\}+\max\limits_{\substack{k \in \mathcal{K}\\ y_{k}=1}}\left\{L_{k}^{\text{A}}\right\}
        \end{aligned}
        , n\ge 2.
    \end{cases}
\end{equation}
For $n=1$, the retained prefix only consists of the input prompt tokens. For $n\ge 2$, it further includes the accepted draft tokens and additional tokens retained from previous iterations.

Following $\Gamma_{l}^{\text{L,a}}\left(L^{\text{I}}\right)$, we define the SLM workload for generating one token for a single input prompt at the $l$-th decoding step of iteration $n$, given $L_{n}^{\text{P}}$ prefix tokens available at the beginning of iteration $n$, as $\Gamma_{n,l}^{\text{S,a}}\left(L_{n}^{\text{P}}\right)=M^{\text{S}}\left[
            \begin{aligned}
                8D^{\text{S,m}}D^{\text{S,h}}H^{\text{S}}+4\left(L_{n}^{\text{P}}+l\right)D^{\text{S,h}}H^{\text{S}}
                +4D^{\text{S,m}}D^{\text{S,f}}
            \end{aligned}\right].$
Thus, the SLM workload at decoding step $l$ of iteration $n$ is 
\begin{equation}
    \begin{aligned}
    W_{n,l}^{\text{sd,a}}
    =\sum\limits_{k\in\mathcal{K}}y_{k}\mathbb{I}_{\left\{n\le N_{k}\right\}}\Gamma_{n,l}^{\text{S,a}}\left(L_{n}^{\text{sd,P}}\right),
    \end{aligned}
\end{equation}
where users with $n>N_{k}$ are excluded in iteration $n$. We define the LLM workload for verifying $L^{\text{D}}$ tokens in iteration $n$ for a single input prompt, given $L_{n}^{\text{P}}$ prefix tokens available at the beginning of iteration $n$, as $\Gamma_{n}^{\text{L,v}}\left(L_{n}^{\text{P}},L^{\text{D}}\right)=M^{\text{L}}$
$\left[
8L^{\text{D}}D^{\text{L,m}}D^{\text{L,h}}H^{\text{L}}
            +4\sum\limits_{l=1}^{L^{\text{D}}}\left(L_{n}^{\text{P}}+l\right)D^{\text{L,h}}H^{\text{L}}
            +4L^{\text{D}}D^{\text{L,m}}D^{\text{L,f}}
\right]$.
Thus, the LLM workload for verification in iteration $n$ is
\begin{equation}\label{eq_f_n_k_v}
\begin{aligned}
    W_{n}^{\text{sd,v}}=
    \begin{cases}
        \sum\limits_{k\in\mathcal{K}}y_{k}\mathbb{I}_{\left\{n\le N_{k}\right\}}\Gamma_{1}^{\text{L,v}}\left(L_{1}^{\text{sd,P}},L^{\text{D}}\right),\ n=1,\\
        \sum\limits_{k\in\mathcal{K}}y_{k}\mathbb{I}_{\left\{n\le N_{k}\right\}}\Gamma_{n}^{\text{L,v}}\left(L_{n}^{\text{sd,P}},L^{\text{D}}+1\right),\ n\ge 2.
    \end{cases}
\end{aligned}
\end{equation}
In the first iteration, the LLM verifies $L^{\text{D}}$ draft tokens for each request. When $n\ge 2$, the verification further includes the additional token generated in the previous iteration.

\subsection{Memory Model}
Following previous works~\cite{10.5555/3618408.3619696}, we consider the memory required for storing model weights and the KV cache during inference when analyzing memory consumption.
\subsubsection{Memory for hosting the LLM and SLM}
The GPU memory required to store the LLM and SLM weights is 
\begin{equation}
q^{\text{L}}=M^{\text{L}}\theta\left(4D^{\text{L,m}}D^{\text{L,h}}H^{\text{L}}+2D^{\text{L,m}}D^{\text{L,f}}\right),
\end{equation}
and
\begin{equation}
    q^{\text{S}}=M^{\text{S}}\theta\left(4D^{\text{S,m}}D^{\text{S,h}}H^{\text{S}}+2D^{\text{S,m}}D^{\text{S,f}}\right),
\end{equation}
respectively, where $\theta$ denotes the number of bytes required to represent one floating-point number~\cite{10.5555/3618408.3619696}.

\subsubsection{Memory for the AD mode} 
In the AD mode, the required KV cache memory for prefilling is $
    q^{\text{ad,p}}=M^{\text{L}}\theta\left(\sum\limits_{k\in\mathcal{K}}x_{k}L_{\max}^{\text{ad,I}}\right)2D^{\text{L,h}}H^{\text{L}}$,
where $2D^{\text{L,h}}H^{\text{L}}$ denotes the number of KV cache elements required by one token across all attention heads in each layer. During decoding, the peak KV cache memory is denoted by $q^{\text{ad,a}}    =M^{\text{L}}\theta\left[\sum\limits_{k\in\mathcal{K}}x_{k}\left(L_{\max}^{\text{ad,I}}+L_{k}^{\text{O}}\right)\right]2D^{\text{L,h}}H^{\text{L}}$. Therefore, the peak KV cache memory in the AD mode is given by
\begin{equation}
    q^{\text{ad}} = \max\left\{q^{\text{ad,p}}  ,q^{\text{ad,a}}  \right\}=q^{\text{ad,a}}.
\end{equation}

\subsubsection{Memory for the SD mode}
In the SD mode, the required KV cache memory for prefilling is given by $q^{\text{sd,p}} = q^{\text{sd,L,p}} + q^{\text{sd,S,p}}$, where $q^{\text{sd,L,p}}$ and $q^{\text{sd,S,p}}$ denote the KV cache memory required by the LLM and SLM during prefilling, respectively. Following $q^{\text{ad,p}}$, $q^{\text{sd,L,p}}=M^{\text{L}}\theta\left(\sum\limits_{k\in\mathcal{K}}y_{k}L_{\max}^{\text{sd,I}}\right)2D^{\text{L,h}}H^{\text{L}}$ and $q^{\text{sd,S,p}}=M^{\text{S}}\theta\left(\sum\limits_{k\in\mathcal{K}}y_{k}L_{\max}^{\text{sd,I}}\right)2D^{\text{S,h}}H^{\text{S}}$. 
Following $q^{\text{ad,a}}$, the peak SLM KV cache memory during draft generation is given by $q^{\text{sd,a}}=M^{\text{S}}\theta\left[\sum\limits_{k\in\mathcal{K}}y_{k}\left(L_{N_{k}}^{\text{sd,P}}+L^{\text{D}}+1\right)\right]2D^{\text{S,h}}H^{\text{S}}$.
During verification, the peak LLM KV cache  is given by
$q^{\text{sd,v}}
        =M^{\text{L}}\theta\left[\sum\limits_{k\in\mathcal{K}}y_{k}\left(L_{N_{k}}^{\text{sd,P}}+L^{\text{D}}+1\right)\right]2D^{\text{L,h}}H^{\text{L}}$.
Therefore, the peak KV cache memory in the SD mode is
\begin{equation}
    q^{\text{sd}} = \max\left\{q^{\text{sd,p}},q^{\text{sd,a}}+q^{\text{sd,v}}\right\}=q^{\text{sd,a}}+q^{\text{sd,v}}.
\end{equation}

\subsection{User E2E latency}
To model the inference latency under shared GPU computing resources, we consider the computing resource allocation between the AD and SD modes. Let $z$ denote the fraction of GPU computing resources allocated to the AD mode, and thus $1-z$ is allocated to the SD mode\footnote{In practice, GPU computing resource allocation can be implemented by partitioning GPU resources, e.g., streaming multiprocessors and GPU engines, across multiple tasks, which has already been supported by NVIDIA~\cite{nvidia_mig}. For simplicity, we refer to it as ``GPU computing resource allocation" in this paper. Such allocation has been adopted in existing studies~\cite{xu2022igniter,qu2026slide}.}. 

\subsubsection{Latency of the AD mode}
Following previous studies~\cite{9843917,xu2022igniter,wu2023graft}, the LLM prefilling time in the AD mode is 
$\tau^\text{ad,p}=\frac{\alpha^{\text{L}}W^\text{ad,p}}{z}+\beta^{\text{L}}$,
where $\alpha^{\text{L}}$ and $\beta^{\text{L}}$ are model- and hardware-dependent constants measured offline. The time of decoding step $l$ in the AD mode is $\tau_{l}^\text{ad,a}= \frac{\alpha^{\text{L}}W_{l}^\text{ad,a}}{z}+\beta^{\text{L}}$. Thus, the time from prefilling to the completion of step $l$ in the AD mode is 
\begin{equation}\label{eq_tau_l_ad}
    \tau_{l}^{\text{ad}}= \tau^\text{ad,p}+\sum\limits_{l'=1}^{l}\tau_{l'}^\text{ad,a}.
\end{equation}

\subsubsection{Latency of the SD mode}
The prefilling times of the LLM and the SLM in the SD mode are given by $\tau^\text{sd,L,p}=\frac{\alpha^{\text{L}}W^\text{sd,L,p}}{1-z}+\beta^{\text{L}}$ and $\tau^\text{sd,S,p}=\frac{\alpha^{\text{S}}W^\text{sd,S,p}}{1-z}+\beta^{\text{S}}$, respectively. For iteration $n$, the time required by the SLM to generate $L^{\text{D}}$ draft tokens for all users admitted to the SD mode is
\begin{equation}
    \tau_{n}^{\text{sd,a}}= 
    \begin{cases}
        \sum\limits_{l=1}^{L^{\text{D}}-1}\left(\frac{\alpha^{\text{S}} W_{n,l}^{\text{sd,a}}}{1-z}+\beta^{\text{S}}\right),\ n=1,\\
        \sum\limits_{l=1}^{L^{\text{D}}}\left(\frac{\alpha^{\text{S}} W_{n,l}^{\text{sd,a}}}{1-z}+\beta^{\text{S}}\right),\ 2\le n\le N_{\max}^{\text{sd}},
    \end{cases}
\end{equation}
where the first iteration requires only $L^{\text{D}}-1$ steps due to the prefilling. Moreover, the LLM verification time in iteration $n$ is denoted by
$\tau_{n}^\text{sd,v}=\frac{\alpha^{\text{L}}W_{n}^{\text{sd,v}}}{1-z}+\beta^{\text{L}}$.
Thus, the time from prefilling to the completion of iteration $n$ in the SD mode is 
\begin{equation}\label{tau_n_sd}
    \tau_{n}^{\text{sd}}= \tau^\text{sd,L,p}+\tau^\text{sd,S,p}+\sum\limits_{n'=1}^{n}\left(\tau_{n'}^{\text{sd,a}}+\tau_{n'}^{\text{sd,v}}\right).
\end{equation}

\subsubsection{E2E latency}
Based on~\eqref{eq_tau_ad_up}, \eqref{eq_tau_sd_up}, \eqref{eq_tau_down}, \eqref{eq_tau_l_ad}, and \eqref{tau_n_sd}, the E2E latency of user $k$ is expressed as
\begin{equation}
    t_{k}^{\text{E}}=x_{k}\left(\tau^{\text{ad,up}}+\tau_{L_{\max}^{\text{ad,O}}-1}^{\text{ad}}\right)+y_{k}\left(\tau^{\text{sd,up}}+\tau_{N_{\max}^{\text{sd}}}^{\text{sd}}\right)+\tau_{k}^{\text{down}}.
\end{equation}

\subsection{Problem Formulation}\label{sec:problem}
To maximize the task throughput of BALANCE, i.e., the number of served users satisfying their latency requirements, we formulate the following optimization problem.
\begin{subequations}
	\begin{equation}\label{p1_o}
		{\mathcal{P}1}:\mathop{\max}\limits_{{\bf{X}},{\bf{Y}},z}\ r\left({\bf{X}}+{\bf{Y}}\right)=\sum\limits_{k\in\mathcal{K}}x_{k}+y_{k}
	\end{equation}	
        \begin{equation}\label{const_1}
		{\rm{s.t.}} \ x_{k}+y_{k}\le 1,\ \forall k \in\mathcal{K},
	\end{equation}	
    \begin{equation}\label{const_4}
        t_{k}^{\text{E}}\le T_{k},\ \forall k\in\mathcal{K},
    \end{equation}
    \begin{equation}\label{const_5}
        \mathbb{I}_{\left\{z<1\right\}}q^{\text{S}}+q^{\text{L}}+q^{\text{ad}}+q^{\text{sd}}\le Q, 
    \end{equation}
        \begin{equation}\label{const_6}
		x_{k}\in\left\{0,1\right\},\ \forall k \in\mathcal{K},
	\end{equation}	
	\begin{equation}\label{const_7}
		y_{k}\in\left\{0,1\right\},\ \forall k \in\mathcal{K},
	\end{equation}	
    \begin{equation}\label{const_9}
		z\in{\bf{Z}}.
	\end{equation}	
\end{subequations}
Here, $x_{k}\in{\bf{X}}$, $y_{k}\in{\bf{Y}}$, and $z$ are the decision variables for AD-mode user admission, SD-mode user admission, and GPU computing resource allocation, respectively. In~\eqref{p1_o}, ${\bf{X}}+{\bf{Y}}$ denotes the element-wise sum of ${\bf{X}}$ and ${\bf{Y}}$. Constraint~\eqref{const_1} assigns each user to at most one inference mode. Constraints~\eqref{const_4} and~\eqref{const_5} enforce the user latency and GPU memory capacity constraints, respectively. Here, $Q$ is the available GPU memory capacity, and the indicator function $\mathbb{I}_{\left\{z<1\right\}}$ equals 1 if $z<1$ and 0 otherwise. When $z=1$, the SD mode is inactive, so the SLM weights are not loaded into GPU memory. Constraints~\eqref{const_6} and \eqref{const_7} are binary constraints. Constraint~\eqref{const_9} restricts $z$ to ${\bf{Z}}=\left\{0,z_{1},\dots,z_{I},1\right\}$, where $0<z_{1}<\dots<z_{I}<1$. The values in ${\bf{Z}}$ represent the GPU computing resource allocation ratios supported by the system.

\section{Algorithm Design}\label{sec:algorithm}
This section proves the NP-hardness of $\mathcal{P}1$ and develops a polynomial-time constant-approximation algorithm.
\subsection{Problem Mapping}

The NP-hardness of $\mathcal{P}1$ is stated below.
\begin{proposition}\label{proposition_1}
    $\mathcal{P}1$ is an NP-hard problem. 
\end{proposition}
\begin{proof}
    The proof is provided in Appendix~\ref{proof_proposition_1}.
\end{proof}

\subsection{Equivalent Problem Transformation}
To facilitate the solution design, we introduce an auxiliary variable $\lambda$ to represent the fraction of available GPU memory allocated to the AD mode. Since the memory consumption of the AD and SD modes can be characterized separately, $\mathcal{P}1$ is equivalently transformed into the following $\mathcal{P}2$.
\begin{subequations}
	\begin{equation}
		{\mathcal{P}2}:\mathop{\max}\limits_{{\bf{X}},{\bf{Y}},z,\lambda}\ r\left({\bf{X}}+ {\bf{Y}}\right)=\sum\limits_{k\in\mathcal{K}}x_{k}+y_{k}
	\end{equation}	
        \begin{equation}
		{\rm{s.t.}} \ \eqref{const_1},\eqref{const_4},\eqref{const_6},\eqref{const_7},\eqref{const_9},
	\end{equation}	
    \begin{equation}\label{const_p2_2}
        q^{\text{ad}}\le Q_{1}, 
    \end{equation}
    \begin{equation}\label{const_p2_3}
        q^{\text{sd}}\le Q_{2}, 
    \end{equation}
    \begin{equation}\label{const_p2_4}
        \lambda\in\Lambda. 
    \end{equation}
\end{subequations}
Here, $Q_{1}=\lambda\left(Q-q^{\text{L}}-\mathbb{I}_{\left\{z<1\right\}}q^{\text{S}}\right)$ and $Q_{2}=\left(1-\lambda\right)\left(Q-q^{\text{L}}-\mathbb{I}_{\left\{z<1\right\}}q^{\text{S}}\right)$ denote the memory reserved for the AD and SD modes, respectively. Constraint~\eqref{const_p2_4} restricts $\lambda$ to $\Lambda=\left\{0,\lambda_{1},\dots,\lambda_{J},1\right\}$, where $0<\lambda_{1}<\dots<\lambda_{J}<1$. The set $\Lambda$ contains all supported ratios for partitioning GPU memory between the AD and SD modes, ensuring that every feasible memory split can be represented by some $\lambda\in\Lambda$.

Next, we state that solving $\mathcal{P}2$ is equivalent to solving $\mathcal{P}1$. 
\begin{proposition}\label{proposition_2}
    Solving $\mathcal{P}1$ is equivalent to solving $\mathcal{P}2$.
\end{proposition}
\begin{proof}
    The proof is shown in Appendix~\ref{proof_proposition_2}.
\end{proof}

To solve $\mathcal{P}2$, we fix $z$ and $\lambda$, and have the below proposition.

\begin{proposition}\label{proposition_3}
    Given $z$ and $\lambda$, solving $\mathcal{P}2$ is equivalent to solving $\mathcal{P}3$.
\end{proposition}
    
\begin{subequations}
	\begin{equation}
		{\mathcal{P}3}:\mathop{\max}\limits_{{\bf{X}},{\bf{Y}}}\ r\left({\bf{X}}+ {\bf{Y}}\right)=\sum\limits_{k\in\mathcal{K}}x_{k}+y_{k}
	\end{equation}	
        \begin{equation}
		{\rm{s.t.}} \ \eqref{const_1},\eqref{const_4},\eqref{const_6},\eqref{const_7},\eqref{const_p2_2},\eqref{const_p2_3}.
	\end{equation}	
\end{subequations}
\begin{proof}
    The proof is shown in Appendix~\ref{proof_proposition_3}.
\end{proof}

In Proposition~\ref{proposition_2}, the auxiliary variable $\lambda$ decouples the GPU memory constraint into two separate memory constraints for the AD and SD modes. Proposition~\ref{proposition_3} further implies that, when $z$ and $\lambda$ are determined, the problem reduces to user admission under fixed resource budgets. Although Propositions~\ref{proposition_2} and~\ref{proposition_3} simplify the problem, $\mathcal{P}3$ remains \textit{NP-hard}, which follows the proof of Proposition~\ref{proposition_1}. 

\subsection{Problem Decomposition}
In $\mathcal{P}3$, the AD and SD modes have separate resource budgets, and each user can be admitted to at most one inference mode. Therefore, we decompose $\mathcal{P}3$ into the following two sub-problems, $\mathcal{P}3.1$ and $\mathcal{P}3.2$, which admit users to the AD and SD modes, respectively. We have
\begin{subequations}
	\begin{equation}
		{\mathcal{P}3.1}:\mathop{\max}\limits_{{\bf{X}}}\ r\left({\bf{X}}\right)=\sum\limits_{k\in\mathcal{K}}x_{k}
	\end{equation}	
    \begin{equation}
        {\rm{s.t.}} \ \eqref{const_6},\eqref{const_p2_2}
    \end{equation}
    \begin{equation}
        x_{k}t^{\text{ad}}\le T'_{k},\ \forall k\in\mathcal{K},
    \end{equation}
\end{subequations}
where $t^{\text{ad}}=\tau^{\text{ad,up}}+\tau_{L_{\max}^{\text{ad,O}}-1}^{\text{ad}}$ denotes the latency of the AD mode excluding output downloading, and $T'_{k}=T_{k}-\Delta\tau_{k}^{\text{down}}$ denotes the remaining latency budget of user $k$ after reserving the output downloading latency. Also, we have
\begin{subequations}
	\begin{equation}
		{\mathcal{P}3.2}:\mathop{\max}\limits_{{\bf{Y}}}\ r'\left({\bf{Y}}\right)=\sum\limits_{k\in\mathcal{K}\setminus\mathcal{K}^{\text{ad}}}y_{k}
	\end{equation}	
    \begin{equation}
        {\rm{s.t.}} \ \eqref{const_p2_3},
    \end{equation}
    \begin{equation}
        y_{k}t^{\text{sd}}\le T'_{k},\ \forall k\in\mathcal{K}\setminus\mathcal{K}^{\text{ad}},
    \end{equation}
    \begin{equation}
        y_{k}\in\left\{0,1\right\},\ \forall k \in\mathcal{K}\setminus\mathcal{K}^{\text{ad}}, 
    \end{equation}
\end{subequations}
where $t^{\text{sd}}=\tau^{\text{sd,up}}+\tau_{N_{\max}^{\text{sd}}}^{\text{sd}}$ denotes the latency of the SD mode excluding output downloading, and $\mathcal{K}^{\text{ad}}=\left\{k \ \middle| \ x_{k}=1\right\}$ is the set of users admitted to the AD mode.

Based on this decomposition, the following subsections solve $\mathcal{P}3.1$ and $\mathcal{P}3.2$ sequentially to determine the optimal user admission decisions for the AD and SD modes, respectively. The performance gap between the sequential solution and the optimal solution to $\mathcal{P}1$ will be analyzed in Section~\ref{sec:algorithm_outline}.

\subsection{Solution to $\mathcal{P}3.1$: User Admission for the AD Mode}
We obtain an optimal solution to $\mathcal{P}3.1$ by enumerating the feasible values of $L_{\max}^{\text{ad,I}}$, $L_{\max}^{\text{ad,O}}$, $t^{\text{ad}}$, and $\tau^{\text{ad,up}}$, and identifying the served users under each setting. When $z=0$ or $\lambda=0$, the AD mode is inactive, and thus the optimal solution to $\mathcal{P}3.1$ is $\hat{{\bf{X}}}={\bf{0}}$. The following analysis considers $z>0$ and $\lambda>0$.

First, given $L_{\max}^{\text{ad,I}}=L_{1}$, $L_{\max}^{\text{ad,O}}=L_{2}$, $t^{\text{ad}}=t_{1}$, and $\tau^{\text{ad,up}}=\tau_{1}$, the candidate user set for the AD mode is 
\begin{equation}\label{eq_K1}
    \mathcal{K}_{1}=\left\{k\ \middle | \ k\in\mathcal{K},  L_{k}^{\text{I}}\le L_{1}, L_{k}^{\text{O}}\le L_{2}, t_{1}\le T'_{k},\Delta\tau_{k}^{\text{up}}\le \tau_{1}\right\}.
\end{equation}
For each candidate user $k\in\mathcal{K}_{1}$, its latency contribution is
\begin{equation}\label{eq_t_k_li}
    \tilde{t}_{k}^{\text{ad}}
    =\frac{\alpha^{\text{L}}\tilde{W}_{k}^{\text{ad,p}}}{z}
    +\sum\limits_{l=1}^{L_{2}-1}\frac{\alpha^{\text{L}}\tilde{W}_{l,k}^{\text{ad,a}}}{z}.
\end{equation}
Here, $\tilde{W}_{k}^\text{ad,p}$ and $\tilde{W}_{l,k}^\text{ad,a}$ denote the contributions of user $k$ to the LLM workloads during prefilling and at the $l$-th decoding step in the AD mode, respectively, where $\tilde{W}_{k}^\text{ad,p}=\Gamma^{\text{L,p}}\left(L_{1}\right)$ and $\tilde{W}_{l,k}^\text{ad,a}
        =\mathbb{I}_{\left\{l\le L_{k}^{\text{O}}-1\right\}}\Gamma_{l}^{\text{L,a}}\left(L_{1}\right)$. 
Moreover, the contribution of user $k\in\mathcal{K}_{1}$ to the total memory consumption of the AD mode is given by 
\begin{equation}
    \tilde{q}_{k}^{\text{ad}}=\tilde{q}^{\text{ad}}=M^{\text{L}}\theta\left(L_{1}+L_{2}\right)2D^{\text{L,h}}H^{\text{L}}.
\end{equation}

Second, let $\Pi_{1}=\left\{1,\dots,d_{1}\dots,\left|\mathcal{K}_{1}\right|\right\}$ denote the index mapping of the users in $\mathcal{K}_{1}$ sorted in a non-decreasing order of $\tilde{t}_{k}^{\text{ad}}$, where $\Pi_{1}\left(d_{1}\right)$ gives the user index at the $d_{1}$-th position and $\tilde{t}_{\Pi_{1}\left(d_{1}\right)}^{\text{ad}}\le \tilde{t}_{\Pi_{1}\left(d_{1}+1\right)}^{\text{ad}}$. Therefore, the number of served users under $L_{1}$, $L_{2}$, $t_{1}$, and $\tau_{1}$ is given by
\begin{equation}\label{eq_d1}
    d_{1}^{*} = \max\left\{d_{1}\ \middle | \ d_{1}\le \lfloor\frac{Q_{1}}{\tilde{q}^{\text{ad}}}\rfloor,\sum\limits_{d'_{1}=1}^{d_{1}}\tilde{t}_{\Pi_{1}\left(d'_{1}\right)}^{\text{ad}}\le\Delta t_{1}\right\},
\end{equation}
where $\Delta t_{1}=t_{1}-\tau_{1}-\Delta\tau_{1}$ and $\Delta\tau_{1}=L_{2}\beta^{\text{L}}$ denotes the fixed latency term of one forward pass in prefilling and $L_{2}-1$ forward passes in decoding. Based on \eqref{eq_d1}, the served user set for the AD mode under $L_{1}$, $L_{2}$, $t_{1}$, and $\tau_{1}$ is $\hat{\mathcal{K}}^{\text{ad}}=\left\{\Pi_{1}\left(1\right),\dots,\Pi_{1}\left(d_{1}^{*}\right)\right\}$, and the corresponding user admission decision $\hat{x}_{k}\in\hat{{\bf{X}}}$ is 
\begin{equation}\label{eq_hat_x}
    \hat{x}_{k}=
    \begin{cases}
        1,\text{ if }k\in\hat{\mathcal{K}}^{\text{ad}},\\
        0,\text{ otherwise}.
    \end{cases}
\end{equation}

By enumerating all feasible $L_{1}$, $L_{2}$, $t_{1}$, and $\tau_{1}$, we can obtain the maximum number of served users and the corresponding user admission decision under given $z$ and $\lambda$ for $\mathcal{P}3.1$. Algorithm~\ref{algorithm_li} summarizes the above procedures, and its optimality is established in the following proposition.
\begin{proposition}\label{proposition_li}
    Algorithm~\ref{algorithm_li} obtains an optimal solution to $\mathcal{P}3.1$ under given $z$ and $\lambda$.
\end{proposition}

\begin{proof}
    The proof is provided in Appendix~\ref{proof_proposition_li}.
\end{proof}

\begin{algorithm}[!t]
	\caption{AD-Mode User Admission Algorithm} 
	\label{algorithm_li}
	\LinesNumbered
	\KwIn{$z$ and $\lambda$.}
	\KwOut{$\hat{{\bf{X}}}$ and $\hat{\mathcal{K}}^{\text{ad}}$.} 
    {\bf Initialize:} $r_{1}=0$,  $\hat{{\bf{X}}}={\bf{0}}$, $\hat{\mathcal{K}}^{\text{ad}}=\emptyset$.\\
    \If{$z=0$ or $\lambda=0$}
    {
        Return $\hat{{\bf{X}}}={\bf{0}}$ and $\hat{\mathcal{K}}^{\text{ad}}=\emptyset$.\\
    }
    \For{$L_{1}\in\left\{0\right\}\bigcup\left\{L_{k}^{\text{I}} \middle| k\in\mathcal{K} \right\}$, $L_{2}\in\left\{0\right\}\bigcup\left\{L_{k}^{\text{O}} \middle| k\in\mathcal{K} \right\}$, $t_{1}\in\left\{0\right\}\bigcup\left\{T'_{k} \middle| k\in\mathcal{K} \right\}$}
    {\label{line:li_for1_start}
        \For{$\tau_{1}\in\left\{0\right\}\bigcup\left\{\Delta\tau_{k}^{\text{up}}\ \middle|\ k\in\mathcal{K} \right\}$}
        {\label{line:li_for2_start}
            Calculate $d_{1}^{*}$ with \eqref{eq_d1}.\label{line:li_d1}\\
            \If{$d_{1}^{*}>r_{1}$}
            {
                $r_{1}=d_{1}^{*}$ and $\hat{\mathcal{K}}^{\text{ad}}=\left\{\Pi_{1}\left(1\right),\dots,\Pi_{1}\left(d_{1}^{*}\right)\right\}$. \\
            }
        }
    }
    Calculate $\hat{{\bf{X}}}$ with \eqref{eq_hat_x}.\\
\end{algorithm}

\subsection{Solution to $\mathcal{P}3.2$: User Admission for the SD Mode}
We solve $\mathcal{P}3.2$ similarly by enumerating the feasible values of $L_{\max}^{\text{sd,I}}$, $\max\left\{L_{k}^{\text{A}}\right\}=L_{\max}^{\text{sd,A}}$, $N_{\max}^{\text{sd}}$, $t^{\text{sd}}$, and $\tau^{\text{sd,up}}$, and identifying the served users under each setting. When $z=1$ or $\lambda=1$, the SD mode is inactive, and thus the optimal solution to $\mathcal{P}3.2$ is $\hat{{\bf{Y}}}={\bf{0}}$. The following analysis considers $z<1$ and $\lambda<1$.

First, given $L_{\max}^{\text{sd,I}}=L_{3}$, $L_{\max}^{\text{sd,A}}=L_{4}$, $N_{\max}^{\text{sd}}=\tilde{N}_{1}$, $t^{\text{sd}}=t_{2}$, and $\tau^{\text{sd,up}}=\tau_{2}$, the candidate user set for the SD mode is
\begin{equation}\label{eq_K2}
    \mathcal{K}_{2}=\left\{k\ \middle | \ 
    \begin{aligned}
        &k\in\mathcal{K}',  L_{k}^{\text{I}}\le L_{3}, L_{k}^{\text{A}}\le L_{4}, N_{k}\le \tilde{N}_{1},\\
        &t_{2}\le T'_{k},\Delta\tau_{k}^{\text{up}}\le \tau_{2}
    \end{aligned}
    \right\},
\end{equation}
where $\mathcal{K}'=\mathcal{K}\setminus\hat{\mathcal{K}}^{\text{ad}}$.
The contribution of each candidate user $k\in\mathcal{K}_{2}$ to the total latency of the SD mode is
\begin{equation}\label{eq_tilde_t_k_sd}
    \begin{aligned}
        \tilde{t}_{k}^{\text{sd}}
        &=\frac{\alpha^{\text{L}}\tilde{W}_{k}^{\text{sd,L,p}}}{1-z}+\frac{\alpha^{\text{S}}\tilde{W}_{k}^{\text{sd,S,p}}}{1-z}\\
        &+\sum\limits_{l=1}^{L^{\text{D}}-1}\frac{\alpha^{\text{S}}\tilde{W}_{1,l,k}^{\text{sd,a}}}{1-z}+\sum\limits_{n=2}^{\tilde{N}_{1}}\sum\limits_{l=1}^{L^{\text{D}}}\frac{\alpha^{\text{S}}\tilde{W}_{n,l,k}^{\text{sd,a}}}{1-z}+\sum\limits_{n=1}^{\tilde{N}_{1}}\frac{\alpha^{\text{L}}\tilde{W}_{n,k}^{\text{sd,v}}}{1-z}.
    \end{aligned}
\end{equation}
Here, $\tilde{W}_{k}^\text{sd,L,p}$ and $\tilde{W}_{k}^\text{sd,S,p}$ are the contributions of user $k$ to the total LLM and SLM workloads during prefilling in the SD mode, respectively, where $\tilde{W}_{k}^\text{sd,L,p}=\Gamma^{\text{L,p}}\left(L_{3}\right)$ 
and $\tilde{W}_{k}^\text{sd,S,p}=\Gamma^{\text{S,p}}\left(L_{3}\right)$. 
Additionally, $\tilde{W}_{n,l,k}^{\text{sd,a}}$ and $\tilde{W}_{n,k}^{\text{sd,v}}$ denote the contributions of user $k$ to the total SLM workload at the $l$-th decoding step of iteration $n$ and the total LLM workload during verification in iteration $n$, respectively. 
Let $\tilde{L}_{n}^{\text{sd,P}}$ denote the prefix length of the retained tokens before draft generation in the $n$-th iteration under $L_{3}$ and $L_{4}$, which is given by
\begin{equation}
    \tilde{L}_{n}^{\text{sd,P}}=
    \begin{cases}
        L_{3}, \ n=1,\\
        L_{3}+\left(n-2\right)\left(L_{4}+1\right)+L_{4}, \ n\ge 2.
    \end{cases}
\end{equation}
Thus, we have $\tilde{W}_{n,l,k}^{\text{sd,a}}
    =\mathbb{I}_{\left\{n\le N_{k}\right\}}\Gamma_{n,l}^{\text{S,a}}\left(\tilde{L}_{n}^{\text{sd,P}}\right)$ and
\begin{equation}
    \tilde{W}_{n,k}^{\text{sd,v}}=
    \begin{cases}
        \mathbb{I}_{\left\{n\le N_{k}\right\}}\Gamma_{1}^{\text{L,v}}\left(\tilde{L}_{1}^{\text{sd,P}},L^{\text{D}}\right), \ n=1,\\
        \mathbb{I}_{\left\{n\le N_{k}\right\}}\Gamma_{n}^{\text{L,v}}\left(\tilde{L}_{n}^{\text{sd,P}},L^{\text{D}}+1\right), \ n\ge 2.
    \end{cases}
\end{equation}
Moreover, the contribution of user $k$ to the total memory consumption of the SD mode is given by
\begin{equation}\label{eq_tilde_q_k_sd}
    \tilde{q}_{k}^{\text{sd}}=\tilde{q}_{k}^{\text{sd,a}}+\tilde{q}_{k}^{\text{sd,v}}.
\end{equation}
Here, $\tilde{q}_{k}^{\text{sd,a}}=M^{\text{S}}\theta\left(\tilde{L}_{N_{k}}^{\text{sd,P}}+L^{\text{D}}+1\right)2D^{\text{S,h}}H^{\text{S}}$ denotes the contribution of user $k$ to the total memory consumption during SLM draft generation. $\tilde{q}_{k}^{\text{sd,v}}$ denotes the contribution of user $k$ to the total memory consumption during LLM draft verification, where $\tilde{q}_{k}^{\text{sd,v}}=M^{\text{L}}\theta\left(\tilde{L}_{N_{k}}^{\text{sd,P}}+L^{\text{D}}+1\right)2D^{\text{L,h}}H^{\text{L}}$.

Second, let $\Pi_{2}=\left\{1,\dots,d_{2}\dots,\left|\mathcal{K}_{2}\right|\right\}$ denote the index mapping of the users in $\mathcal{K}_{2}$ sorted in a non-decreasing order of $\tilde{t}_{k}^{\text{sd}}$, where $\Pi_{2}\left(d_{2}\right)$ gives the user index at the $d_{2}$-th position and $\tilde{t}_{\Pi_{2}\left(d_{2}\right)}^{\text{sd}}\le \tilde{t}_{\Pi_{2}\left(d_{2}+1\right)}^{\text{sd}}$. Since both $\tilde{t}_{k}^{\text{sd}}$ and $\tilde{q}_{k}^{\text{sd}}$ are monotonically non-decreasing with $N_{k}$, sorting users by $\tilde{t}_{k}^{\text{sd}}$ also preserves the non-decreasing order of $\tilde{q}_{k}^{\text{sd}}$. Therefore, the number of served users under $L_{3}$, $L_{4}$, $\tilde{N}_{1}$, $t_{2}$, and $\tau_{2}$ is
\begin{equation}\label{eq_d2}
    d_{2}^{*} = \max\left\{d_{2}\ \middle | \ \sum\limits_{d'_{2}=1}^{d_{2}}\tilde{q}_{\Pi_{2}\left(d'_{2}\right)}^{\text{sd}}\le Q_{2},\sum\limits_{d'_{2}=1}^{d_{2}}\tilde{t}_{\Pi_{2}\left(d'_{2}\right)}^{\text{sd}}\le\Delta t_{2}\right\},
\end{equation}
where $\Delta t_{2}=t_{2}-\tau_{2}-\Delta\tau_{2}$ and $\Delta\tau_{2}=\left(\tilde{N}_{1}+1\right)\beta^{\text{L}}+\tilde{N}_{1}L^{\text{D}}\beta^{\text{S}}$ denotes the fixed latency term in the SD mode. Based on~\eqref{eq_d2}, the served user set of the SD mode under $L_{3}$, $L_{4}$, $\tilde{N}_{1}$, $t_{2}$, and $\tau_{2}$ is $\hat{\mathcal{K}}^{\text{sd}}=\left\{\Pi_{2}\left(1\right),\dots,\Pi_{2}\left(d_{2}^{*}\right)\right\}$, and the corresponding user admission decision $\hat{y}_{k}\in\hat{{\bf{Y}}}$ is
\begin{equation}\label{eq_hat_y}
    \hat{y}_{k}=
    \begin{cases}
        1,\text{ if }k\in\hat{\mathcal{K}}^{\text{sd}},\\
        0,\text{ otherwise}.
    \end{cases}
\end{equation}

By enumerating all feasible $L_{3}$, $L_{4}$, $\tilde{N}_{1}$, $t_{2}$, and $\tau_{2}$, we can identify the maximum number of served users and the corresponding user admission decision under given $z$ and $\lambda$ for $\mathcal{P}3.2$. Algorithm~\ref{algorithm_sd} summarizes the above procedures, and its optimality is established in the following proposition.
\begin{proposition}\label{proposition_sd}
    Algorithm~\ref{algorithm_sd} obtains an optimal solution to $\mathcal{P}3.2$ under given $z$, $\lambda$, and $\hat{\mathcal{K}}^{\text{ad}}$.
\end{proposition}
\begin{proof}
    The proof is provided in Appendix~\ref{proof_proposition_sd}.
\end{proof}

\begin{algorithm}[!t]
	\caption{SD-Mode User Admission Algorithm} 
	\label{algorithm_sd}
	\LinesNumbered
	\KwIn{$z$, $\lambda$, and $\hat{\mathcal{K}}^{\text{ad}}$.}
	\KwOut{$\hat{{\bf{Y}}}$ and $\hat{\mathcal{K}}^{\text{sd}}$.} 
    {\bf Initialize:} $\mathcal{K}'=\mathcal{K}\setminus\hat{\mathcal{K}}^{\text{ad}}$, $r_{2}=0$, $\hat{{\bf{Y}}}={\bf{0}}$, $\hat{\mathcal{K}}^{\text{sd}}=\emptyset$.\\
    \If{$z=1$ or $\lambda=1$}
    {
        Return $\hat{{\bf{Y}}}={\bf{0}}$ and $\hat{\mathcal{K}}^{\text{sd}}=\emptyset$.\\
    }
    \For{$L_{3}\in\left\{0\right\}\bigcup\left\{L_{k}^{\text{I}} \middle| k\in\mathcal{K}' \right\}$, $L_{4}\in\left\{0\right\}\bigcup\left\{L_{k}^{\text{A}} \middle| k\in\mathcal{K}'\right\}$, $\tilde{N}_{1}\in\left\{0\right\}\bigcup\left\{N_{k} \middle| k\in\mathcal{K}'\right\}$, $t_{2}\in\left\{0\right\}\bigcup\left\{T'_{k}\ \middle|\ k\in\mathcal{K}'\right\}$}
    {\label{line:sd_for1_start}
        \For{$\tau_{2}\in\left\{0\right\}\bigcup\left\{\Delta\tau_{k}^{\text{up}}\ \middle|\ k\in\mathcal{K}' \right\}$}
        {\label{line:sd_for2_start}
            Calculate $d_{2}^{*}$ with \eqref{eq_d2}.\label{line:sd_d2}\\
            \If{$d_{2}^{*}>r_{2}$}
            {
                $r_{2}=d_{2}^{*}$ and $\hat{\mathcal{K}}^{\text{sd}}=\left\{\Pi_{2}\left(1\right),\dots,\Pi_{2}\left(d_{2}^{*}\right)\right\}$. \\
            }
        }
    }
    Calculate $\hat{{\bf{Y}}}$ with \eqref{eq_hat_y}.\\
\end{algorithm}

\subsection{Algorithm Outline}\label{sec:algorithm_outline}
Algorithm~\ref{algorithm_all} summarizes the overall procedure for solving $\mathcal{P}1$. It enumerates all feasible pairs of $\left(z,\lambda\right)$. For each pair, it first solves $\mathcal{P}3.1$ using Algorithm~\ref{algorithm_li} to admit users to the AD mode, and then solves $\mathcal{P}3.2$ using Algorithm~\ref{algorithm_sd} to admit the remaining users to the SD mode. The solution that serves the maximum number of users over all pairs gives the user admission decision $\dot{{\bf{X}}}$ for the AD mode, user admission decision $\dot{{\bf{Y}}}$ for the SD mode, and GPU computing resource allocation $\dot{z}$. The approximation guarantee of Algorithm~\ref{algorithm_all} is characterized in the following theorem.
\begin{theorem}\label{theorem_1}
    Algorithm~\ref{algorithm_all} achieves a constant approximation guarantee. Specifically, the produced solution $\dot{{\bf{X}}}$ and $\dot{{\bf{Y}}}$ satisfy $r\left(\dot{{\bf{X}}}+\dot{{\bf{Y}}}\right)\ge \frac{1}{2}r\left({\bf{X}}^{*}+{\bf{Y}}^{*}\right)$, where ${\bf{X}}^{*}$ and ${\bf{Y}}^{*}$ denote the optimal user admission decisions of $\mathcal{P}1$.
\end{theorem}
\begin{proof}
    The proof is shown in Appendix~\ref{proof_theorem_1}.
\end{proof}

\begin{algorithm}[!t]
	\caption{BALANCE User Admission Algorithm} 
	\label{algorithm_all}
	\LinesNumbered
	\KwIn{$\mathcal{K}$, $Q$, and $T_{k}$.}
	\KwOut{$\dot{{\bf{X}}}$, $\dot{{\bf{Y}}}$, $\dot{z}$, and $\dot{\lambda}$.} 
    {\bf Initialize:} $r_{3}=0$, $\dot{{\bf{X}}}={\bf{0}}$, $\dot{{\bf{Y}}}={\bf{0}}$, $\dot{z}=0$, $\dot{\lambda}=0$.\\
    \For{$z\in{\bf{Z}}$, $\lambda\in\Lambda$}
    {\label{line:all_for1_start}
        Obtain $\hat{{\bf{X}}}$ and $\hat{\mathcal{K}}^{\text{ad}}$ using Algorithm~\ref{algorithm_li} with $z$ and $\lambda$.\label{line:all_li}\\
        Obtain $\hat{{\bf{Y}}}$ and $\hat{\mathcal{K}}^{\text{sd}}$ using Algorithm~\ref{algorithm_sd} with $z$, $\lambda$, and $\hat{\mathcal{K}}^{\text{ad}}$.\label{line:all_sd}\\
        \If{$\left|\hat{\mathcal{K}}^{\text{ad}}\right|+\left|\hat{\mathcal{K}}^{\text{sd}}\right|>r_{3}$}
        {
            $r_{3} = \left|\hat{\mathcal{K}}^{\text{ad}}\right|+\left|\hat{\mathcal{K}}^{\text{sd}}\right|$, $\dot{{\bf{X}}}=\hat{{\bf{X}}}$, $\dot{{\bf{Y}}}=\hat{{\bf{Y}}}$, $\dot{z}=z$, and $\dot{\lambda}=\lambda$.\\
            
        }
    }
\end{algorithm}

The time complexity of Algorithm~\ref{algorithm_all} is stated below.
\begin{theorem}\label{theorem_2}
    Algorithm~\ref{algorithm_all} has a polynomial time complexity of $O\left(K^{6}\log K\right)$, where $K$ is the total number of users in $\mathcal{K}$.
\end{theorem}
\begin{proof}
The proof is shown in Appendix~\ref{proof_theorem_2}.
\end{proof}
\section{Numerical Results}\label{sec:exp}
This section evaluates the performance of BALANCE. We first describe the experimental setup and then examine the effectiveness of BALANCE under varying system settings. Next, we compare the algorithm running time and conduct ablation studies. We further analyze the impact of inaccurate predictions of output length and acceptance rate on system performance. Finally, we evaluate the performance of BALANCE using different SLM-LLM pairs.

\subsection{Experimental Setup}
We adopt a hybrid testing method that combines device measurements and numerical simulations to evaluate the performance of the proposed BALANCE framework. Specifically, computing latencies are measured on a real server, while the communication latencies in multi-user scenarios are obtained through simulations. The inference is performed on an edge server using an AMD Ryzen Threadripper PRO 5975WX CPU, an NVIDIA GeForce RTX 4090 GPU with 24 GB of GPU memory, and 256 GB of DDR4 RAM. TinyLlama-1.1B and Llama-2-7B are used as the SLM and LLM, respectively, with FP16 precision. Accordingly, $Q$ = 24 GB.

The coverage radius of the edge server is set to 250 m. The uplink and downlink bandwidths allocated to user $k$ are set to $B_{k}=\text{1 MHz}$ and $\bar{B}_{k}=\text{2 MHz}$, respectively. The transmit power spectral densities of the users for uplink transmission and of the edge server for downlink transmission are set to -50 dBm/Hz and -29 dBm/Hz, respectively. The number of users $K$ ranges from 20 to 70. The user latency requirement satisfies $T_{k}\in\left[T_{\min},T_{\max}\right]$, ranging from 0.5 s to 13 s. The input length $L_{k}^{\text{I}}$ and output length $L_{k}^{\text{O}}$ are set within $\left[L_{\min}^{\text{I}},L_{\max}^{\text{I}}\right]$ and $\left[L_{\min}^{\text{O}},L_{\max}^{\text{O}}\right]$, respectively, both ranging from 1 to 1000 tokens. For SD, the draft token length $L^{\text{D}}$ varies from 1 to 8, and the acceptance rate $A_{k}\in\left[A_{\min},A_{\max}\right]$ ranges from 0.6 to 0.95. 

We compare BALANCE  with the following baselines.
\begin{itemize}
    \item \textbf{AD-only}~\cite{zhang2024beyond}. In this baseline, all users are served only by the AD mode, and user admission is determined by Algorithm~\ref{algorithm_li}. By Proposition~\ref{proposition_li}, this baseline achieves the maximum task throughput under the AD-only mode.
    \item \textbf{SD-only}~\cite{zhu2025efficient}. In this baseline, all users are served only by the SD mode. User admission is determined by Algorithm~\ref{algorithm_sd}. By Proposition~\ref{proposition_sd}, this baseline achieves the maximum task throughput under the SD-only mode.
    \item \textbf{Exhaustive search}. This baseline solves the original problem $\mathcal{P}1$ by exhaustively searching all feasible solutions, thereby incurring exponential time complexity. We use this baseline as an exact solution method to evaluate the running time efficiency of the proposed algorithm.
\end{itemize}

To evaluate system performance, we use normalized task throughput, defined as the ratio of the number of served users to the total number of users.

\subsection{Performance Evaluation}

\begin{figure}[!t]
    \centering
	\subfigure[Results vs. number of users $K$.]{\includegraphics[height=3.1cm, keepaspectratio]{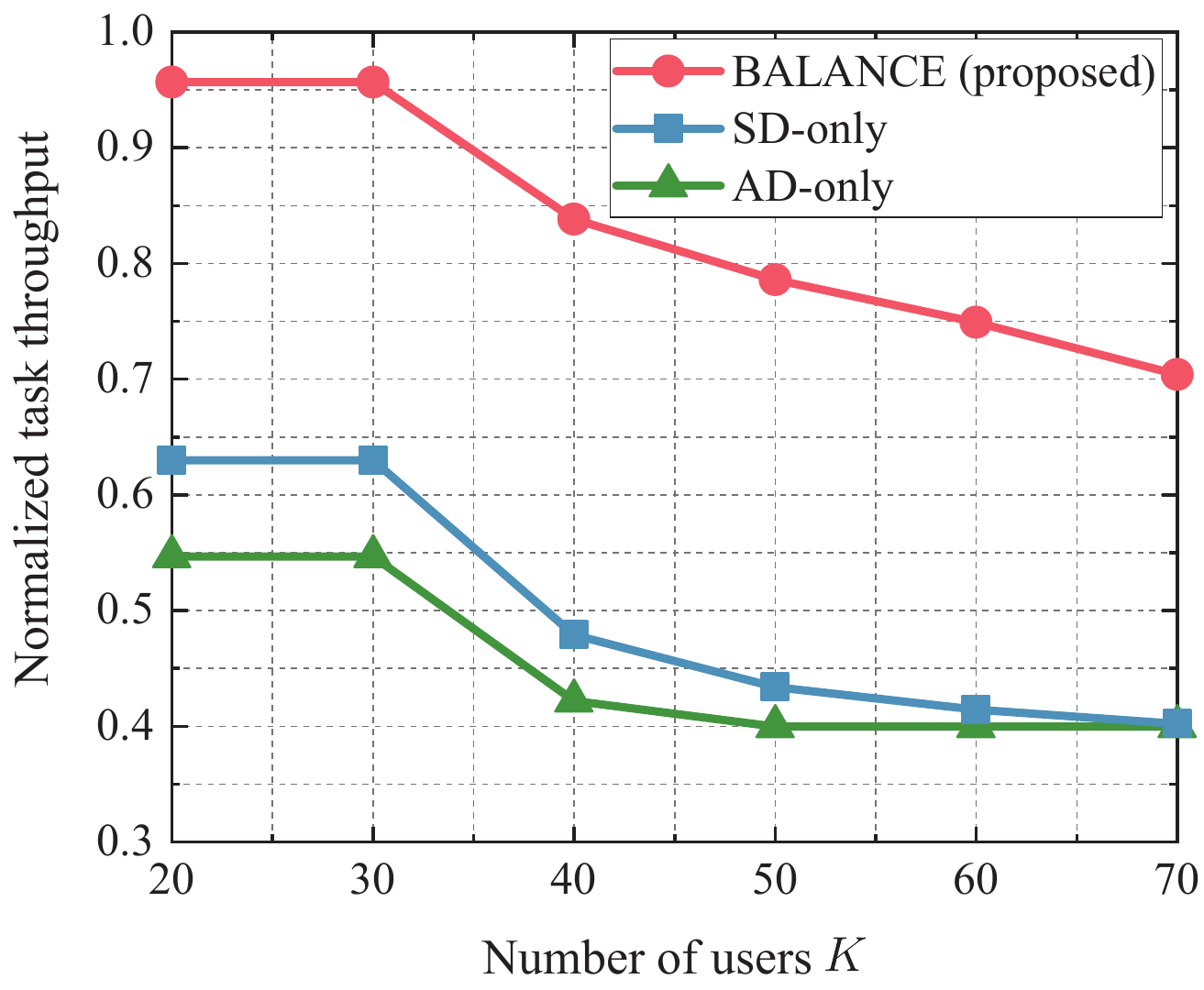}\label{fig:result_user}}
	\quad
	\subfigure[Results vs. maximum latency requirement $T_{\max}$.]{\includegraphics[height=3.1cm, keepaspectratio]{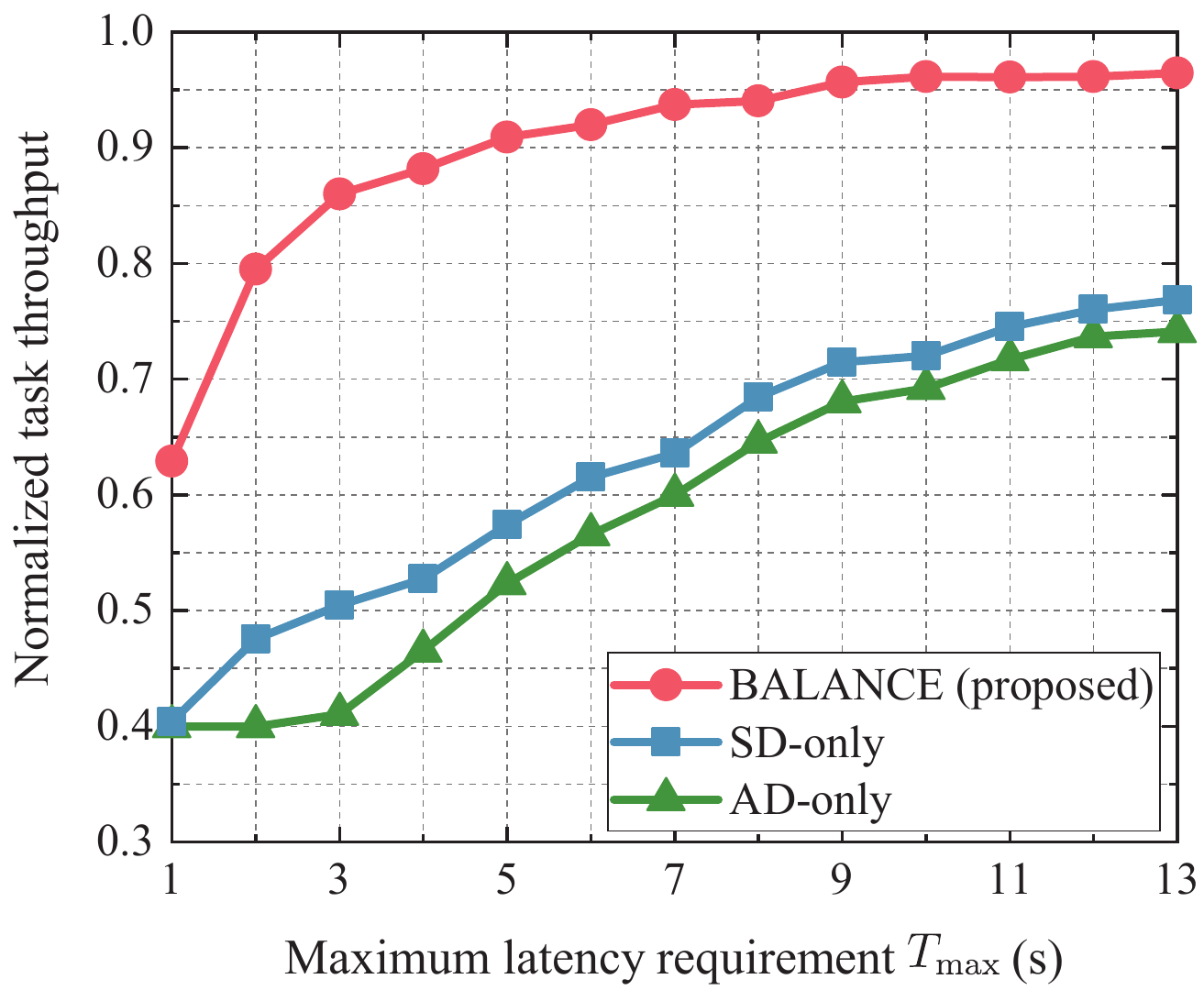}\label{fig:result_ddl}}
    \vspace{-0.25cm}
 \caption{Performance of BALANCE under varying $K$ and $T_{\max}$. The default values of $K$, $T_{\min}$, $T_{\max}$ are set to 30, 0.5, and 4, respectively. $L_{\min}^{\text{I}}$, $L_{\max}^{\text{I}}$, $L_{\min}^{\text{O}}$, $L_{\max}^{\text{O}}$, $L^{\text{D}}$, $A_{\min}$, $A_{\max}$, and $Q$ are set to 1, 300, 1, 100, 1, 0.6, 0.8, and 24 GB, respectively.}
 \label{fig:result1}
 \vspace{-10pt}
\end{figure}

\begin{figure}[!t]
    \centering
    \subfigure[Results vs. maximum input length $L_{\max}^{\text{I}}$.]{\includegraphics[height=3.1cm, keepaspectratio]{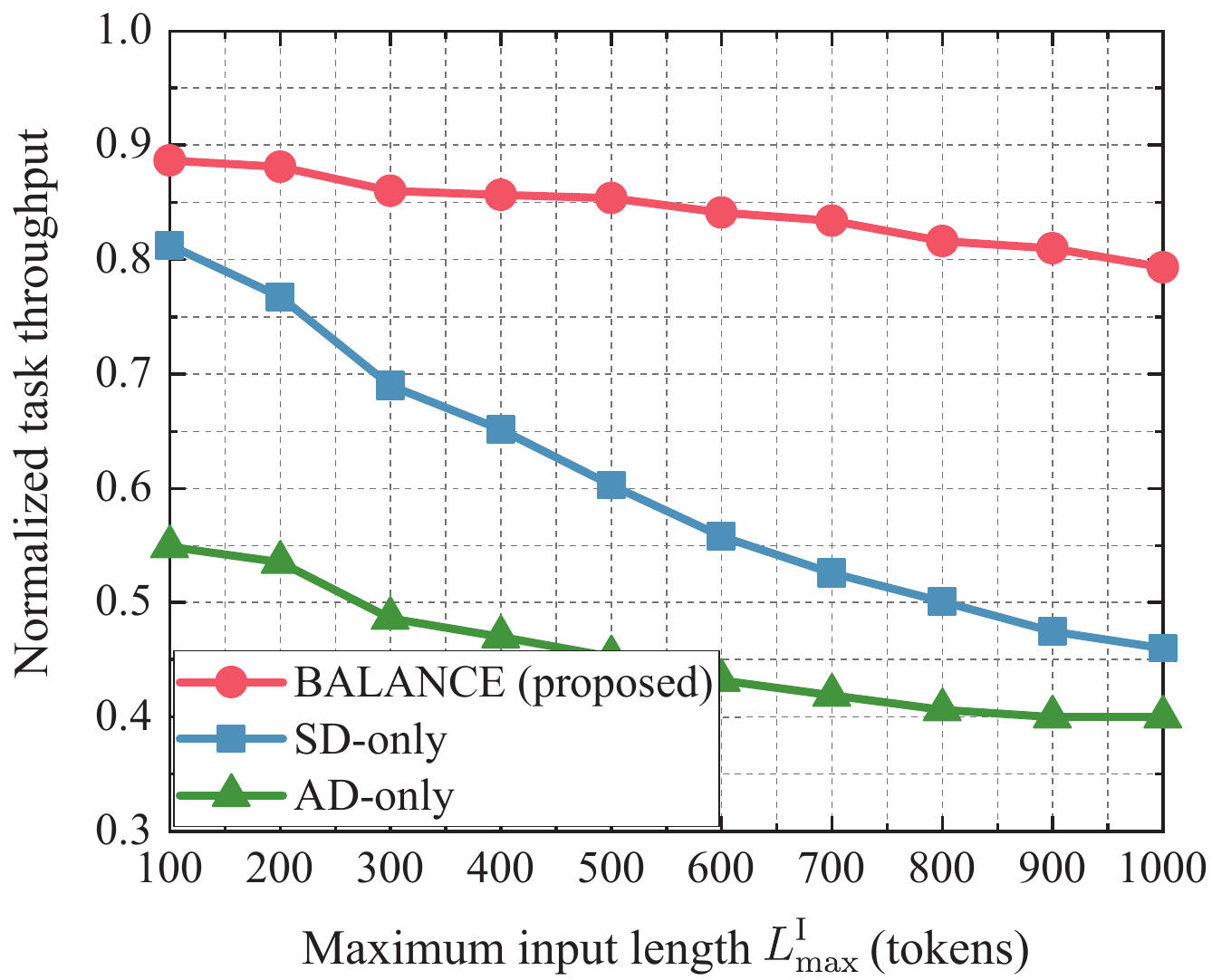}\label{fig:result_input}}
	\quad
	\subfigure[Results vs. maximum output length $L_{\max}^{\text{O}}$.]{\includegraphics[height=3.1cm, keepaspectratio]{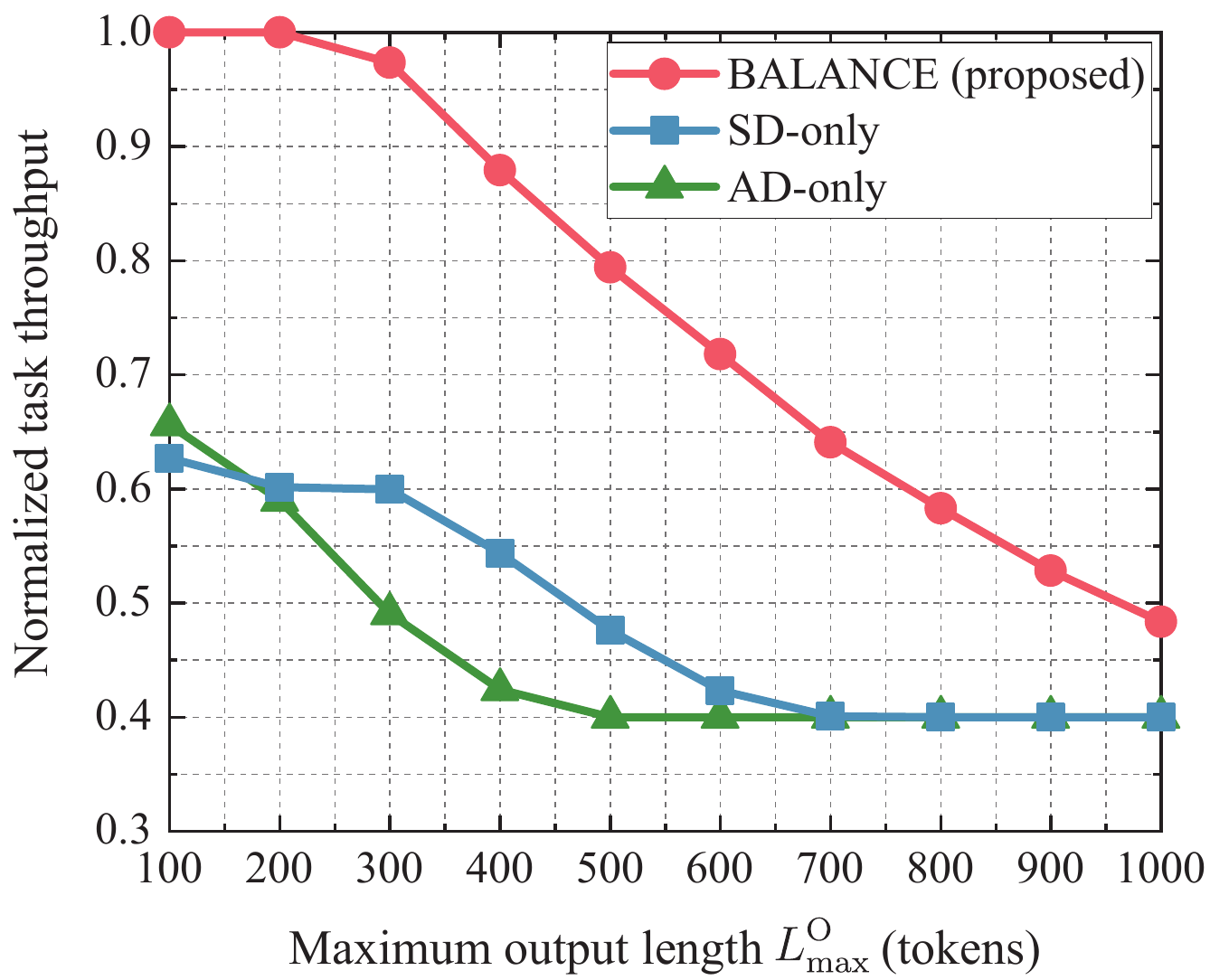}\label{fig:result_output}}
    \vspace{-0.25cm}
 \caption{Performance of BALANCE under varying $L_{\max}^{\text{I}}$ and $L_{\max}^{\text{O}}$. The default values of $L_{\min}^{\text{I}}$, $L_{\max}^{\text{I}}$, $L_{\min}^{\text{O}}$, $L_{\max}^{\text{O}}$ are set to 1, 1000, 1, and 500, respectively. $T_{\min}$ and $T_{\max}$ are set to 2 s and 12 s, respectively. The default values of $K$, $L^{\text{D}}$, $A_{\min}$, $A_{\max}$, and $Q$ are same as Fig.~\ref{fig:result1}.}
 \label{fig:result2}
 \vspace{-10pt}
\end{figure}

\begin{figure*}[!t]
    \centering
    \subfigure[Results vs. GPU memory capacity $Q$.]{\includegraphics[height=3.4cm, keepaspectratio]{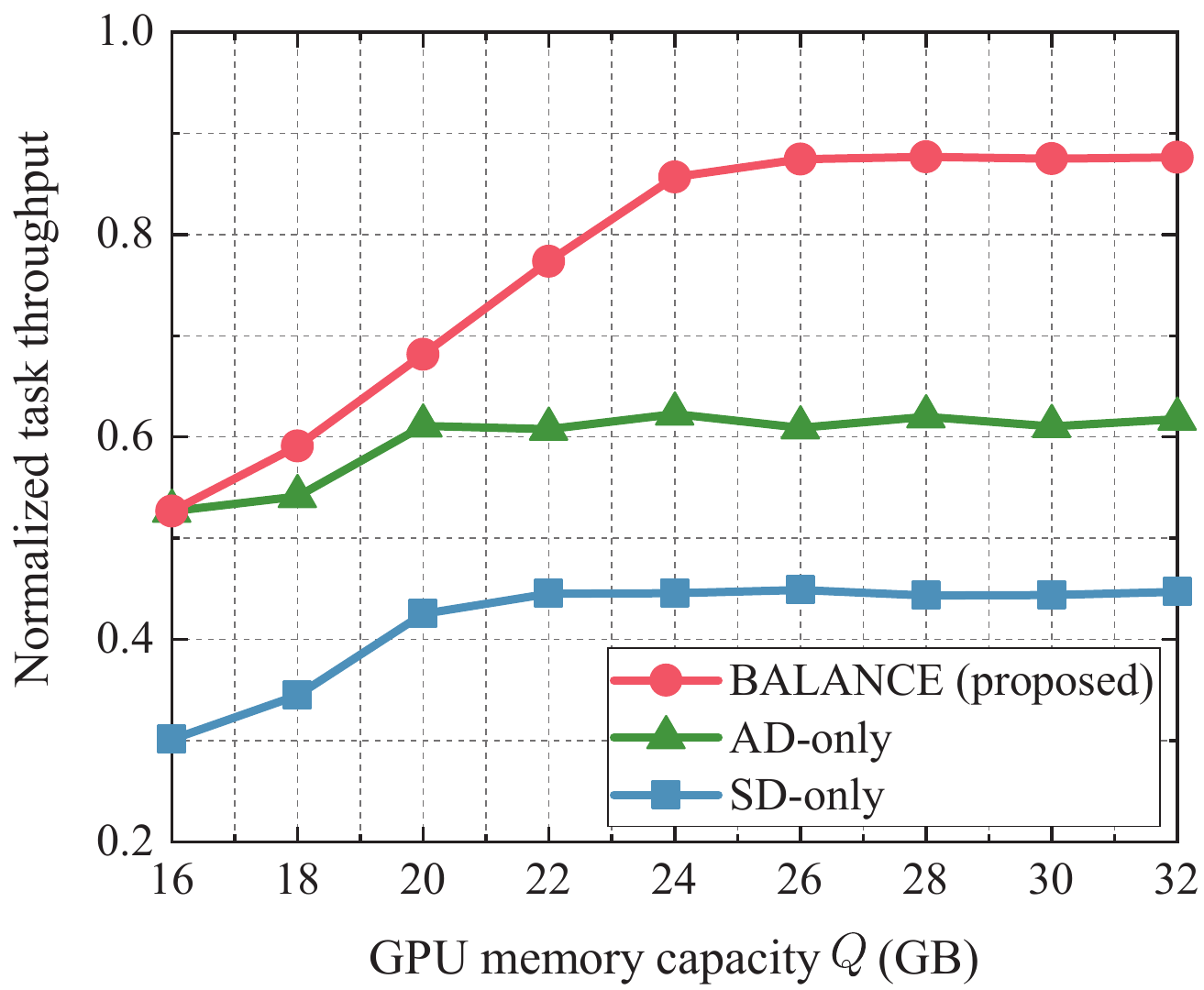}\label{fig:result_memory}}
	\quad
	\subfigure[Results vs. maximum acceptance rate $A_{\max}$.]{\includegraphics[height=3.4cm, keepaspectratio]{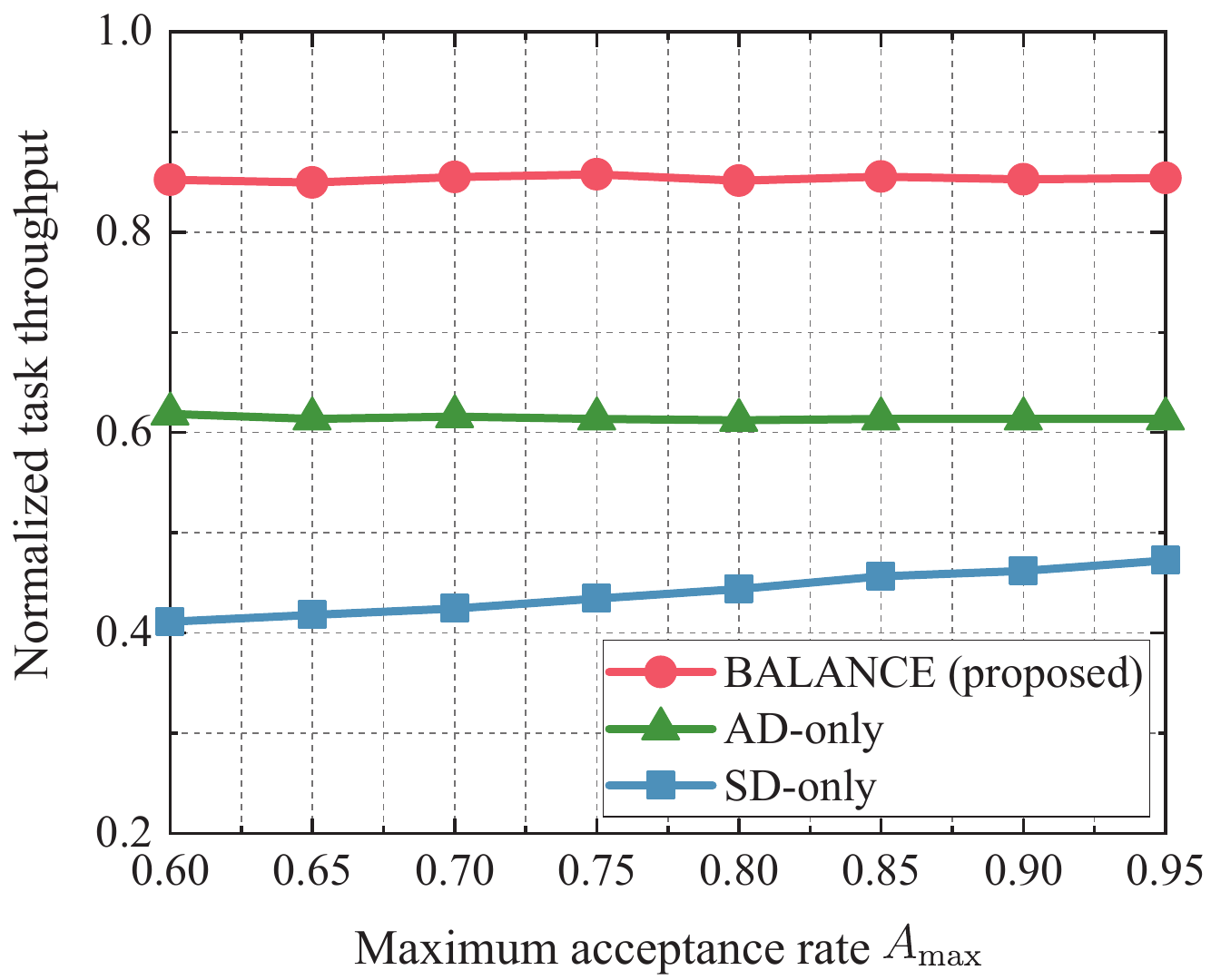}\label{fig:result_arate}}
    \quad
    \subfigure[Results vs. draft length $L^{\text{D}}$.]{\includegraphics[height=3.4cm, keepaspectratio]{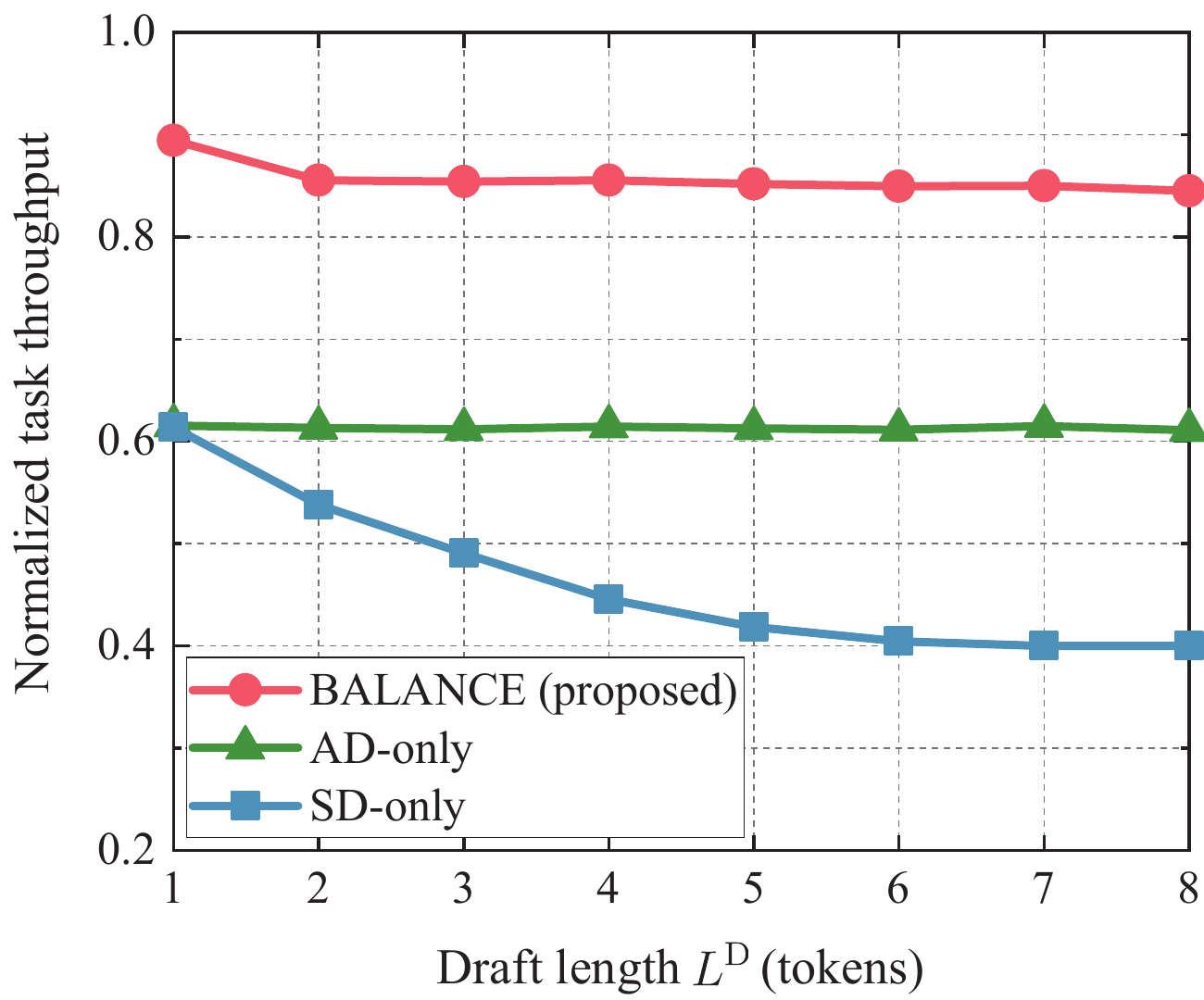}\label{fig:result_draft}}
    \vspace{-0.25cm}
 \caption{Performance of BALANCE under varying $Q$, $A_{\max}$, and $L^{\text{D}}$. The default values of $Q$, $A_{\min}$, $A_{\max}$, and $L^{\text{D}}$ are set to 24 GB, 0.6, 0.8, and 4, respectively. $K$, $L_{\min}^{\text{I}}$, $L_{\max}^{\text{I}}$, $L_{\min}^{\text{O}}$, and $L_{\max}^{\text{O}}$ are set to 40, 300, 700, 50, and 150, respectively. $T_{\min}$ and $T_{\max}$ follow those in Fig.~\ref{fig:result2}.}
 \label{fig:result3}
 \vspace{-10pt}
\end{figure*}

Figs.~\ref{fig:result1},~\ref{fig:result2}, and~\ref{fig:result3} evaluate the normalized task throughput of the proposed BALANCE framework under varying system settings. BALANCE consistently outperforms both AD-only and SD-only baselines, demonstrating its effectiveness in maximizing task throughput. 

In Fig.~\ref{fig:result1}, the throughput decreases with the number of users and increases under looser latency requirements. On average, BALANCE improves throughput over AD-only and SD-only baselines by 37.9\% and 33.4\%, respectively, under varying $K$, and by 31.5\% and 27.3\%, respectively, under varying $T_{\max}$. When $K=70$, BALANCE serves 70.4\% of users, exceeding AD-only and SD-only baselines by 30.4\% and 30.2\%, respectively. Additionally, when $T_{\max}=1$ s, BALANCE achieves a task throughput of 62.9\%, which is 22.9\% and 22.5\% higher than those of AD-only and SD-only baselines, respectively. These results demonstrate the effectiveness of BALANCE in improving task throughput, especially under high user density and stringent latency requirements.

Fig.~\ref{fig:result2} further evaluates BALANCE under different input and output length settings. Specifically, Fig.~\ref{fig:result_input} shows average gains of 38.8\% and 23.9\% over AD-only and SD-only baselines, respectively, while Fig.~\ref{fig:result_output} shows corresponding gains of 30.4\% and 27.3\%. In addition, Fig.~\ref{fig:result3} evaluates BALANCE under varying GPU memory capacity, acceptance rate, and draft token length. Compared with the AD-only and SD-only baselines, BALANCE improves the average throughput by 17.4\% and 35.4\% in Fig.~\ref{fig:result_memory}, by 23.9\% and 41.3\% in Fig.~\ref{fig:result_arate}, and by 24.4\% and 39.3\% in Fig.~\ref{fig:result_draft}, respectively.

\subsection{Algorithm Running Time Comparisons}\label{sec:exp_algorithm_time}

\begin{figure}[!t]
    \centering
    \subfigure[Running time vs. number of users $K$.]{\includegraphics[height=3.1cm, keepaspectratio]{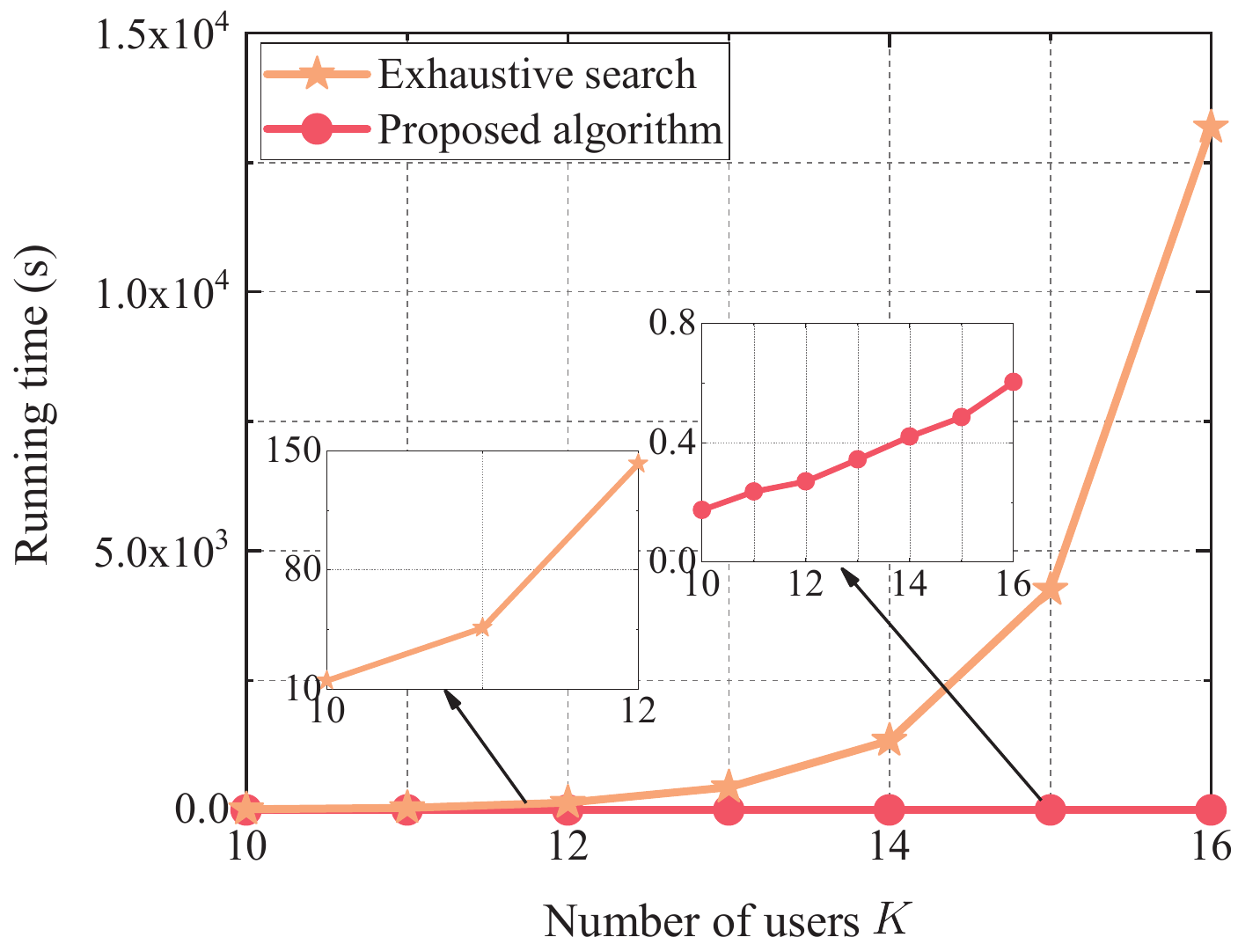}\label{fig:result_time}}
	\quad
	\subfigure[Normalized task throughput vs. number of users $K$.]{\includegraphics[height=3.1cm, keepaspectratio]{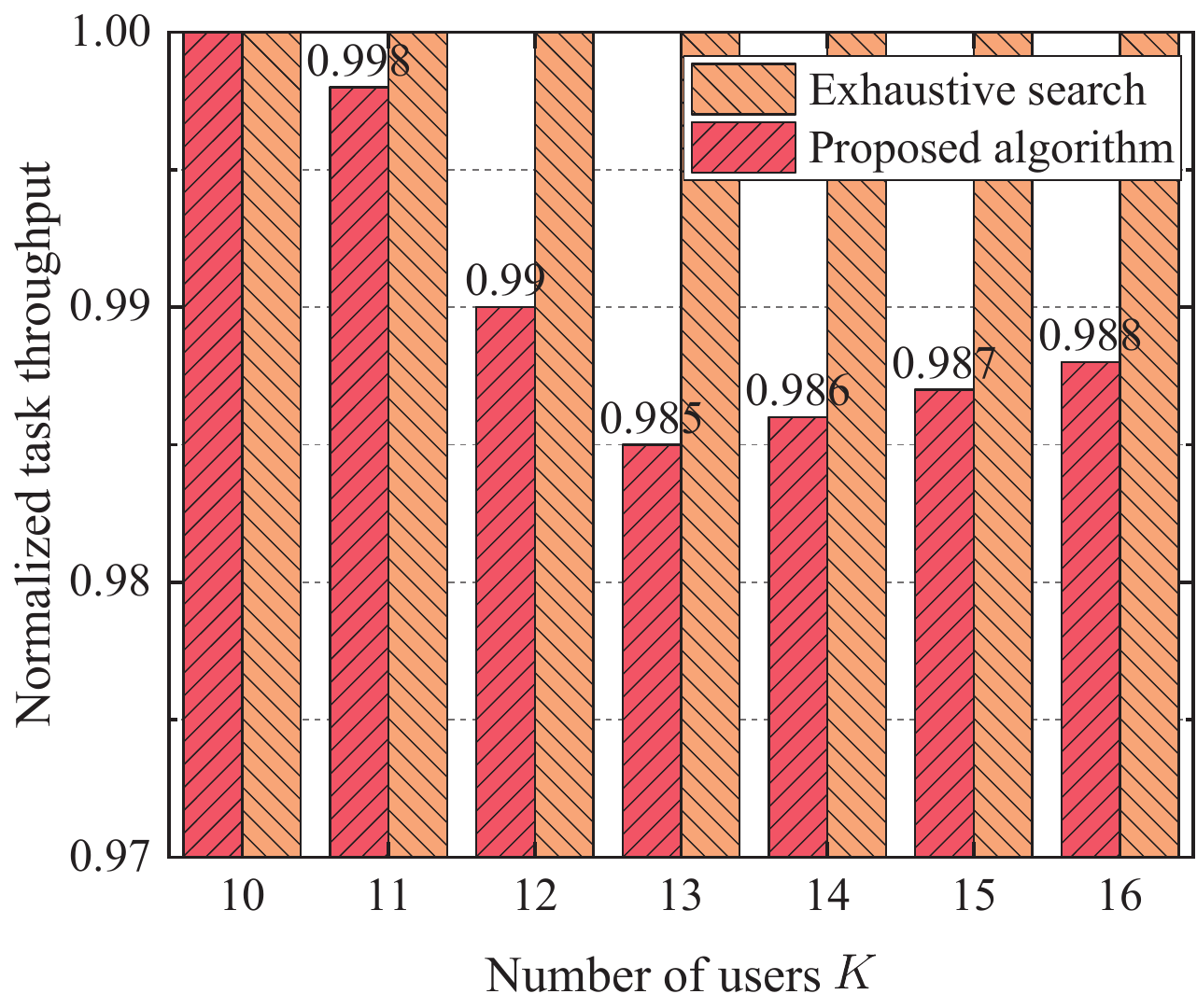}\label{fig:result_time_throughput}}
    \vspace{-0.25cm}
 \caption{Comparisons of algorithm running time and task throughput between the proposed algorithm and the exhaustive search under varying $K$. The system parameters are the same as those in Fig.~\ref{fig:result_user}.}
 \label{fig:result_time_and_throughput}
 \vspace{-10pt}
\end{figure}

Fig.~\ref{fig:result_time_and_throughput} compares the algorithm running time and task throughput of the proposed algorithm with exhaustive search under varying $K$. To run the exhaustive search efficiently, $K$ is varied from 10 to 16. As shown in Fig.~\ref{fig:result_time}, the running time of the proposed algorithm grows nearly linearly with $K$, while exhaustive search increases nearly exponentially. The proposed algorithm achieves an average acceleration of about 5,100 times and up to 21,800 times at $K=16$. Moreover, Fig.~\ref{fig:result_time_throughput} shows that although the proposed algorithm produces a sub-optimal solution to the original problem, its task throughput remains close to that of exhaustive search, with an average degradation of only about 0.9\%.

\subsection{Ablation Study}
\begin{figure*}[!t]
    \centering
    \subfigure[Ablation results vs. number of users $K$.]{\includegraphics[height=3.1cm, keepaspectratio]{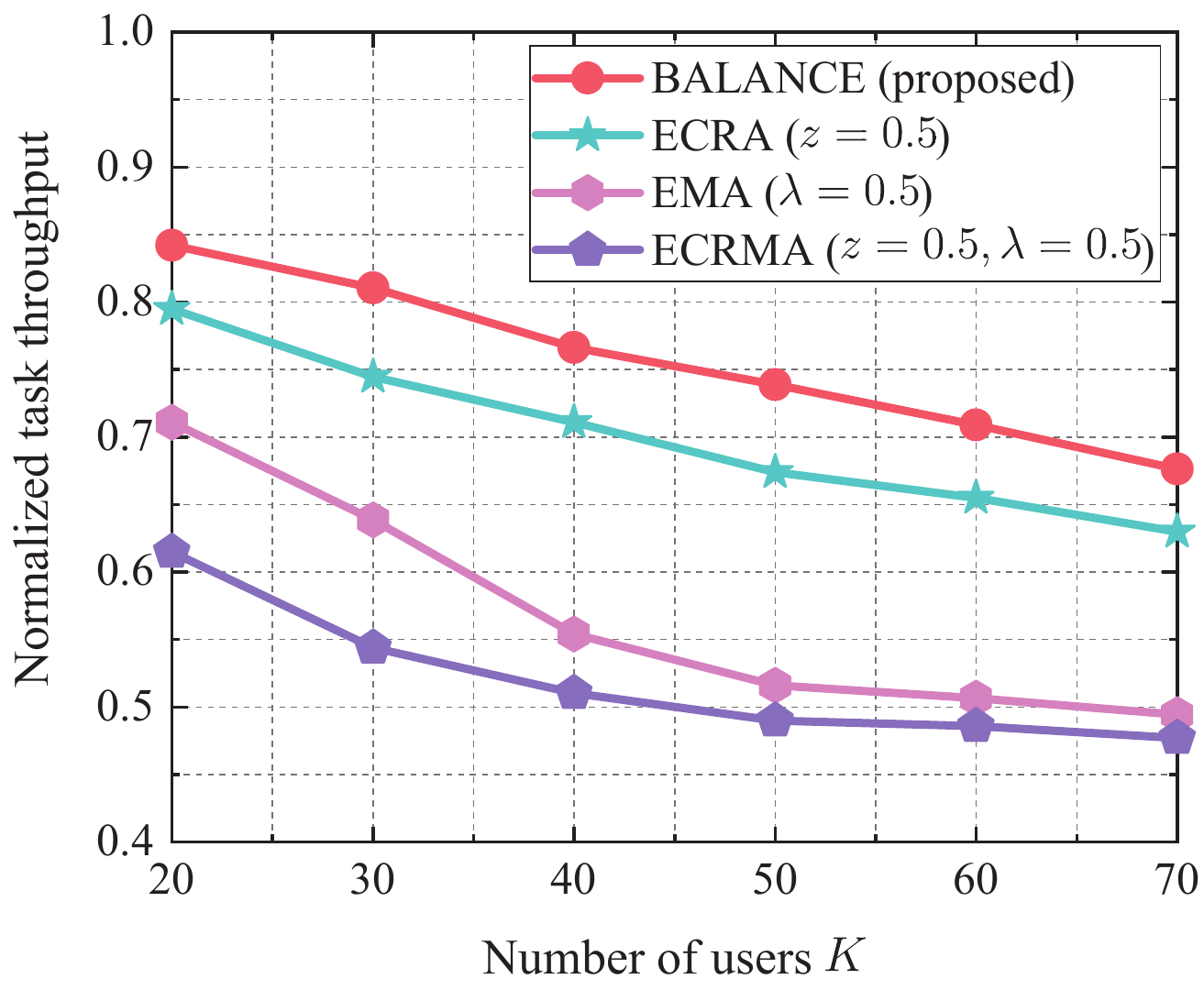}\label{fig:result_ablation_user}}
	\quad
	\subfigure[Ablation results vs. maximum latency requirement $T_{\max}$.]{\includegraphics[height=3.1cm, keepaspectratio]{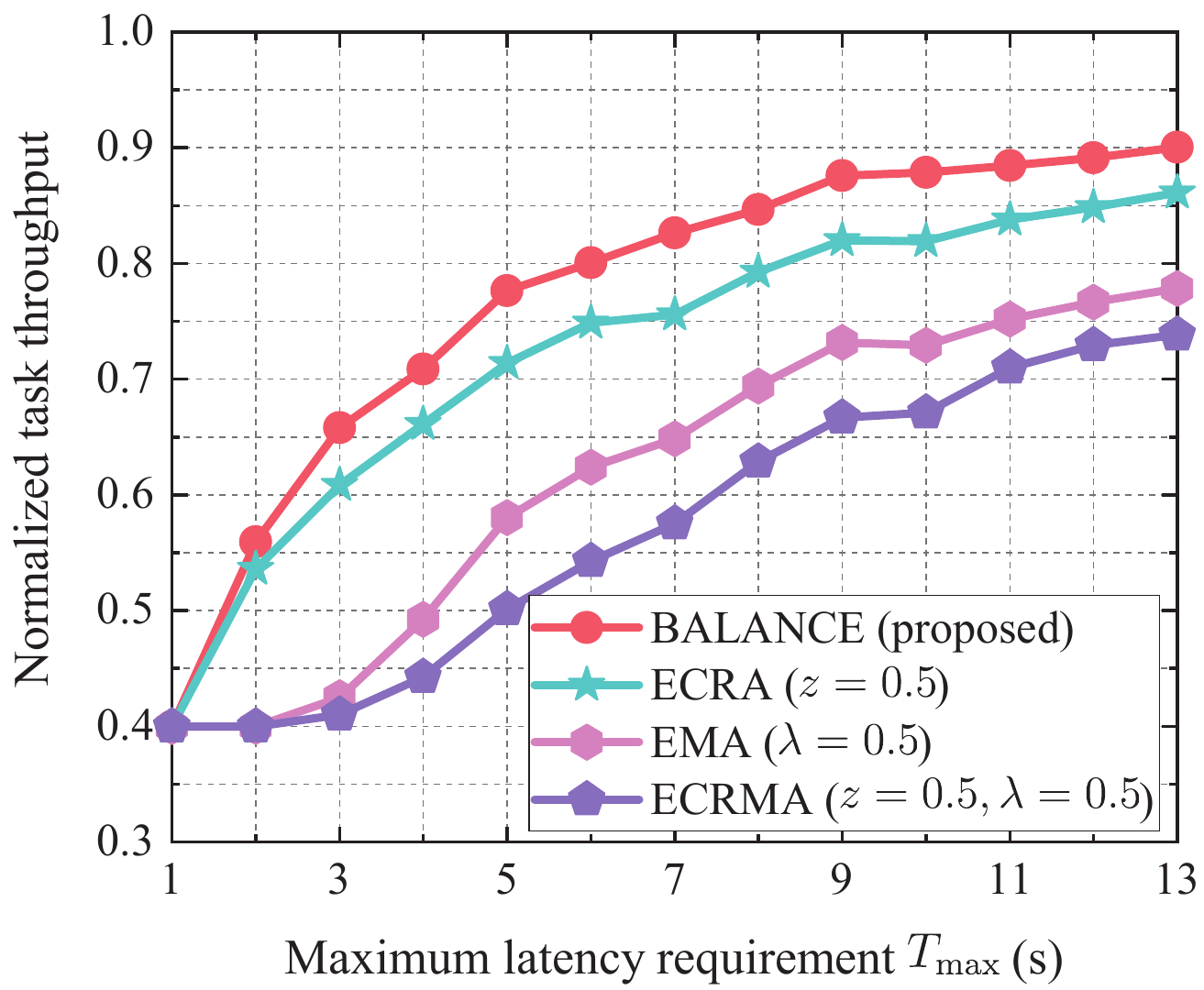}\label{fig:result_ablation_ddl}}
    \quad
	\subfigure[Ablation results vs. GPU memory capacity $Q$.]{\includegraphics[height=3.1cm, keepaspectratio]{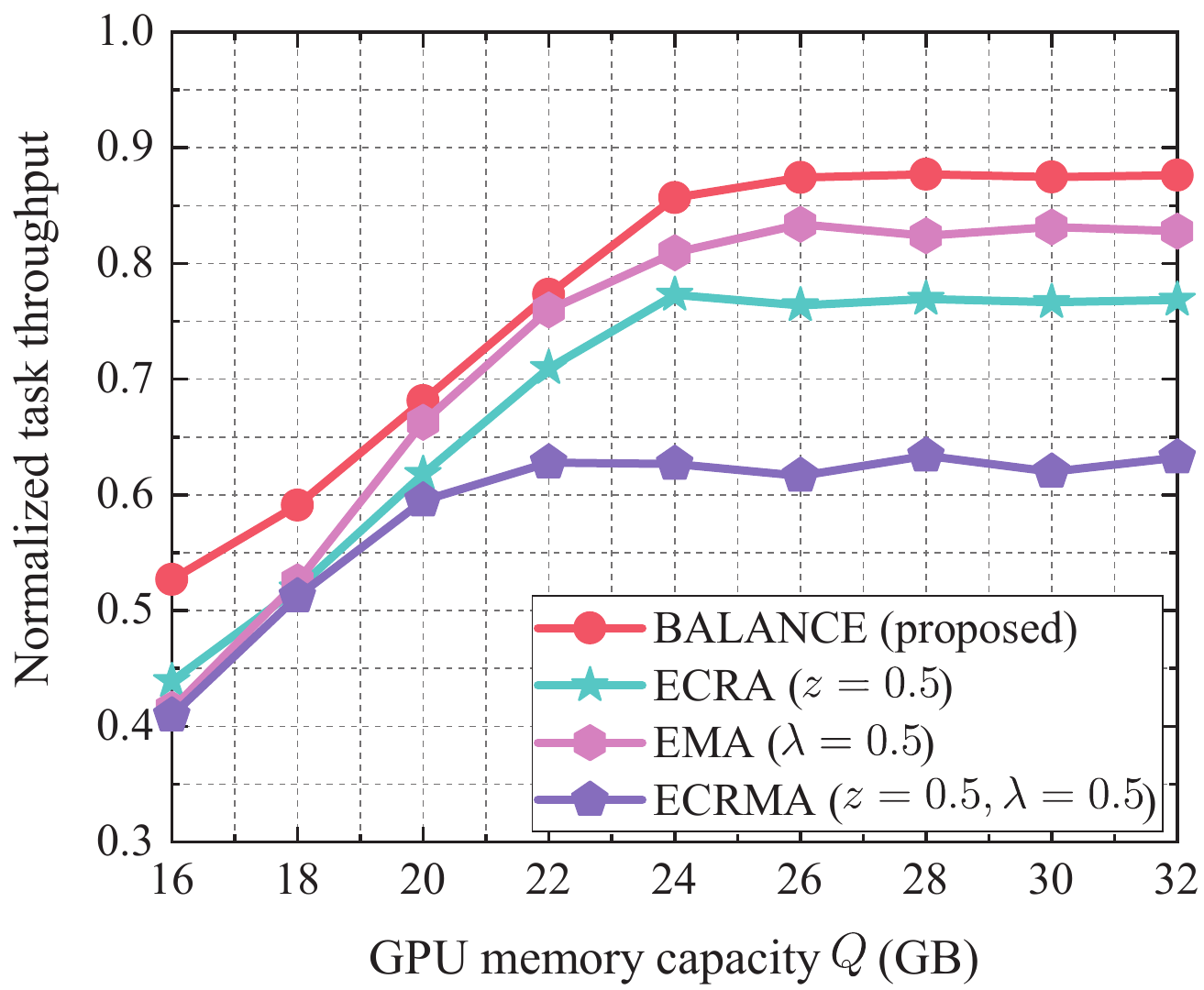}\label{fig:result_ablation_memory}}
    \caption{Ablation study results for the proposed algorithm. The system parameters of Fig.~\ref{fig:result_ablation_user}, Fig.~\ref{fig:result_ablation_ddl}, and Fig.~\ref{fig:result_ablation_memory} are the same as those in Fig.~\ref{fig:result_user}, Fig.~\ref{fig:result_ddl}, and Fig.~\ref{fig:result_memory}, respectively.}
 \label{fig:result_ablation}
 \vspace{-10pt}
\end{figure*}

We conduct ablation studies to evaluate the contribution of each component in the proposed algorithm. We compare the proposed algorithm with three variants, including equal computing resource allocation (ECRA), equal memory allocation (EMA), and equal computing resource and memory allocation (ECRMA). In ECRA, $z=0.5$, while $\lambda$ and user admission are optimized using Algorithm~\ref{algorithm_all}. In EMA, $\lambda=0.5$, while $z$ and user admission are optimized using Algorithm~\ref{algorithm_all}. In ECRMA, both $z$ and $\lambda$ are 0.5, and only user admission is optimized using Algorithm~\ref{algorithm_all}. Fig.~\ref{fig:result_ablation} illustrates the performance gains of the proposed algorithm over ECRA, EMA, and ECRMA under varying $K$, $T_{\max}$, and $Q$. Specifically, in Fig.~\ref{fig:result_ablation_user}, the proposed algorithm improves the task throughput by 5.6\%, 18.7\%, and 23.7\% on average compared with ECRA, EMA, and ECRMA, respectively. The corresponding average performance improvements are 9.0\%, 4.9\%, and 18.4\% in Fig.~\ref{fig:result_ablation_ddl}, and 4.6\%, 15.3\%, and 19.9\% in Fig.~\ref{fig:result_ablation_memory}. These results confirm the effectiveness and necessity of the joint optimization in the proposed algorithm. 

\subsection{Impact of Inaccurate Predictions of Output Length and Acceptance Rate}\label{exp_prediction_impact}
This subsection studies the impact of inaccurate predictions of the output length $L_{k}^{\text{O}}$ and acceptance rate $A_{k}$ on the performance of the proposed BALANCE framework. In Section~\ref{sec:balance_procedure}, user admission and resource allocation are determined based on the expected output length and acceptance rate. However, during the inference process, the actual output length and acceptance rate may deviate from their expected values. Such deviations cause mismatches in the numbers of AD decoding steps, SD iterations, and accepted draft tokens, which may degrade task throughput. To quantify the performance degradation, we first determine the user admission and computing resource allocation using Algorithm~\ref{algorithm_all} based on the expected output length and acceptance rate, and then evaluate the task throughput using the actual values. Following previous works, the actual output length is generated from a log-t distribution~\cite{zheng2026scheduling}, and the actual acceptance rate is generated from a Beta distribution~\cite{ferrari2004beta}. Their means are set equal to the corresponding expected values. The coefficient of variation (CV) characterizes the variability of the actual acceptance rate and output length, and the degrees-of-freedom parameter $\nu$ characterizes the tail heaviness of the log-t output length distribution. A larger CV indicates higher variability, and a smaller $\nu$ indicates a heavier tail.

\begin{figure}[!t]
    \centering
    \subfigure[Results vs. number of users $K$.]{\includegraphics[height=3.4cm, keepaspectratio]{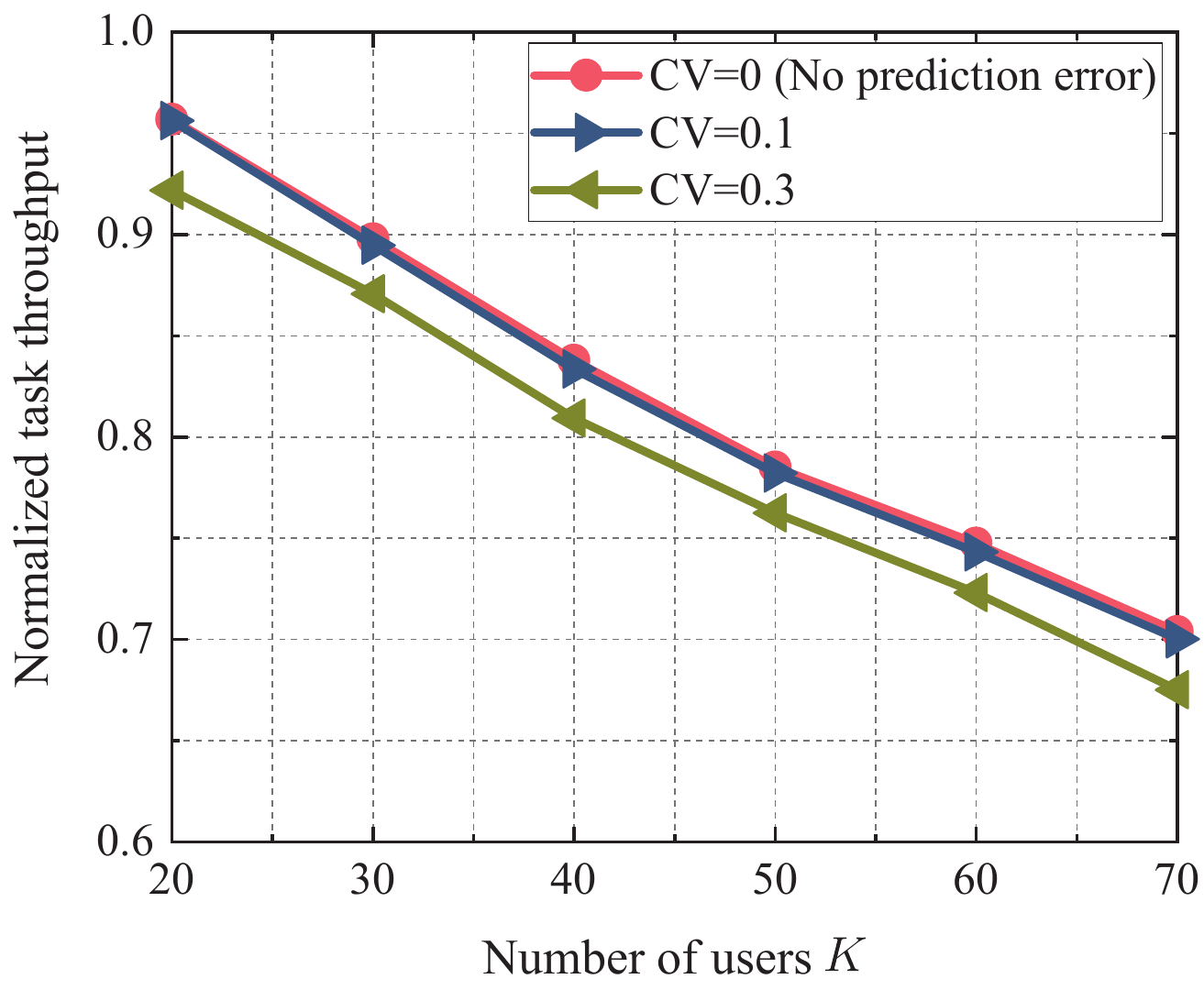}\label{fig:result_dy_rate_user}}
	\quad
	\subfigure[Results vs. maximum output length $L_{\max}^{\text{O}}$.]{\includegraphics[height=3.4cm, keepaspectratio]{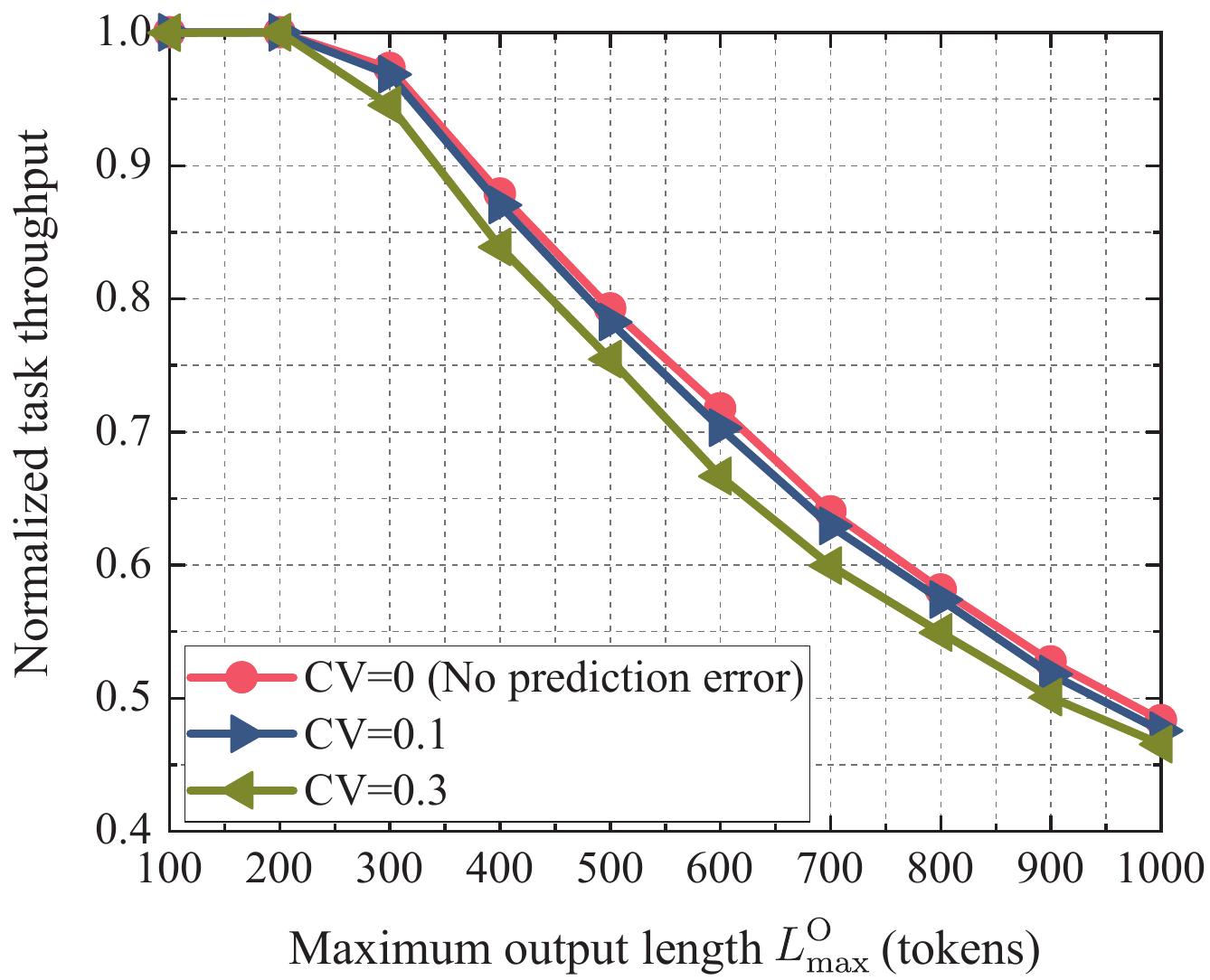}\label{fig:result_dy_rate_output}}
    \vspace{-0.25cm}
 \caption{Performance evaluation under inaccurate acceptance rate prediction. The system parameters of Fig.~\ref{fig:result_dy_rate_user} and Fig.~\ref{fig:result_dy_rate_output} are the same as those in Fig.~\ref{fig:result_user} and Fig.~\ref{fig:result_output}, respectively.}
 \label{fig:result_dy_rate}
 \vspace{-10pt}
\end{figure}

Fig.~\ref{fig:result_dy_rate} shows that BALANCE is robust to inaccurate acceptance rate prediction under varying CV values, where CV=0 corresponds to the case without prediction errors. As shown in Figs.~\ref{fig:result_dy_rate_user} and~\ref{fig:result_dy_rate_output}, when CV=0.1, the task throughput decreases by only 0.3\% and 0.7\%, respectively, compared with CV=0. Even when CV increases to 0.3, the corresponding degradation remains limited to 2.8\% and 4.3\%, respectively. 

\begin{figure}[!t]
    \centering
    \subfigure[Results vs. number of users $K$.]{\includegraphics[height=3.4cm, keepaspectratio]{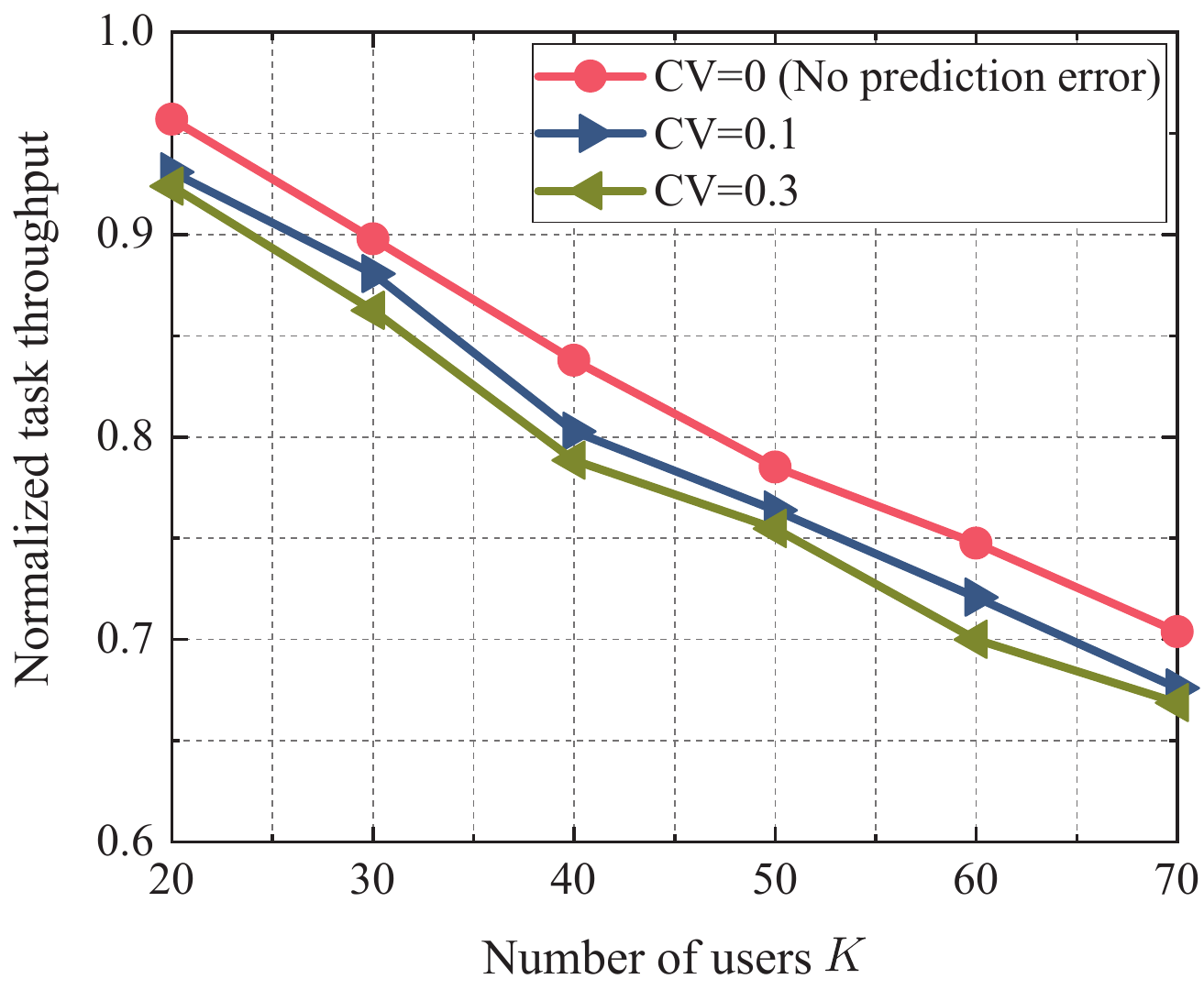}\label{fig:result_dy_output_nu_19_user}}
	\quad
	\subfigure[Results vs. maximum output length $L_{\max}^{\text{O}}$.]{\includegraphics[height=3.4cm, keepaspectratio]{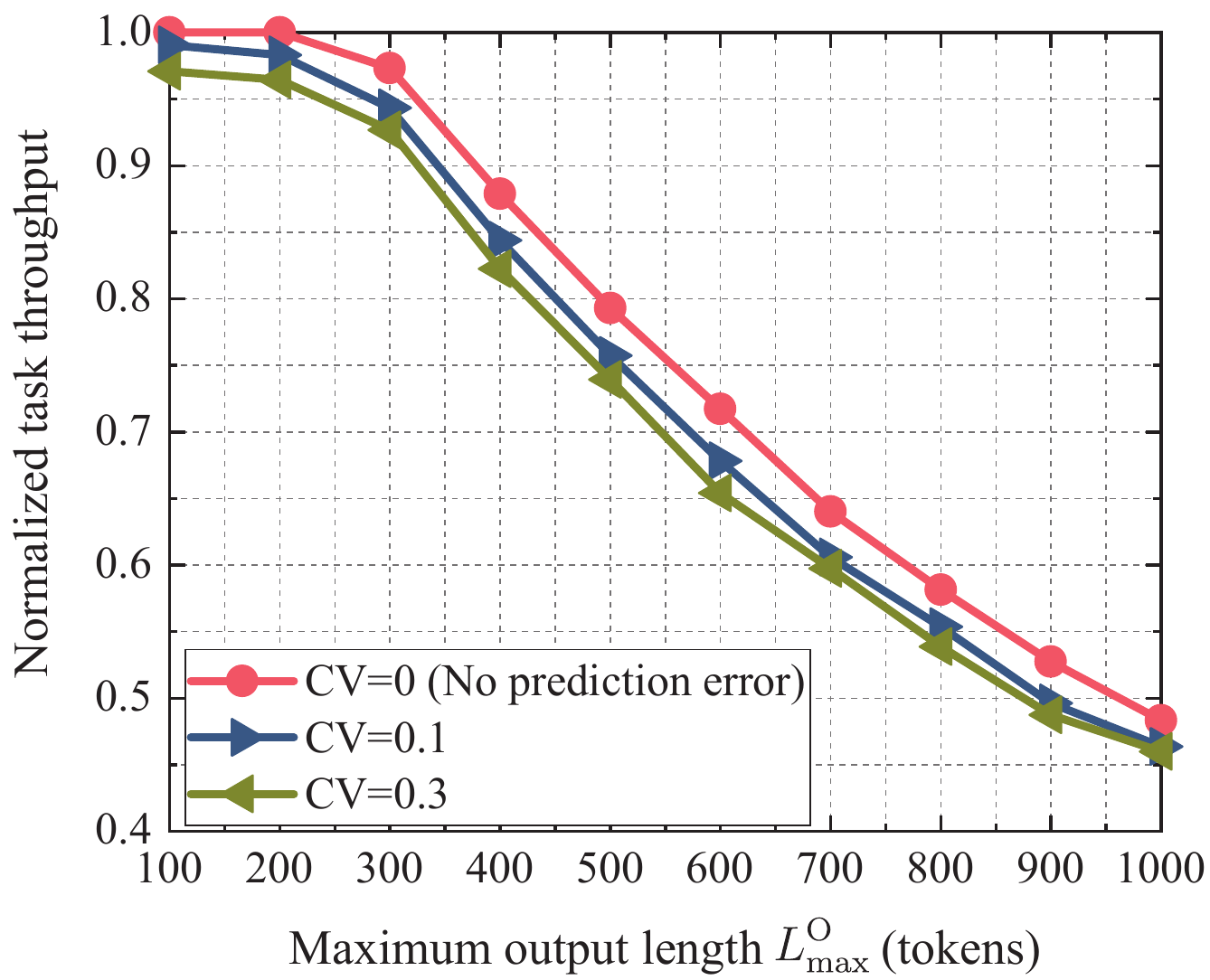}\label{fig:result_dy_output_nu_19_output}}
    \vspace{-0.25cm}
 \caption{Performance evaluation under inaccurate output length prediction with fixed $\nu$. The system parameters of Fig.~\ref{fig:result_dy_output_nu_19_user} and Fig.~\ref{fig:result_dy_output_nu_19_output} follow those in Fig.~\ref{fig:result_user} and Fig.~\ref{fig:result_output}, respectively.}
 \label{fig:result_dy_output_nu_19}
 \vspace{-10pt}
\end{figure}

\begin{figure}[!t]
    \centering
    \subfigure[Results vs. number of users $K$.]{\includegraphics[height=3.4cm, keepaspectratio]{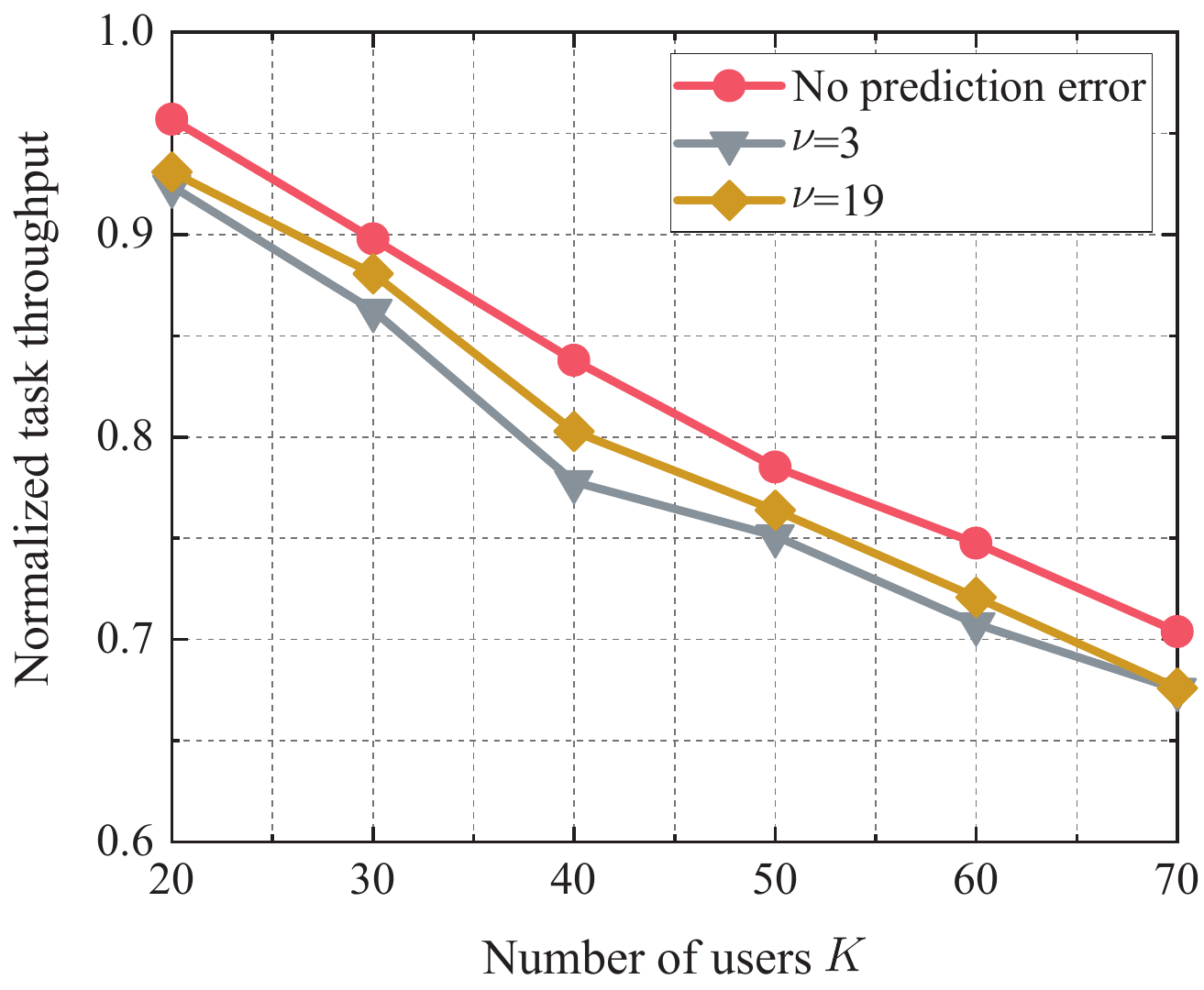}\label{fig:result_dy_output_cv_0.1_user}}
	\quad
	\subfigure[Results vs. maximum output length $L_{\max}^{\text{O}}$.]{\includegraphics[height=3.4cm, keepaspectratio]{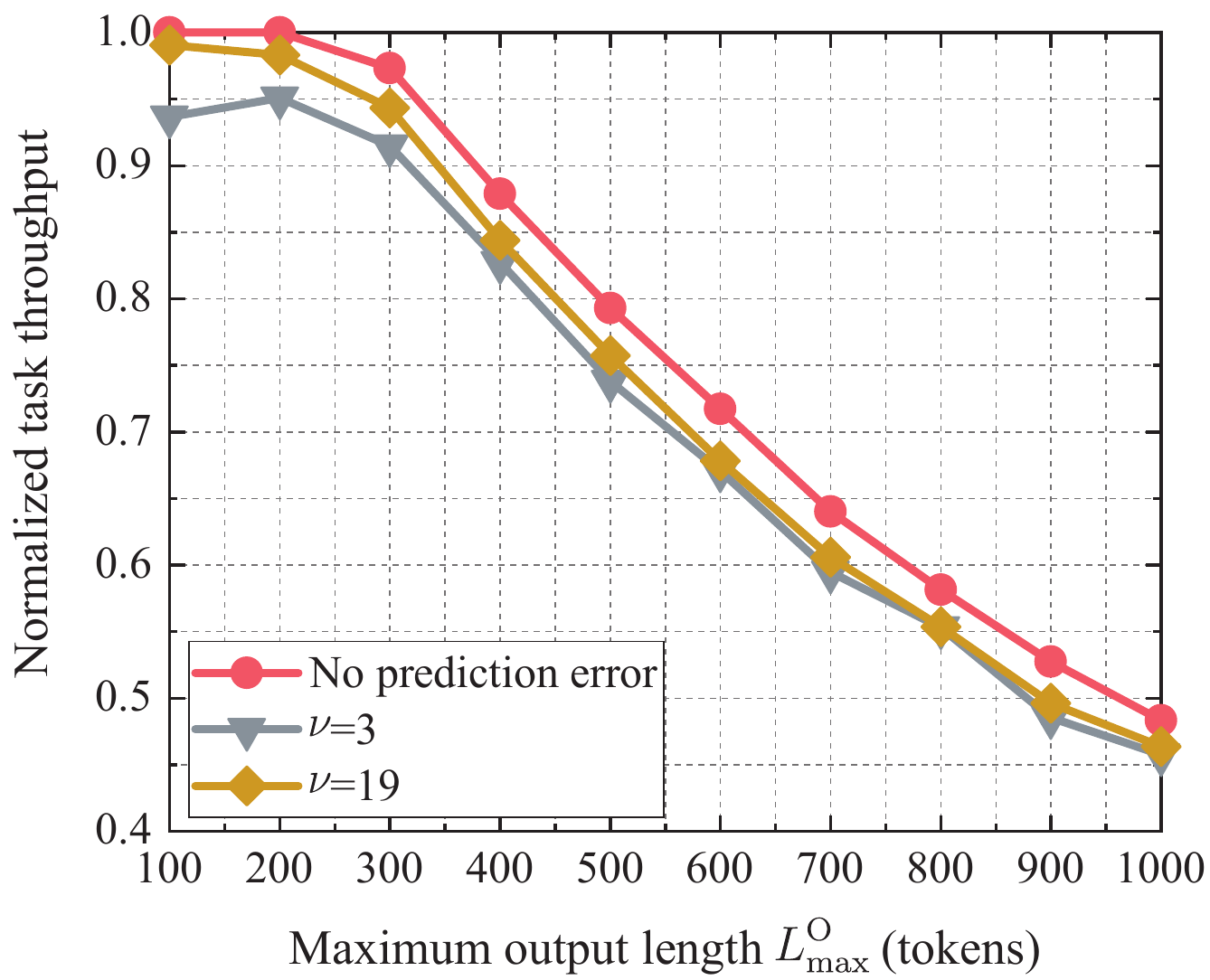}\label{fig:result_dy_output_cv_0.1_output}}
    \vspace{-0.25cm}
 \caption{Performance evaluation under inaccurate output length prediction with fixed CV. The system parameters of Fig.~\ref{fig:result_dy_output_cv_0.1_user} and Fig.~\ref{fig:result_dy_output_cv_0.1_output} are the same as those in Fig.~\ref{fig:result_user} and Fig.~\ref{fig:result_output}, respectively.}
 \label{fig:result_dy_output_cv_0.1}
 \vspace{-10pt}
\end{figure}

Figs.~\ref{fig:result_dy_output_nu_19} and~\ref{fig:result_dy_output_cv_0.1} further show that BALANCE maintains stable throughput under output length prediction errors with different variability and tail heaviness. Fig.~\ref{fig:result_dy_output_nu_19} studies the impact of output length variability with fixed $\nu=19$. As shown in Figs.~\ref{fig:result_dy_output_nu_19_user} and~\ref{fig:result_dy_output_nu_19_output}, when CV=0.1, the task throughput decreases by only 2.6\% and 2.8\%, respectively, compared with CV=0. Even when CV=0.3, the corresponding degradation is only 3.9\% and 3.6\%, respectively. 
Fig.~\ref{fig:result_dy_output_cv_0.1} then investigates the impact of tail heaviness with fixed CV=0.1. As shown in Figs.~\ref{fig:result_dy_output_cv_0.1_user} and~\ref{fig:result_dy_output_cv_0.1_output}, when $\nu=19$, the task throughput decreases by only 2.6\% and 2.9\%, respectively, compared with the case without prediction errors. Even under a heavier-tailed distribution with $\nu=3$, the degradation remains 3.8\% and 4.7\%, respectively. 

\subsection{Performance Evaluation with Other SLM-LLM Pairs}\label{sec:pair_result}
\begin{figure}[!t]
    \centering
    \subfigure[Results vs. number of users $K$.]{\includegraphics[height=3.4cm, keepaspectratio]{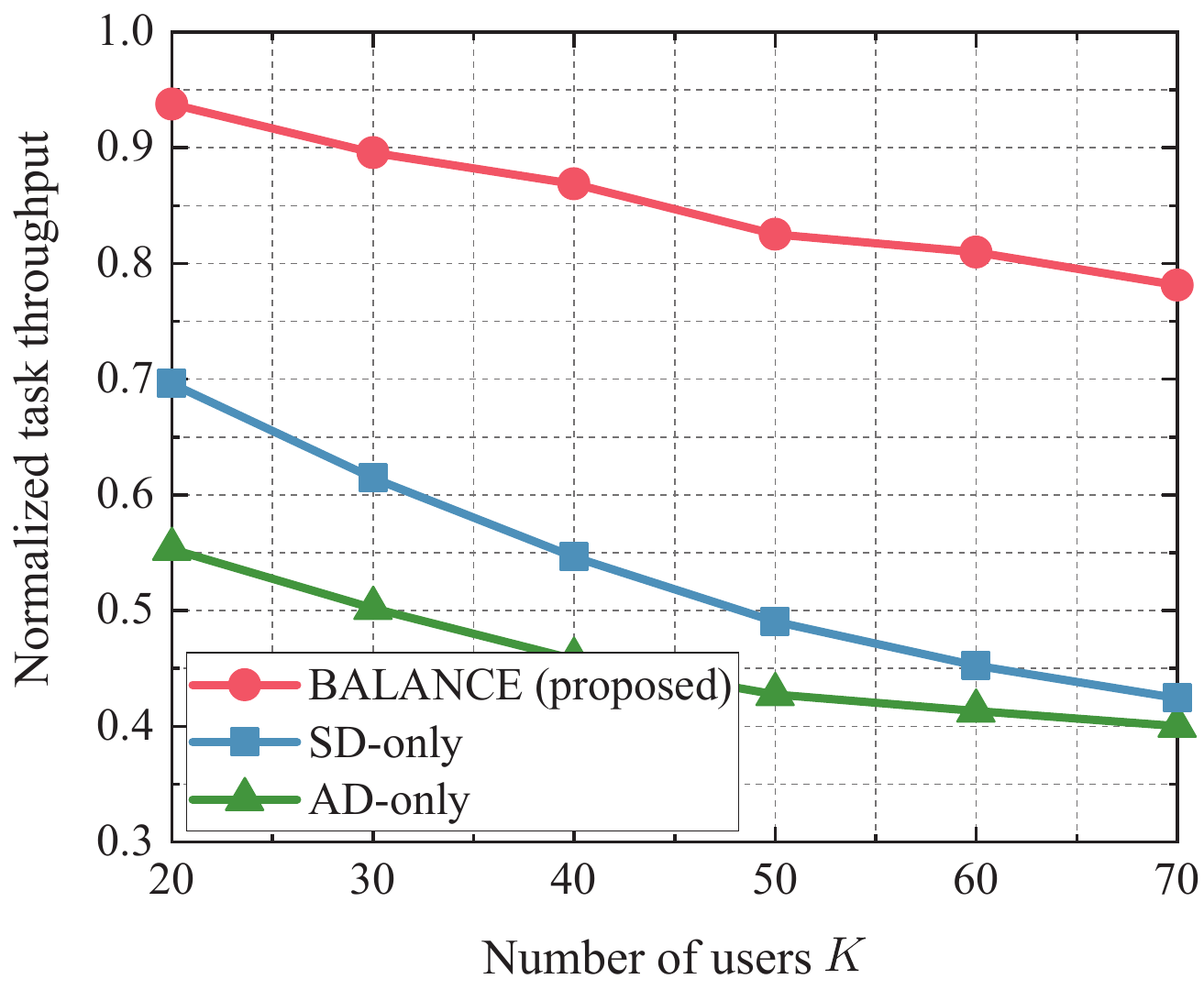}\label{fig:result_1B_3B_user}}
	\quad
    \subfigure[Results vs. GPU memory capacity $Q$.]{\includegraphics[height=3.4cm, keepaspectratio]{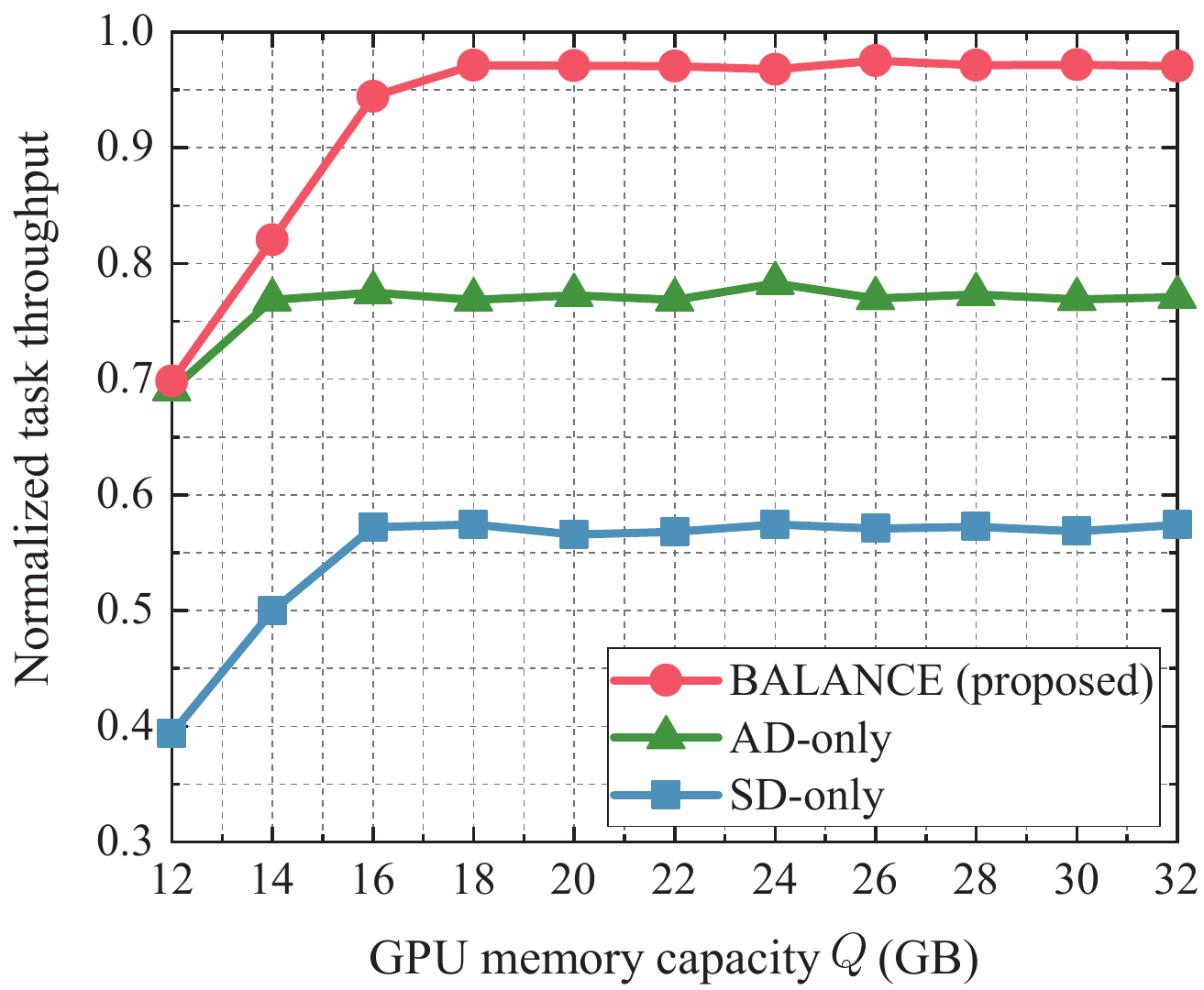}\label{fig:result_1B_3B_memory}}
    \vspace{-0.25cm}
    \caption{Performance evaluation with TinyLlama-1.1B as the SLM and Llama-3.2-3B as the LLM. $A_{\min}$ and $A_{\max}$ are set to 0.65 and 0.85, respectively. The other system parameters of Fig.~\ref{fig:result_1B_3B_user} and Fig.~\ref{fig:result_1B_3B_memory} are the same as those in Fig.~\ref{fig:result_user} and Fig.~\ref{fig:result_memory}, respectively.}
 \label{fig:result_1B_3B}
 \vspace{-10pt}
\end{figure}

\begin{figure}[!t]
    \centering
    \subfigure[Results vs. number of users $K$.]{\includegraphics[height=3.4cm, keepaspectratio]{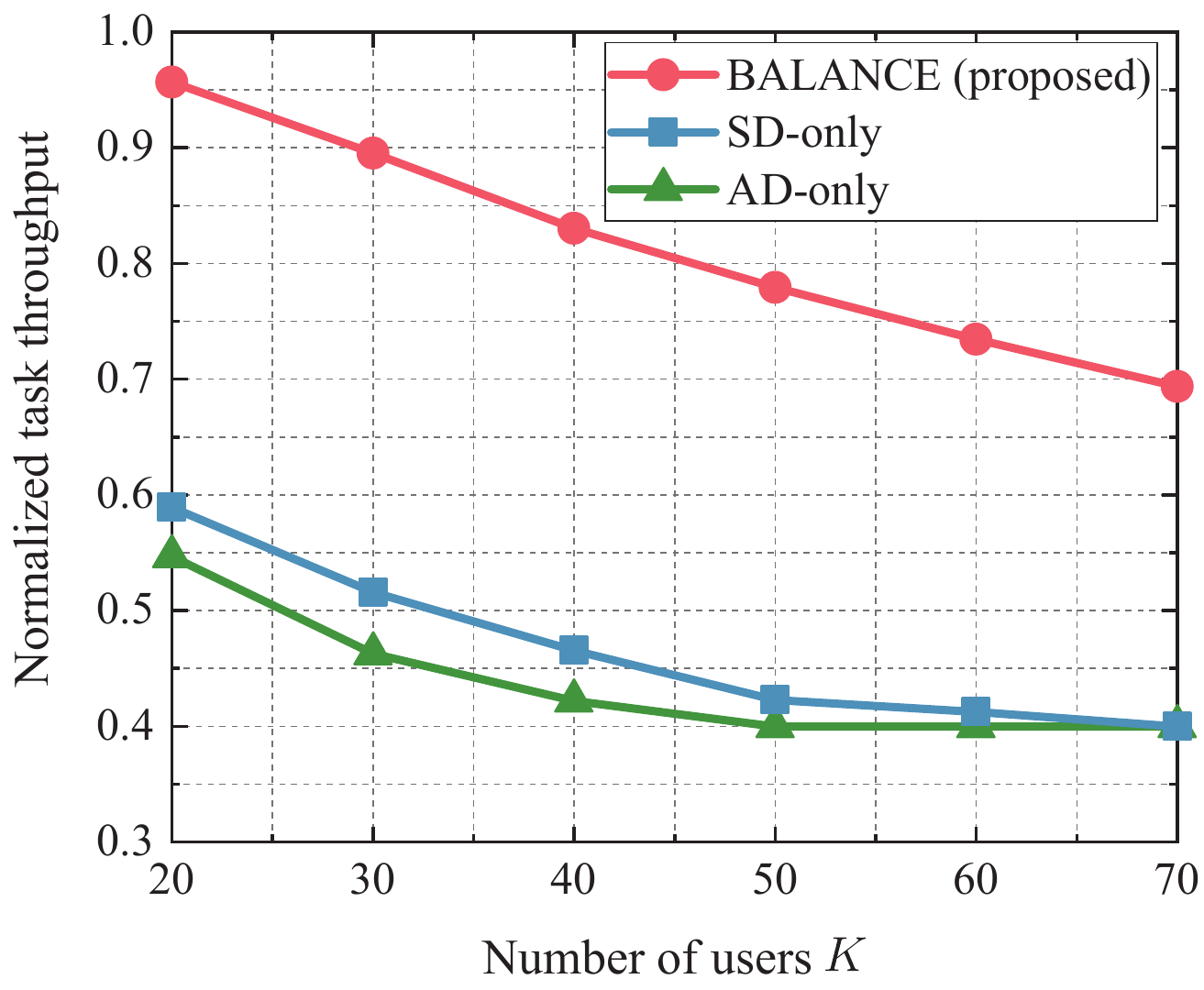}\label{fig:result_3B_7B_user}}
    \quad
	\subfigure[Results vs. GPU memory capacity $Q$.]{\includegraphics[height=3.4cm, keepaspectratio]{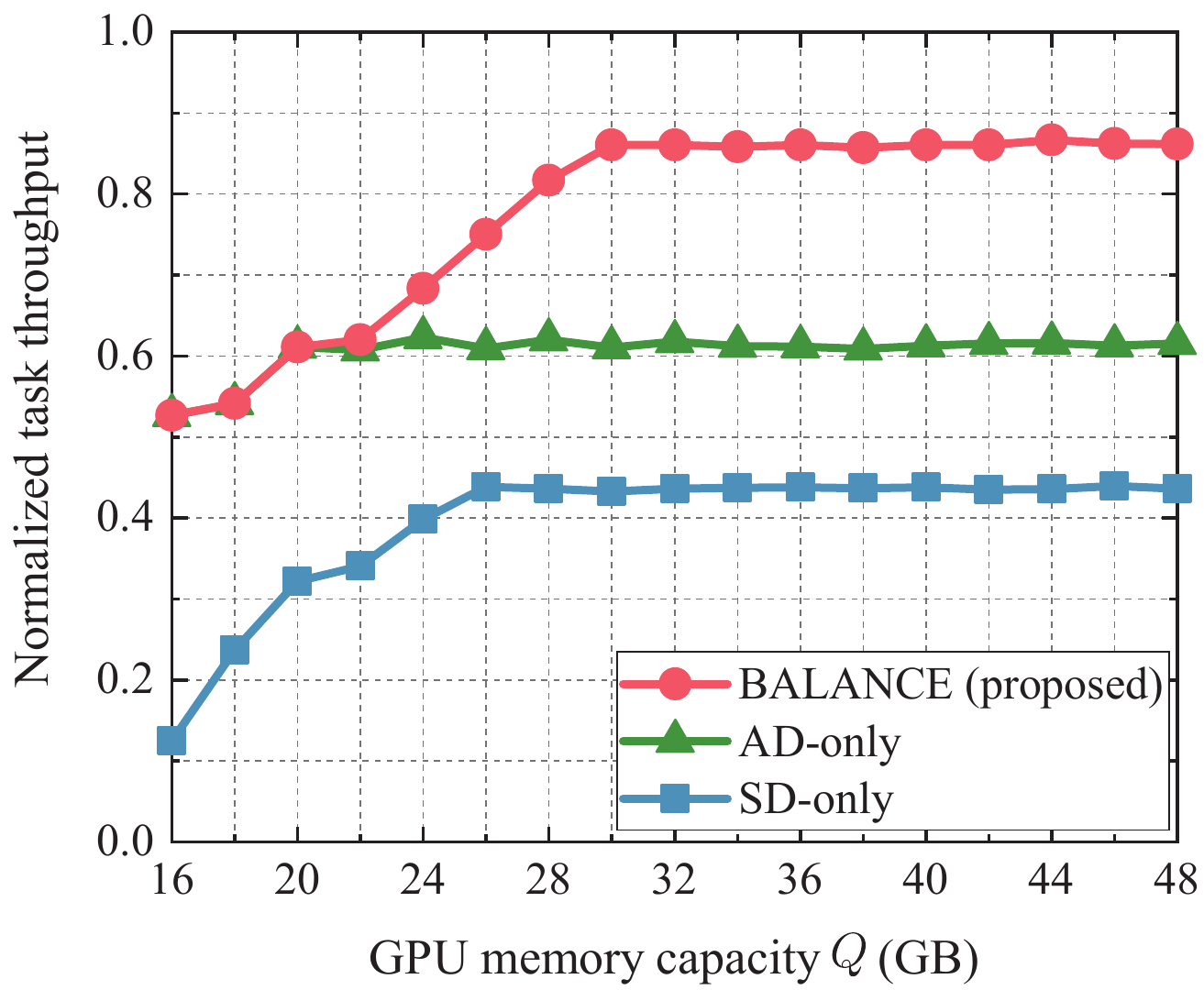}\label{fig:result_3B_7B_memory}}
    \vspace{-0.25cm}
    \caption{Performance evaluation with Llama-3.2-3B as the SLM and Llama-2-7B as the LLM. To accommodate this SLM-LLM pair, the GPU memory capacity $Q$ is set to 32 GB. $A_{\min}$ and $A_{\max}$ are set to 0.7 and 0.9, respectively. The other system parameters of Fig.~\ref{fig:result_3B_7B_user} and Fig.~\ref{fig:result_3B_7B_memory} are the same as those in Fig.~\ref{fig:result_user} and Fig.~\ref{fig:result_memory}, respectively.}
 \label{fig:result_3B_7B}
 \vspace{-10pt}
\end{figure}

This subsection examines the performance of the proposed BALANCE framework with additional SLM-LLM pairs. We first evaluate BALANCE with TinyLlama-1.1B as the SLM and Llama-3.2-3B as the LLM. As shown in Fig.~\ref{fig:result_1B_3B}, BALANCE achieves average task throughput gains of 39.4\% and 31.6\% over the AD-only and SD-only baselines, respectively, under varying $K$. When $Q$ varies, the corresponding gains are 16.6\% and 38.2\%, respectively.
We then consider Llama-3.2-3B as the SLM and Llama-2-7B as the LLM. In Fig.~\ref{fig:result_3B_7B}, BALANCE improves the task throughput over the AD-only and SD-only baselines by 37.6\% and 34.7\% on average, respectively, under varying $K$, and by 17.0\% and 38.2\% on average, respectively, under varying $Q$. These results demonstrate that BALANCE consistently improves task throughput across different SLM-LLM pairs. 

\section{Conclusion}\label{sec:conclusion}
In this paper, we investigated a hy\underline{b}rid \underline{a}utoregressive-specu\underline{la}tive infere\underline{nce} (BALANCE) framework for edge LLM inference to balance the latency-memory tradeoff between conventional autoregressive decoding (AD) and speculative decoding (SD). In BALANCE, an edge server hosts both a small language model and a large language model, and simultaneously supports AD and SD for different users with heterogeneous demands. To maximize the number of served users, we formulated a joint user admission and computing resource allocation problem under end-to-end latency requirements and server memory constraints. Since this problem is an NP-hard problem, we transformed it into two sub-problems and developed a polynomial-time algorithm with a constant approximation guarantee. The experimental results showed that BALANCE consistently outperforms the conventional AD and SD schemes and achieves significant improvements in task throughput. Overall, BALANCE provides an effective and practical solution for resource-constrained edge LLM inference systems.

\bibliographystyle{IEEEtran}
\bibliography{IEEEabrv,reference}

\begin{thebibliography}{10}
\providecommand{\url}[1]{#1}
\csname url@samestyle\endcsname
\providecommand{\newblock}{\relax}
\providecommand{\bibinfo}[2]{#2}
\providecommand{\BIBentrySTDinterwordspacing}{\spaceskip=0pt\relax}
\providecommand{\BIBentryALTinterwordstretchfactor}{4}
\providecommand{\BIBentryALTinterwordspacing}{\spaceskip=\fontdimen2\font plus
\BIBentryALTinterwordstretchfactor\fontdimen3\font minus \fontdimen4\font\relax}
\providecommand{\BIBforeignlanguage}[2]{{%
\expandafter\ifx\csname l@#1\endcsname\relax
\typeout{** WARNING: IEEEtran.bst: No hyphenation pattern has been}%
\typeout{** loaded for the language `#1'. Using the pattern for}%
\typeout{** the default language instead.}%
\else
\language=\csname l@#1\endcsname
\fi
#2}}
\providecommand{\BIBdecl}{\relax}
\BIBdecl

\bibitem{10.1145/3809166}
Z.~Chen, B.~Zhu, J.~Wang, H.~Shin, A.~Nallanathan, and D.~T. Niyato, ``Network edge inference for large language models: Principles, techniques, and opportunities,'' \emph{ACM Comput. Surv.}, vol.~58, no.~12, pp. 1--35, May 2026.

\bibitem{chen2026siftmoe}
Q.~Chen, X.~Chen, and K.~Huang, ``{SiftMoE}: Similarity-aware energy-efficient expert selection for wireless distributed {MoE} inference,'' \emph{arXiv preprint arXiv:2603.23888}, 2026.

\bibitem{qu2024mobile}
G.~Qu, Q.~Chen, W.~Wei, Z.~Lin, X.~Chen, and K.~Huang, ``Mobile edge intelligence for large language models: A contemporary survey,'' \emph{{IEEE} Commun. Surveys Tuts.}, vol.~27, no.~6, pp. 3820--3860, Dec. 2025.

\bibitem{huang2024d}
J.~Huang, K.~Yuan, C.~Huang, and K.~Huang, ``{D}$^{2}$-{JSCC}: Digital deep joint source-channel coding for semantic communications,'' \emph{{IEEE} J. Sel. Areas Commun.}, vol.~43, no.~4, pp. 1246--1261, Apr. 2025.

\bibitem{11044591}
M.~Hu, Q.~He, and D.~Wu, ``{QLLMS}: Quantization-adaptive {LLM} scheduling for partially informed edge serving systems,'' in \emph{Proc. IEEE Int. Conf. Comput. Commun. (INFOCOM)}, May 2025, pp. 1--10.

\bibitem{11262690}
S.~Jang and R.~Morabito, ``Edge-first language model inference: Models, metrics, and tradeoffs,'' in \emph{Proc. IEEE 45th Int. Conf. Distrib. Comput. Syst. Workshops (ICDCSW)}, Jul. 2025, pp. 309--314.

\bibitem{11123401}
D.~Kafetzis, R.~Khalili, and I.~Koutsopoulos, ``Large language model partitioning for low-latency inference at the edge,'' in \emph{Proc. 23rd Int. Symp. Model. Optim. Mobile, Ad Hoc, Wireless Netw. (WiOpt)}, Aug. 2025, pp. 1--8.

\bibitem{qu2024trimcaching}
G.~Qu, Z.~Lin, Q.~Chen, J.~Li, F.~Liu, X.~Chen, and K.~Huang, ``{TrimCaching}: Parameter-sharing edge caching for {AI} model downloading,'' \emph{{IEEE} Trans. Netw.}, vol.~34, pp. 5207--5222, Apr. 2026.

\bibitem{10.5555/3691938.3691945}
A.~Agrawal, N.~Kedia, A.~Panwar, J.~Mohan, N.~Kwatra, B.~S. Gulavani, A.~Tumanov, and R.~Ramjee, ``Taming throughput-latency tradeoff in {LLM} inference with sarathi-serve,'' in \emph{Proc. 18th USENIX Conf. Oper. Syst. Des. Implement. (OSDI)}, Jul. 2024.

\bibitem{vaswani2017attention}
A.~Vaswani, N.~Shazeer, N.~Parmar, J.~Uszkoreit, L.~Jones, A.~N. Gomez, {\L}.~Kaiser, and I.~Polosukhin, ``Attention is all you need,'' in \emph{Proc. Adv. Neural Inform. Process. Syst. (NeurIPS)}, Long Beach, CA, USA, Dec. 2017, pp. 5998--6008.

\bibitem{jaiswal-etal-2024-ffn}
A.~K. Jaiswal, B.~Hu, L.~Yin, Y.~Ro, T.~Chen, S.~Liu, and A.~Akella, ``{FFN}-{S}kip{LLM}: A hidden gem for autoregressive decoding with adaptive feed forward skipping,'' in \emph{Proc. Conf. Empirical Methods Natural Lang. Process. (EMNLP)}, Miami, Florida, USA, Nov. 2024, pp. 16\,943--16\,956.

\bibitem{xia-etal-2024-unlocking}
H.~Xia, Z.~Yang, Q.~Dong, P.~Wang, Y.~Li, T.~Ge, T.~Liu, W.~Li, and Z.~Sui, ``Unlocking efficiency in large language model inference: A comprehensive survey of speculative decoding,'' in \emph{Proc. Findings Assoc. Comput. Linguistics: ACL 2024}, Aug. 2024, pp. 7655--7671.

\bibitem{yin2024theoretical}
M.~Yin, M.~Chen, K.~Huang, and M.~Wang, ``A theoretical perspective for speculative decoding algorithm,'' in \emph{Proc. Adv. Neural Inform. Process. Syst. (NeurIPS)}, vol.~37, Dec. 2024, pp. 128\,082--128\,117.

\bibitem{yan-etal-2025-decoding}
M.~Yan, S.~Agarwal, and S.~Venkataraman, ``Decoding speculative decoding,'' in \emph{Proc. Conf. Nations Americas Chapter Assoc. Comput. Linguistics: Human Lang. Technol.}, Albuquerque, New Mexico, Apr. 2025, pp. 6460--6473.

\bibitem{svirschevski2024specexec}
R.~Svirschevski, A.~May, Z.~Chen, B.~Chen, Z.~Jia, and M.~Ryabinin, ``{SpecExec}: Massively parallel speculative decoding for interactive {LLM} inference on consumer devices,'' in \emph{Proc. Adv. Neural Inform. Process. Syst. (NeurIPS)}, vol.~37, Dec. 2024, pp. 16\,342--16\,368.

\bibitem{pmlr-v267-tiwari25b}
R.~Tiwari, H.~Xi, A.~Tomar, C.~R.~C. Hooper, S.~Kim, M.~Horton, M.~Najibi, M.~W. Mahoney, K.~Keutzer, and A.~Gholami, ``{Q}uant{S}pec: Self-speculative decoding with hierarchical quantized {KV} cache,'' in \emph{Proc. 42nd Int. Conf. Mach. Learn. (ICML)}, Jul. 2025, pp. 59\,668--59\,686.

\bibitem{zhang-etal-2024-draft}
J.~Zhang, J.~Wang, H.~Li, L.~Shou, K.~Chen, G.~Chen, and S.~Mehrotra, ``Draft {\&} verify: Lossless large language model acceleration via self-speculative decoding,'' in \emph{Proc. 62nd Annu. Meeting Assoc. Comput. Linguistics (ACL)}, Bangkok, Thailand, Aug. 2024, pp. 11\,263--11\,282.

\bibitem{xia2024swift}
H.~Xia, Y.~Li, J.~Zhang, C.~Du, and W.~Li, ``{SWIFT}: On-the-fly self-speculative decoding for {LLM} inference acceleration,'' in \emph{Proc. Int. Conf. Learn. Represent. (ICLR)}, Apr. 2025, pp. 1--24.

\bibitem{11493570}
J.~Jiang, Z.~Chen, H.~Shin, and A.~Nallanathan, ``Edge inference for large language models with pipeline parallelism and batching,'' \emph{{IEEE} Trans. Commun.}, vol.~74, pp. 8390--8406, Apr. 2026.

\bibitem{qu2025partialloading}
G.~Qu, Q.~Chen, X.~Chen, K.~Huang, and Y.~Fang, ``{PartialLoading}: User scheduling and bandwidth allocation for parameter-sharing edge inference,'' \emph{arXiv preprint arXiv:2503.22982}, 2025.

\bibitem{zhang2024beyond}
X.~Zhang, J.~Nie, Y.~Huang, G.~Xie, Z.~Xiong, J.~Liu, D.~Niyato, and X.~Shen, ``Beyond the cloud: Edge inference for generative large language models in wireless networks,'' \emph{{IEEE} Trans. Wireless Commun.}, vol.~24, no.~1, pp. 643--658, Jan. 2025.

\bibitem{10683673}
C.~Liu and J.~Zhao, ``Resource allocation in large language model integrated {6G} vehicular networks,'' in \emph{Proc. IEEE 99th Veh. Technol. Conf. (VTC)}, Sep. 2024, pp. 1--6.

\bibitem{11045182}
K.~Zhang, H.~He, S.~Song, J.~Zhang, and K.~B. Letaief, ``Communication-efficient distributed on-device {LLM} inference over wireless networks,'' \emph{{IEEE} J. Sel. Topics Signal Process.}, vol.~19, no.~7, pp. 1301--1317, Oct. 2025.

\bibitem{11395617}
Q.~Chen, X.~Chen, and K.~Huang, ``{SlimCaching}: Edge caching of mixture-of-experts for distributed inference,'' \emph{{IEEE} Trans. Mobile Comput.}, vol.~25, no.~7, pp. 10\,924--10\,938, Jul. 2026.

\bibitem{zhu2025efficient}
B.~Zhu, Z.~Chen, L.~Zhao, H.~Shin, and A.~Nallanathan, ``Efficient {LLM} inference over heterogeneous edge networks with speculative decoding,'' \emph{arXiv preprint arXiv:2510.11331}, 2025.

\bibitem{xu2026dip}
Y.~Xu, S.~Zhou, and Z.~Niu, ``{DiP-SD}: Distributed pipelined speculative decoding for efficient {LLM} inference at the edge,'' \emph{arXiv preprint arXiv:2604.20919}, 2026.

\bibitem{10.1145/3769102.3770608}
X.~Li, D.~Spatharakis, S.~Ghafouri, J.~Fan, H.~Vandierendonck, D.~John, B.~Ji, and D.~S. Nikolopoulos, ``{SLED}: A speculative {LLM} decoding framework for efficient edge serving,'' in \emph{Proc. 10th ACM/IEEE Symp. Edge Comput. (SEC)}, Dec. 2025.

\bibitem{wang2025opt}
J.~Wang, Y.~Su, J.~Li, Q.~Xia, Z.~Ye, X.~Duan, Z.~Wang, and M.~Zhang, ``{OPT}-tree: Speculative decoding with adaptive draft tree structure,'' \emph{Transactions of the Association for Computational Linguistics}, vol.~13, pp. 188--199, Feb. 2025.

\bibitem{liu2025pearl}
T.~Liu, Y.~Li, Q.~Lv, K.~Liu, J.~Zhu, W.~Hu, and X.~Sun, ``{PEARL}: Parallel speculative decoding with adaptive draft length,'' in \emph{Proc. Int. Conf. Learn. Represent. (ICLR)}, vol. 2025, May 2025, pp. 1085--1104.

\bibitem{11274878}
A.~Albaseer, E.~Bentafat, M.~Hamood, M.~Abdallah, A.~Al-Fuqaha, and M.~Hamdi, ``Think fast, infer smart: A hybrid distributed {LLMs} inference at the wireless edge,'' in \emph{Proc. IEEE 36th Int. Symp. Pers., Indoor Mobile Radio Commun. (PIMRC)}, Sep. 2025, pp. 1--6.

\bibitem{ding2024hybrid}
D.~Ding, A.~Mallick, C.~Wang, R.~Sim, S.~Mukherjee, V.~Ruhle, L.~V. Lakshmanan, and A.~H. Awadallah, ``Hybrid {LLM}: Cost-efficient and quality-aware query routing,'' in \emph{Proc. Int. Conf. Learn. Represent. (ICLR)}, Vienna Austria, May 2024, pp. 1--19.

\bibitem{oh2026communication}
S.~Oh, J.~Kim, J.~Park, S.-W. Ko, J.~Choi, T.~Q. Quek, and S.-L. Kim, ``Communication-efficient hybrid language model via uncertainty-aware opportunistic and compressed transmission,'' \emph{{IEEE} Trans. Commun.}, early access 2026.

\bibitem{huang2025jakiro}
H.~Huang, F.~Yang, Z.~Liu, Y.~Xu, J.~Li, Y.~Liu, X.~Yin, D.~Li, P.~Ren, and E.~Barsoum, ``Jakiro: Boosting speculative decoding with decoupled multi-head via {MoE},'' \emph{arXiv preprint arXiv:2502.06282}, 2025.

\bibitem{fang2025and}
M.~Fang, Z.~Fu, Q.~Zhao, and J.~Wang, ``When, what, and how: Rethinking retrieval-enhanced speculative decoding,'' \emph{arXiv preprint arXiv:2511.01282}, 2025.

\bibitem{10.1145/3620665.3640383}
H.~Oh, K.~Kim, J.~Kim, S.~Kim, J.~Lee, D.-s. Chang, and J.~Seo, ``{ExeGPT}: Constraint-aware resource scheduling for {LLM} inference,'' in \emph{in Proc. 29th ACM Int. Conf. Archit. Support Program. Lang. Operating Syst. (ASPLOS)}, Apr. 2024, p. 369–384.

\bibitem{aiops2024qiu}
H.~Qiu, W.~Mao, A.~Patke, S.~Cui, S.~Jha, C.~Wang, H.~Franke, Z.~T. Kalbarczyk, T.~Ba\c{s}ar, and R.~K. Iyer, ``Efficient interactive {LLM} serving with proxy model-based sequence length prediction,'' in \emph{Proc. 5th Int. Workshop Cloud Intell./AIOps}, vol.~5, San Diego, CA, USA, Apr. 2024, pp. 1--7.

\bibitem{10.1145/3676641.3716011}
R.~Gong, S.~Bai, S.~Wu, Y.~Fan, Z.~Wang, X.~Li, H.~Yang, and X.~Liu, ``Past-future scheduler for {LLM} serving under {SLA} guarantees,'' in \emph{Proc. 30th ACM Int. Conf. Archit. Support Program. Lang. Oper. Syst. (ASPLOS)}, Mar. 2025, pp. 798--813.

\bibitem{yu2022orca}
G.-I. Yu, J.~S. Jeong, G.-W. Kim, S.~Kim, and B.-G. Chun, ``Orca: A distributed serving system for $\{$Transformer-Based$\}$ generative models,'' in \emph{Proc. 16th USENIX Symp. Operating Syst. Design Implement. (OSDI)}, Jul. 2022, pp. 521--538.

\bibitem{khattab2020colbert}
O.~Khattab and M.~Zaharia, ``{ColBERT}: Efficient and effective passage search via contextualized late interaction over {BERT},'' in \emph{Proc. 43rd Int. ACM SIGIR Conf. Res. Develop. Inf. Retrieval (SIGIR)}, Mar. 2020, pp. 39--48.

\bibitem{padding_hugging}
\BIBentryALTinterwordspacing
{Hugging Face}, ``Padding and truncation.'' [Online]. Available: \url{https://huggingface.co/docs/transformers/pad_truncation}
\BIBentrySTDinterwordspacing

\bibitem{padding_pytorch}
\BIBentryALTinterwordspacing
{PyTorch}, ``Reference {API}.'' [Online]. Available: \url{https://docs.pytorch.org/docs/stable/generated/torch.nn.utils.rnn.pad_sequence.html}
\BIBentrySTDinterwordspacing

\bibitem{10.5555/3618408.3619696}
Y.~Sheng, L.~Zheng, B.~Yuan, Z.~Li, M.~Ryabinin, B.~Chen, P.~Liang, C.~R\'{e}, I.~Stoica, and C.~Zhang, ``{FlexGen}: High-throughput generative inference of large language models with a single {GPU},'' in \emph{{Proc. 40th Int. Conf. Mach. Learn. (ICML)}}, Honolulu, Hawaii, USA, Jul. 2023, pp. 31\,094--31\,116.

\bibitem{nvidia_mig}
\BIBentryALTinterwordspacing
{NVIDIA}, ``{NVIDIA} multi-instance {GPU} user guide release r580,'' 2025. [Online]. Available: \url{https://docs.nvidia.com/datacenter/tesla/pdf/MIG_User_Guide.pdf}
\BIBentrySTDinterwordspacing

\bibitem{xu2022igniter}
F.~Xu, J.~Xu, J.~Chen, L.~Chen, R.~Shang, Z.~Zhou, and F.~Liu, ``{iGniter}: Interference-aware {GPU} resource provisioning for predictable {DNN} inference in the cloud,'' \emph{{IEEE} Trans. Parallel Distrib. Syst.}, vol.~34, no.~3, pp. 812--827, Mar. 2023.

\bibitem{qu2026slide}
G.~Qu, T.~Li, Q.~Chen, X.~Chen, and S.~Zhou, ``{SLIDE}: Simultaneous model downloading and inference at the wireless network edge,'' \emph{{IEEE} Trans. Mobile Comput.}, pp. 1--12, early access 2026.

\bibitem{9843917}
W.~Shi, S.~Zhou, Z.~Niu, M.~Jiang, and L.~Geng, ``Multiuser co-inference with batch processing capable edge server,'' \emph{{IEEE} Trans. Wireless Commun.}, vol.~22, no.~1, pp. 286--300, Jan. 2023.

\bibitem{wu2023graft}
J.~Wu, L.~Wang, Q.~Jin, and F.~Liu, ``Graft: Efficient inference serving for hybrid deep learning with {SLO} guarantees via {DNN} re-alignment,'' \emph{{IEEE} Trans. Parallel Distrib. Syst.}, vol.~35, no.~2, pp. 280--296, Feb. 2024.

\bibitem{zheng2026scheduling}
H.~Zheng, Y.~Zhang, F.~Fu, X.~Zhou, H.~Luo, H.~Zhu, Y.~Zhu, H.~Wang, X.~Yan, and J.~Jiang, ``Scheduling {LLM} inference with uncertainty-aware output length predictions,'' \emph{arXiv preprint arXiv:2604.00499}, 2026.

\bibitem{ferrari2004beta}
S.~Ferrari and F.~Cribari-Neto, ``Beta regression for modelling rates and proportions,'' \emph{J. Appl. Stat.}, vol.~31, no.~7, pp. 799--815, Aug. 2004.

\bibitem{shojaei2013fast}
H.~Shojaei, T.~Basten, M.~Geilen, and A.~Davoodi, ``A fast and scalable multidimensional multiple-choice knapsack heuristic,'' \emph{ACM Trans. Design Autom. Electron. Syst.}, vol.~18, no.~4, pp. 1--32, Oct. 2013.

\bibitem{hifi2004heuristic}
M.~Hifi, M.~Michrafy, and A.~Sbihi, ``Heuristic algorithms for the multiple-choice multidimensional knapsack problem,'' \emph{J. Oper. Res. Soc.}, vol.~55, no.~12, pp. 1323--1332, Jul. 2004.

\end{thebibliography}

\newpage
\clearpage
\setcounter{page}{1}
\begin{appendices}

\section{Proof of Proposition~\ref{proposition_1}}\label{proof_proposition_1}
We prove the NP-hardness of $\mathcal{P}1$ by considering a special case where $z$ is fixed. In this case, the edge server only needs to determine the user admission decisions under the latency requirements and the GPU memory constraint. This special case can be reduced from a multi-dimensional multiple-choice knapsack problem (MMKP). 
In a general MMKP, items are partitioned into multiple groups, and at most one item can be selected from each group. Each selected item yields a profit and consumes multiple types of resources, and the objective is to maximize the total profit subject to multiple resource-capacity constraints. 
In the considered special case of $\mathcal{P}1$, each user corresponds to one group in the MMKP, and each candidate service option of this user corresponds to one item in the corresponding group. Specifically, the three candidate items of each user represent being served by the AD mode, being served by the SD mode, and not being served. Selecting an AD or SD item admits the corresponding user to the associated inference mode and yields a unit profit, whereas selecting the non-service item yields zero profit. Meanwhile, each selected service item is associated with latency and GPU memory costs, which correspond to the multi-dimensional resource weights of the item in the MMKP. The latency requirements and the memory budget correspond to the resource constraints in the MMKP. Therefore, the considered special case has the same structure as the MMKP.
Since the MMKP is NP-hard~\cite{shojaei2013fast,hifi2004heuristic}, this special case of $\mathcal{P}1$ is NP-hard. Consequently, the general problem $\mathcal{P}1$ is also NP-hard. This completes the proof.

\section{Proof of Proposition~\ref{proposition_2}}\label{proof_proposition_2}
After reserving the GPU memory for the required model weights, the remaining GPU memory is used to store the KV caches of the AD and SD modes. Since these two modes maintain separate KV caches, their memory consumptions can be separated and allocated according to $\lambda$. Therefore, any feasible solution to $\mathcal{P}1$ corresponds to a feasible solution to $\mathcal{P}2$ with the same user admission decisions and objective value. Conversely, any feasible solution to $\mathcal{P}2$ satisfies the original GPU memory constraint in $\mathcal{P}1$, since $Q_{1}$ and $Q_{2}$ are obtained by partitioning the remaining GPU memory. All other constraints and the objective function remain unchanged. Therefore, $\mathcal{P}1$ and $\mathcal{P}2$ are equivalent. This completes the proof.

\section{Proof of Proposition~\ref{proposition_3}}\label{proof_proposition_3}
When $z$ and $\lambda$ are given, constraints~\eqref{const_9} and~\eqref{const_p2_4}, which specify the feasible ranges of $z$ and $\lambda$, can be removed. The remaining objective function and constraints are exactly those of $\mathcal{P}3$. This completes the proof. 

\section{Proof of Proposition~\ref{proposition_li}}\label{proof_proposition_li}
First, under given $z$ and $\lambda$, the memory budget $Q_{1}$ for the AD mode is fixed.

Second, any feasible solution to $\mathcal{P}3.1$ induces a tuple $\left(L_{1},L_{2},t_{1},\tau_{1}\right)$, which specifies $L_{\max}^{\text{ad,I}}$, $L_{\max}^{\text{ad,O}}$, $t^{\text{ad}}$, and $\tau^{\text{ad,up}}$, respectively. For any fixed tuple $\left(L_{1},L_{2},t_{1},\tau_{1}\right)$, the users that can possibly be admitted to the AD mode are exactly those in the candidate set $\mathcal{K}_{1}$ shown in~\eqref{eq_K1}. Moreover, under this tuple, each candidate user is associated with a determined latency contribution $\tilde{t}_{k}^{\text{ad}}$ and the same memory contribution $\tilde{q}^{\text{ad}}$. Therefore, maximizing the number of served users under the memory budget $Q_{1}$ and the latency budget $\Delta t_{1}$ is equivalent to selecting the maximum number of users from $\mathcal{K}_{1}$ such that the two constraints in~\eqref{eq_d1} are satisfied. Since each selected user yields the same unit profit and has the same memory contribution, the optimal selection under the fixed tuple is obtained by choosing users with the smallest latency contributions first. Therefore, under the fixed tuple, the optimal served user set is the prefix $\left\{\Pi_{1}\left(1\right),\dots,\Pi_{1}\left(d_{1}^{*}\right)\right\}$, where $d_{1}^{*}$ is given in~\eqref{eq_d1}. The corresponding user admission decision is then given by \eqref{eq_hat_x}.

Finally, since Algorithm~\ref{algorithm_li} enumerates all feasible tuples $\left(L_{1},L_{2},t_{1},\tau_{1}\right)$, it must examine the tuple induced by any optimal solution. For each feasible tuple, Algorithm~\ref{algorithm_li} computes the corresponding $d_{1}^{*}$ and keeps $\hat{\mathcal{K}}^{\text{ad}}$ yielding the largest $d_{1}^{*}$. Therefore, the obtained $\hat{{\bf{X}}}$ maximizes the number of served users among all feasible solutions to $\mathcal{P}3.1$ under given $z$ and $\lambda$. This completes the proof.

\section{Proof of Proposition~\ref{proposition_sd}}\label{proof_proposition_sd}
First, under given $z$ and $\lambda$, the memory budget $Q_{2}$ for the SD mode is fixed.

Second, any feasible solution to $\mathcal{P}3.2$ induces a tuple $\left(L_{3},L_{4},\tilde{N}_{1},t_{2},\tau_{2}\right)$, which specifies $L_{\max}^{\text{sd,I}}$, $L_{\max}^{\text{sd,A}}$, $N_{\max}^{\text{sd}}$, $t^{\text{sd}}$, and $\tau^{\text{sd,up}}$, respectively. Consider any fixed tuple $\left(L_{3},L_{4},\tilde{N}_{1},t_{2},\tau_{2}\right)$. Under this tuple and the given remaining user set $\mathcal{K}\setminus\hat{\mathcal{K}}^{\text{ad}}$, the users that can possibly be admitted to the SD mode are exactly those in the candidate set $\mathcal{K}_{2}$ shown in~\eqref{eq_K2}. Moreover, each candidate user is associated with a determined latency contribution $\tilde{t}_{k}^{\text{sd}}$ and a determined memory contribution $\tilde{q}_{k}^{\text{sd}}$. From~\eqref{eq_tilde_t_k_sd} and~\eqref{eq_tilde_q_k_sd}, both $\tilde{t}_{k}^{\text{sd}}$ and $\tilde{q}_{k}^{\text{sd}}$ are monotonically non-decreasing with $N_{k}$ under the fixed tuple. Therefore, sorting users in non-decreasing order of $\tilde{t}_{k}^{\text{sd}}$ also yields a non-decreasing order of $\tilde{q}_{k}^{\text{sd}}$. Under the latency budget $\Delta t_{2}$ and the memory budget $Q_{2}$, the problem reduces to selecting the maximum number of users from $\mathcal{K}_{2}$ such that the two constraints in \eqref{eq_d2} are satisfied. Hence, under the fixed tuple, the optimal served user set is given by the prefix of $\Pi_{2}$ with length $d_{2}^{*}$. The corresponding user admission decision is then given by \eqref{eq_hat_y}. 

Finally, since Algorithm~\ref{algorithm_sd} enumerates all feasible tuples $\left(L_{3},L_{4},\tilde{N}_{1},t_{2},\tau_{2}\right)$, it must examine the tuple induced by any optimal solution. For each feasible tuple, Algorithm~\ref{algorithm_sd} computes the corresponding $d_{2}^{*}$ and keeps $\hat{\mathcal{K}}^{\text{sd}}$ yielding the largest $d_{2}^{*}$. Therefore, the obtained $\hat{{\bf{Y}}}$ maximizes the number of served users among all feasible solutions to $\mathcal{P}3.2$ under given $z$, $\lambda$, and $\hat{\mathcal{K}}^{\text{ad}}$. This completes the proof.

\section{Proof of Theorem~\ref{theorem_1}}\label{proof_theorem_1}
First, for any feasible ${\bf{X}}$ and ${\bf{Y}}$, since ${\bf{X}}$ and ${\bf{Y}}$ are disjoint, we have $r\left({\bf{X}}+{\bf{Y}}\right)=r\left({\bf{X}}\right)+r\left({\bf{Y}}\right)$ and $r'\left({\bf{Y}}\right)=r\left({\bf{Y}}\right)$. Therefore, $r\left(\dot{{\bf{X}}}\right)+r'\left(\dot{{\bf{Y}}}\right)=r\left(\dot{{\bf{X}}}\right)+r\left(\dot{{\bf{Y}}}\right)=r\left(\dot{{\bf{X}}}+\dot{{\bf{Y}}}\right)$.
    
Second, let $z^{*}$ denote the optimal computing resource allocation of $\mathcal{P}1$, and let the auxiliary variable $\lambda^{*}$ determine the GPU memory resource allocation corresponding to the optimal solution in the equivalent transformed problem $\mathcal{P}2$. Since Algorithm~\ref{algorithm_all} enumerates all feasible $z$ and $\lambda$, including $z^{*}$ and $\lambda^{*}$, the obtained solution satisfies
\begin{equation}\label{eq_optimal_1}
    \begin{aligned}
        &r\left(\dot{{\bf{X}}}+\dot{{\bf{Y}}}\right)\\
        &=r\left(\dot{{\bf{X}}}+\dot{{\bf{Y}}}\ \middle | \ \lambda=\dot{\lambda},z=\dot{z}\right)
        =r\left(\hat{{\bf{X}}}+\hat{{\bf{Y}}}\ \middle | \ \lambda=\dot{\lambda},z=\dot{z}\right)\\
        &\ge r\left(\hat{{\bf{X}}}+\hat{{\bf{Y}}}\ \middle | \ \lambda=\lambda^{*},z=z^{*}\right)
    \\
    &=r\left(\hat{{\bf{X}}}\ \middle | \ \lambda=\lambda^{*},z=z^{*}\right)+r\left(\hat{{\bf{Y}}}\ \middle | \ \lambda=\lambda^{*},z=z^{*}\right).
    \end{aligned}
\end{equation}

Third, based on Proposition~\ref{proposition_li}, we have
\begin{equation}\label{eq_optimal_6}
    r\left(\hat{{\bf{X}}}\ \middle | \ \lambda=\lambda^{*},z=z^{*}\right)\ge r\left({\bf{X}}^{*}\ \middle | \ \lambda=\lambda^{*},z=z^{*}\right)=r\left({\bf{X}}^{*}\right).
\end{equation}    
Next, we decompose ${\bf{Y}}^{*}$ into two disjoint components with respect to $\hat{\mathcal{K}}^{\text{ad}}\big|_{\lambda=\lambda^{*},z=z^{*}}$, i.e., ${\bf{Y}}^{*}={\bf{Y}}_{1}^{*}+{\bf{Y}}_{2}^{*}$. Here, ${\bf{Y}}_{1}^{*}$ corresponds to the users admitted according to ${\bf{Y}}^{*}$ that also belong to $\hat{\mathcal{K}}^{\text{ad}}\big|_{\lambda=\lambda^{*},z=z^{*}}$, while ${\bf{Y}}_{2}^{*}$ corresponds to the users admitted according to ${\bf{Y}}^{*}$ that belong to the remaining user set $\mathcal{K}\setminus\hat{\mathcal{K}}^{\text{ad}}\big|_{\lambda=\lambda^{*},z=z^{*}}$. Specifically, for each user $k$, we define
\begin{equation}
    y_{1,k}^{*}=
    \begin{cases}
        y_{k}^{*}, & \text{if } k\in\hat{\mathcal{K}}^{\text{ad}}\big|_{\lambda=\lambda^{*},z=z^{*}},\\
        0, & \text{otherwise},
    \end{cases}
\end{equation}
and
\begin{equation}
    y_{2,k}^{*}=
        \begin{cases}
            y_{k}^{*}, & \text{if } k\in\mathcal{K}\setminus\hat{\mathcal{K}}^{\text{ad}}\big|_{\lambda=\lambda^{*},z=z^{*}},\\
            0, & \text{otherwise},
        \end{cases}
\end{equation}
where $y_{k}^{*}$, $y_{1,k}^{*}$, and $y_{2,k}^{*}$ denote the $k$-th elements of ${\bf{Y}}^{*}$, ${\bf{Y}}_{1}^{*}$, and ${\bf{Y}}_{2}^{*}$, respectively. Furthermore, based on Proposition~\ref{proposition_sd}, we have
\begin{equation}\label{eq_optimal_2}
    r\left(\hat{{\bf{Y}}}\ \middle | \ \lambda=\lambda^{*},z=z^{*}\right)\ge r\left({\bf{Y}}_{2}^{*}\ \middle | \ \lambda=\lambda^{*},z=z^{*}\right)=r\left({\bf{Y}}_{2}^{*}\right).
\end{equation}
Moreover, the number of users in ${\bf{Y}}_{1}^{*}$ can be upper-bounded as
\begin{equation}\label{eq_optimal_3}
    \begin{aligned}
    r\left({\bf{Y}}_{1}^{*}\right)
    &\le \left|\hat{\mathcal{K}}^{\text{ad}}\big|_{\lambda=\lambda^{*},z=z^{*}}\right|
    = r\left(\hat{{\bf{X}}}\ \middle | \ \lambda=\lambda^{*},z=z^{*}\right).
    \end{aligned}
\end{equation}

Fourth, based on~\eqref{eq_optimal_2} and~\eqref{eq_optimal_3}, we have
\begin{equation}\label{eq_optimal_4}
    \begin{aligned}
        r\left({\bf{Y}}^{*}\right)
        &=r\left({\bf{Y}}_{1}^{*}\right)+r\left({\bf{Y}}_{2}^{*}\right)\\
        &\le r\left(\hat{{\bf{X}}}\ \middle | \ \lambda=\lambda^{*},z=z^{*}\right)+r\left(\hat{{\bf{Y}}}\ \middle | \ \lambda=\lambda^{*},z=z^{*}\right)\\
        &=r\left(\hat{{\bf{X}}}+\hat{{\bf{Y}}}\ \middle | \ \lambda=\lambda^{*},z=z^{*}\right).
    \end{aligned}
\end{equation}
Moreover, from~\eqref{eq_optimal_6} and~\eqref{eq_optimal_4}, we obtain
\begin{equation}\label{eq_optimal_5}
    \begin{aligned}
        r\left({\bf{X}}^{*}+{\bf{Y}}^{*}\right)
        &=r\left({\bf{X}}^{*}\right)+r\left({\bf{Y}}^{*}\right)\\
        &\le r\left(\hat{{\bf{X}}}\ \middle | \ \lambda=\lambda^{*},z=z^{*}\right)\\
        &+r\left(\hat{{\bf{X}}}+\hat{{\bf{Y}}}\ \middle | \ \lambda=\lambda^{*},z=z^{*}\right)\\
        &\le 2r\left(\hat{{\bf{X}}}+\hat{{\bf{Y}}}\ \middle | \ \lambda=\lambda^{*},z=z^{*}\right).
    \end{aligned}
\end{equation}

Finally, from~\eqref{eq_optimal_1} and~\eqref{eq_optimal_5}, we have
\begin{equation}
    \begin{aligned}
    r\left({\bf{X}}^{*}+{\bf{Y}}^{*}\right)
    &\le 2r\left(\hat{{\bf{X}}}+\hat{{\bf{Y}}}\ \middle | \ \lambda=\lambda^{*},z=z^{*}\right)\\
    &\le 2r\left(\dot{{\bf{X}}}+\dot{{\bf{Y}}}\right).
    \end{aligned}
\end{equation}
Equivalently, $r\left(\dot{{\bf{X}}}+\dot{{\bf{Y}}}\right)\ge \frac{1}{2}r\left({\bf{X}}^{*}+{\bf{Y}}^{*}\right)$. This completes the proof.

\section{Proof of Theorem~\ref{theorem_2}}\label{proof_theorem_2}
We first analyze the time complexity of Algorithm~\ref{algorithm_li}. In the worst case, each element in the tuple $\left(L_{1},L_{2},t_{1},\tau_{1}\right)$ takes at most $K+1$ possible values, since each element is enumerated from a set consisting of $0$ and at most $K$ corresponding user-specific parameter values. Therefore, the total number of tuples enumerated in Lines~\ref{line:li_for1_start} and~\ref{line:li_for2_start} is $O\left(K^{4}\right)$. For each fixed tuple, the operations in Line~\ref{line:li_d1} require sorting the candidate users to construct $\Pi_{1}$ and then computing $d_{1}^{*}$ by checking the feasible prefix length. Thus, the complexity of Line~\ref{line:li_d1} is $O\left(K\log K+K\right)=O\left(K\log K\right)$. Therefore, the overall time complexity of Algorithm~\ref{algorithm_li} is $O\left(K^{4}\cdot K\log K\right)=O\left(K^{5}\log K\right)$.

Next, we analyze the time complexity of Algorithm~\ref{algorithm_sd}. Similarly, each element in the tuple $\left(L_{3},L_{4},\tilde{N}_{1},t_{2},\tau_{2}\right)$ takes at most $K+1$ possible values in the worst case. Therefore, the total number of tuples enumerated in Lines~\ref{line:sd_for1_start} and~\ref{line:sd_for2_start} is $O\left(K^{5}\right)$. For each fixed tuple, the complexity of Line~\ref{line:sd_d2} is also $O\left(K\log K+K\right)=O\left(K\log K\right)$. Therefore, the overall time complexity of Algorithm~\ref{algorithm_sd} is $O\left(K^{5}\cdot K\log K\right)=O\left(K^{6}\log K\right)$.

Finally, we analyze the time complexity of Algorithm~\ref{algorithm_all}. The enumeration over the pair $\left(z,\lambda\right)$ in Line~\ref{line:all_for1_start} has time complexity $O\left(\left|{\bf{Z}}\right|\left|\Lambda\right|\right)$, where $\left|{\bf{Z}}\right|$ and $\left|\Lambda\right|$ denote the cardinalities of ${\bf{Z}}$ and $\Lambda$, respectively. For each fixed pair $\left(z,\lambda\right)$, Lines~\ref{line:all_li} and~\ref{line:all_sd} require $O\left(K^{5}\log K\right)$ and $O\left(K^{6}\log K\right)$ time, respectively. Since $\left|{\bf{Z}}\right|$ and $\left|\Lambda\right|$ are constants independent of the problem scale, the overall time complexity of Algorithm~\ref{algorithm_all} is $O\left(\left|{\bf{Z}}\right|\left|\Lambda\right|\left(K^{5}\log K+K^{6}\log K\right)\right)=O\left(K^{6}\log K\right)$. This completes the proof.
\end{appendices}

\end{document}